\documentclass[twocolumn,english,aps,prb,showpacs,superscriptaddress]{revtex4}
\usepackage[T1]{fontenc}
\usepackage[latin9]{inputenc}
\usepackage{mathtools}
\usepackage{bm}
\usepackage{amsmath}
\usepackage{amssymb}
\usepackage{stackrel}
\usepackage{graphicx}

\usepackage{ulem}
\usepackage{float}
\usepackage{lipsum}
\usepackage{babel}
\usepackage{epstopdf}
\setcitestyle{numbers, square, comma}
\usepackage{tensor}

\makeatletter
\newsavebox{\@brx}
\newcommand{\llangle}[1][]{\savebox{\@brx}{\(\m@th{#1\langle}\)}%
  \mathopen{\copy\@brx\kern-0.5\wd\@brx\usebox{\@brx}}}
\newcommand{\rrangle}[1][]{\savebox{\@brx}{\(\m@th{#1\rangle}\)}%
  \mathclose{\copy\@brx\kern-0.5\wd\@brx\usebox{\@brx}}}
\makeatother

\makeatother

\usepackage{babel}
\begin{document}
\title{Closed dynamical recursion equations for correlation functions and
the application on the construction of Liouvillian spectrum in Lindbladian systems }
\author{Xueliang Wang}
\thanks{xlwang@iphy.ac.cn }
\affiliation{Beijing National Laboratory for Condensed Matter Physics, Institute
of Physics, Chinese Academy of Sciences, Beijing 100190, China}
\affiliation{School of Physical Sciences, University of Chinese Academy of Sciences,
Beijing 100049, China }
\author{Shu Chen}
\thanks{schen@iphy.ac.cn }
\affiliation{Beijing National Laboratory for Condensed Matter Physics, Institute
of Physics, Chinese Academy of Sciences, Beijing 100190, China}
\affiliation{School of Physical Sciences, University of Chinese Academy of Sciences,
Beijing 100049, China }
\date{\today}
\begin{abstract}
For an open quantum system described by the Lindblad equation, full characterization of its dynamics typically needs the knowledge of
the Liouvillian spectrum and correlation functions. Solving the Liouvillian spectrum and correlation functions are usually formidable tasks, and most previous studies are constrained
to simple models and lower-order correlations. In this work, we derive a closed form of
dynamical recursion equations for all even-order correlation functions associated with
the quadratic Liouvillian superoperator, thus extending the Wick's theorem
and enabling the reconstruction of the corresponding Liouvillian spectrum.
Furthermore, we study the quartic Liouvillian superoperator,
establishing the necessary and sufficient conditions for the closure
of second-order correlation functions. Building on this, the dynamical
expressions for even-order correlation functions under quadratic dissipation
are derived, culminating in a method for constructing the Liouvillian
spectrum in this extended framework.
\end{abstract}
\maketitle

\section{Introduction}

Open quantum systems have garnered widespread attention as a foundational
topic across diverse fields, including quantum information {[}\citep{D. A. Lidar 2013,M. A. Nielsen 2002}{]},
quantum simulation {[}\citep{F. Verstraete 2009,A. J. Daley 2014}{]},
quantum thermodynamics {[}\citep{F. Binder 2018}{]}, and condensed
matter physics {[}\citep{H. P. Breuer 2002}{]}. These systems exhibit
complex non-equilibrium dynamics due to their interactions with the
environment, offering valuable insights into quantum coherence and
decoherence mechanisms {[}\citep{M. Schlosshauer 2019,W. H. Zurek 2003,A. Rivas 2012}{]},
as well as correlations {[}\citep{H. P. Breuer 2002}{]}. Within the
Markovian approximation, the dynamic evolution process of an open
quantum system is governed by the Lindblad master equation {[}\citep{GKS 1976,G. Lindblad 1975}{]}
with the influence of environment described by dissipation operators.
Among the analytical approaches employed in this context, the examination
of the Liouvillian spectrum {[}\citep{T. Prosen 2008,C. Guo 2017,YKZhang,Prosen2,Prosen3,V. V. Albert 2014}{]} and correlation
functions {[}\citep{H. P. Breuer 2002,R. J. Glauber 1963}{]} plays
a pivotal role. The Liouvillian spectrum provides essential information
regarding the stability, long-term behavior, and dissipative characteristics
of the system {[}\citep{CaiZ,Garrahan,Mori,Znidariac,Ciuti,XFeng,Ueda,Barthel-PRA}{]},
thereby serving as a foundational framework for understanding its
dynamical evolution. Meanwhile, correlation functions capture critical
aspects of time evolution and quantum correlations, making them indispensable
tools for elucidating dynamic processes {[}\citep{P. Kos 2017,A. Asadian 2013,T. Prosen 2008a,P. H. Guimaraes 2016,T. Prosen 2011,T. Prosen 2013,H. Froml 2020,QWQang}{]}.

In some scenarios, second-order correlation functions are particularly
effective in characterizing the coherence and correlations of the system, especially
when the initial state is Gaussian and the dissipation process is
relatively straightforward. Under these conditions, second-order correlation
functions are computationally
manageable and can yield rich dynamical information as the higher-order correlation function can be expressed in terms of second-order correlation functions by applying the Wick's theorem {[}\citep{D. Speed 2023,G. Lindblad Gauss 1976,O. Brodier 2010,YKZhang}{]}. However, as the initial state shifts away from Gaussian
forms or as more complex dissipation processes arise, the limitations
of second-order correlation functions become increasingly evident.
In such situations, the applicability of the Wick's theorem breaks down,
which leads to the calculation of higher-order correlation functions
becoming a formidable task, complicating the capture of complete dynamical behaviors of the system. Consequently, developing feasible methods
for extracting high-order correlation functions is imperative, as
they can unveil rich dynamic features and non-Gaussian statistical
properties, supplementing comprehensive information beyond low-order
correlation functions.

While the computation of higher-order correlation functions is physically
significant for understanding complex quantum systems, the associated
challenges are considerable. In particular, obtaining analytic expressions
for these functions often proves difficult when dealing with non-Gaussian
initial states or intricate quadratic dissipation. Additionally, although
great success has been achieved in diagonalizing the quadratic Lidbladian
systems {[}\citep{T. Prosen 2008,C. Guo 2017,YKZhang,Prosen2,Prosen3,katsura2019prb,katsura2020ptep,GuoC2018,yamanaka,ZhengZY,Prosen2010}{]},
constructing the Liouvillian spectrum of dissipative many-body systems
or Lindbladian systems under complex dissipative conditions poses
substantial difficulties {[}\citep{nakagawa2021prl,popkov2021prl}{]}.
Accurately determining the spectral structure of open quantum systems
beyond the quadratic Lidbladians remains an ongoing challenge.

In this work, we aim to solve analytically high-order correlation
functions for general quadratic Liouvillian systems and some quartic Liouvillian systems. To analytically determine high-order correlation
functions, we derive recursive equations for these functions and demonstrate
that even-order correlation functions in general quadratic Liouvillian
systems satisfy closed recursion relations. This allows high-order
correlation functions to be expressed in terms of second-order correlation
functions and the initial state, effectively extending the Wick's
theorem.  Utilizing the analytical expression of high-order correlation functions also allows us to reconstruct
the Liouvillian spectrum. We then extend our method to the quartic Liouvillian superoperator and derive the necessary and sufficient conditions for the closure of the second-order correlation functions. For these specific quartic Liouvillian systems, we can obtain analytical expression of high-order correlation functions and thus construct the Liouvillian spectrum of the systems, which can not be obtained by using the standard third quantization method {[}\citep{T. Prosen 2008,C. Guo 2017,YKZhang}{]}.
Our work builds a systematic framework for calculating high-order correlation functions of general quadratic Liouvillian systems and quadratic quantum systems with quadratic
dissipations, and lays a firm ground for elucidating the dynamical behaviors in these Lindbladian systems.

The organization of the paper is as follows: In section \ref{sec:Eigen-Decomposition},
we present the eigen decomposition of the quadratic Liouvillian superoperator
for the general quadratic Hamiltonian with linear dissipative operators.
In section \ref{sec:Higher-Order-Correlation}, we derive the
expressions for the even-parity high-order correlation functions
of the general quadratic Liouvillian systems and discuss both Gaussian and non-Gaussian
initial states separately. At the end of this section, we reconstruct
the Liouvillian spectrum using the expressions for the high-order
correlation functions. Section \ref{sec:Quartic-Liouvillian-Superoperator}
extends this to the case of the quartic Liouvillian superoperator. After
providing the necessary and sufficient conditions for the closure
of the second-order correlation functions, we further present the
expressions for the higher-order correlation functions in this context.
Based on this, we construct the Liouvillian spectrum when quadratic
dissipation is added. At the end of this section, we illustrate with
a simple model the counterintuitive phenomena observed when considering
high-order correlations. The derivations and proofs of many conclusions
in the paper are placed in the appendix.

\section{\label{sec:Eigen-Decomposition}Quadratic Liouvillian Superoperator:
Eigen-Decomposition}

\subsection{Model}

In open quantum systems, the general Markovian master equation describing
the evolution of density matrix and preserving trace, hermitian and
positive properties during its evolution is the Lindblad equation {[}\citep{GKS 1976,G. Lindblad 1975}{]},
\begin{equation}
\frac{d\rho}{dt}=-i[H,\rho]+\sum_{\mu}\left(2L_{\mu}\rho L_{\mu}^{\dagger}-\{L_{\mu}^{\dagger}L_{\mu},\rho\}\right),\label{eq:lindblad}
\end{equation}
where $\hbar$ has been selected as 1, $H$ is the system Hamiltonian
and $L_{\mu}$ are the Lindblad dissipative operators that represent
the effect of the environment on the system through channel $\mu$.
Here we consider the general quadratic Hamiltonian and the linear
Lindblad dissipative operators,
\begin{equation}
H=\sum_{i,j=1}^{L}(h_{i,j}a_{i}^{\dagger}a_{j}+\Delta_{i,j}a_{i}a_{j}+\Delta_{j,i}^{*}a_{i}^{\dagger}a_{j}^{\dagger}),\label{eq:H}
\end{equation}
\begin{equation}
L_{\mu}=\sum_{j=1}^{L}(l_{\mu,j}a_{j}+g_{\mu,j}a_{j}^{\dagger}),\label{eq:L}
\end{equation}
where $a_{j}(a_{j}^{\dagger})$ is the fermion annihilation(creation)
operator acting on the site $j$, $L$ stands for lattice length.
In order to ensure the hermitian property of the Hamiltonian, the
coefficients $\Delta_{i,j}$ and $\Delta_{j,i}^{*}$ in the superconducting
part are described with opposite angular labels, and the coefficients
$h_{i,j}$ reflecting the hopping part should satisfy $h^{\dagger}=h$.

With $\{a_{i},a_{j}\}=\{a_{i}^{\dagger},a_{j}^{\dagger}\}=0$, we
have $\sum_{i,j=1}^{L}\Delta_{i,j}a_{i}a_{j}=\sum_{i,j=1}^{L}\frac{1}{2}(\Delta_{i,j}-\Delta_{j,i})a_{i}a_{j}$, $\sum_{i,j=1}^{L}\Delta_{j,i}^{*}a_{i}^{\dagger}a_{j}^{\dagger}=\sum_{i,j=1}^{L}\frac{1}{2}(\Delta_{j,i}^{*}-\Delta_{i,j}^{*})a_{i}^{\dagger}a_{j}^{\dagger}$,
then we can replace $\Delta$ in Eq.(\ref{eq:H}) with $\Delta^{\prime}=\frac{1}{2}(\Delta-\Delta^{T})$.
So we can naturally ask that $\Delta$ in Eq.(\ref{eq:H}) satisfies
$\Delta^{T}=-\Delta$.

\subsection{\label{subsec:Eigen-decomposition}The eigen decomposition of the
Liouvillian superoperator}

Here we introduce the Choi-Jamiolkwski isomorphism {[}\citep{M.-D. Choi,Jamiolkowski 1972,J. E. Tyson 2003,M. Zwolak 2003}{]}
which converts a matrix into a vector: $\rho=\sum_{m,n=1}^{2^{L}}\rho_{mn}\left|m\right\rangle \left\langle n\right|\rightarrow|\rho\rrangle=\sum_{m,n=1}^{2^{L}}\rho_{mn}\left|m\right\rangle \otimes\left(\left\langle n\right|\right)^{T}$.
This means that we select a set of base vectors $\{\left|m\right\rangle \left\langle n\right||m,n=1,2,\cdots,2^{L}\}$
for the fock space composed of operators, from which the operator
can be written as a column vector and the superoperator can be written
as a matrix. Define the superoperators $\hat{b}_{j}=a_{j}\otimes\mathbb{I},\ \ \hat{c}_{j}=\mathbb{I}\otimes a_{j},\ \ j=1,2,\cdots,L$,
where $\mathbb{I}$ is the identity operator, and we stipulate $a_{j}$
are real, then we have
\[
\begin{array}{cc}
a_{i}\rho a_{j}\rightarrow\hat{b}_{i}\hat{c}_{j}^{\dagger}|\rho\rrangle, & a_{i}\rho a_{j}^{\dagger}\rightarrow\hat{b}_{i}\hat{c}_{j}|\rho\rrangle,\\
a_{i}^{\dagger}\rho a_{j}\rightarrow\hat{b}_{i}^{\dagger}\hat{c}_{j}^{\dagger}|\rho\rrangle, & a_{i}^{\dagger}\rho a_{j}^{\dagger}\rightarrow\hat{b}_{i}^{\dagger}\hat{c}_{j}|\rho\rrangle.
\end{array}
\]
It is easy to see that $[\hat{b}_{j},\hat{c}_{j}]=0$, which does
not satisfy the Fermi commutation relation. As mentioned in Reference
{[}\citep{C. Guo 2017}{]}, we use a parity superoperator $\mathcal{\hat{P}}=e^{i\pi\sum_{j=1}^{L}(\hat{b}_{j}^{\dagger}\hat{b}_{j}+\hat{c}_{j}^{\dagger}\hat{c}_{j})}$
to construct the true fermionic superoperator:
\begin{equation}
\begin{array}{cc}
\hat{d}_{j}=\hat{b}_{j}, & \hat{d}_{L+j}=\mathcal{\hat{P}}\hat{c}_{j}\end{array},\ \ j=1,2,\cdots,L,\label{eq:bc2d}
\end{equation}
then we have $\{\mathcal{\hat{P}},\hat{b}_{j}\}=\{\mathcal{\hat{P}},\hat{c}_{j}\}=0$
, so that $\{\hat{d}_{i},\hat{d}_{j}^{\dagger}\}=\delta_{i,j}$, $\{\hat{d}_{i},\hat{d}_{j}\}=0=\{\hat{d}_{i}^{\dagger},\hat{d}_{j}^{\dagger}\}$,
$i,j=1,2,\cdots,2L$. We can rewrite Eq.(\ref{eq:lindblad}) as
\begin{equation}
\frac{d}{dt}|\rho\rrangle=\hat{\mathcal{L}}_{quad}|\rho\rrangle,\label{eq:dt_Lquad}
\end{equation}
\begin{widetext}
\begin{equation}
\begin{array}{rl}
\mathcal{\hat{L}}_{quad}= & -i\stackrel[i,j=1]{L}{\sum}\left\{ \left[A_{i,j}\hat{d}_{i}^{\dagger}\hat{d}_{j}-A_{i,j}^{*}\hat{d}_{L+i}^{\dagger}\hat{d}_{L+j}\right]+\left[(B_{a})_{i,j}\hat{d}_{i}\hat{d}_{j}-(B_{a})_{i,j}^{*}\hat{d}_{L+i}\hat{d}_{L+j}\right]+\left[(\widetilde{B}_{a})_{i,j}\hat{d}_{i}^{\dagger}\hat{d}_{j}^{\dagger}-(\widetilde{B}_{a})_{i,j}^{*}\hat{d}_{L+i}^{\dagger}\hat{d}_{L+j}^{\dagger}\right]\right\} \\
 & +2\stackrel[i,j=1]{L}{\sum}\left[-(M_{c})_{j,i}\hat{d}_{i}^{\dagger}\hat{d}_{L+j}+(M_{c})_{i,j}^{*}\hat{d}_{i}\hat{d}_{L+j}^{\dagger}-(M_{l})_{j,i}\hat{d}_{i}\hat{d}_{L+j}+(M_{g})_{j,i}\hat{d}_{i}^{\dagger}\hat{d}_{L+j}^{\dagger}\right]\mathcal{\hat{P}}-2Tr(M_{g}),
\end{array}\label{eq:Lquad}
\end{equation}
\end{widetext} where $(M_{l})_{i,j}=\sum_{\mu}l_{\mu,i}^{*}l_{\mu,j}$,
$(M_{g})_{i,j}=\sum_{\mu}g_{\mu,i}^{*}g_{\mu,j}$, $(M_{c})_{i,j}=\sum_{\mu}l_{\mu,i}^{*}g_{\mu,j}$,
and $A=h-i(M_{l}-M_{g}^{T})$, $B_{a}=\Delta+\frac{i}{2}(M_{c}^{*}-M_{c}^{\dagger})$,
$\widetilde{B}_{a}=\Delta^{\dagger}+\frac{i}{2}(M_{c}^{T}-M_{c})$.
Here, we observe that $\mathcal{\hat{L}}_{quad}$ is a quadratic representation
with respect to the fermionic superoperators $\hat{d}_{j}$ and $\hat{d}_{j}^{\dagger}$,
and we refer to such an $\mathcal{\hat{L}}_{quad}$ as a quadratic
Liouvillian superoperator.

It is easy to know $\mathcal{\hat{P}}=e^{i\pi\sum_{j=1}^{L}(\hat{b}_{j}^{\dagger}\hat{b}_{j}+\hat{c}_{j}^{\dagger}\hat{c}_{j})}=e^{i\pi\sum_{j=1}^{2L}\hat{d}_{j}^{\dagger}\hat{d}_{j}}$
and $[\hat{\mathcal{L}}_{quad},\hat{\mathcal{P}}]=0$. Defining the
projection superoperator $\hat{\mathbb{P}}_{p}$ projected to odd
parity($p=-1$) and even parity($p=1$), then we can replace the parity
superoperator $\mathcal{\hat{P}}$ with its eigenvalue $p=\pm1$,
\begin{equation}
\hat{\mathcal{L}}_{quad}=\hat{\mathbb{P}}_{-1}\hat{\mathcal{L}}_{quad}|_{\mathcal{\hat{P}}\rightarrow-1}\hat{\mathbb{P}}_{-1}+\hat{\mathbb{P}}_{1}\hat{\mathcal{L}}_{quad}|_{\mathcal{\hat{P}}\rightarrow1}\hat{\mathbb{P}}_{1}.
\end{equation}

To simplify the representation of the Liouvillian eigen decomposition, we transform
it to Majorana Fermion subspace. Defining
\begin{equation}
\begin{array}{ll}
\hat{f}_{j}=\frac{1}{\sqrt{2}}(\hat{d}_{j}+\hat{d}_{j}^{\dagger}), & \hat{f}_{L+j}=\frac{i}{\sqrt{2}}(\hat{d}_{j}-\hat{d}_{j}^{\dagger}),\\
\hat{f}_{2L+j}=\frac{1}{\sqrt{2}}(\hat{d}_{L+j}+\hat{d}_{L+j}^{\dagger}), & \hat{f}_{3L+j}=\frac{i}{\sqrt{2}}(\hat{d}_{L+j}-\hat{d}_{L+j}^{\dagger})
\end{array}
\end{equation}
and labeling $\widehat{\mathbf{f}}=[\hat{f}_{1},\hat{f}_{2},\cdots,\hat{f}_{4L}]$,
we have
\[
\{\hat{f}_{i},\hat{f}_{j}\}=\delta_{i,j},
\]
\begin{equation}
\begin{array}{cl}
\mathcal{\hat{L}}_{quad} & =\underset{p=1,-1}{\sum}\mathbb{\hat{P}}_{p}\mathbf{\widehat{f}}F_{p}\widehat{\mathbf{f}}^{T}\mathbb{\hat{P}}_{p}+f_{0}\end{array}\label{eq:L_f}
\end{equation}
with
\begin{equation}
\begin{array}{cc}
F_{p}=\frac{1}{2}\left[\begin{array}{cc|cc}
F_{11} & F_{12} & F_{13} & F_{14}\\
-F_{12}^{T} & F_{22} & F_{23} & -F_{13}^{\dagger}\\
\hline -F_{13}^{T} & F_{23}^{*} & F_{22}^{*} & -F_{12}^{\dagger}\\
F_{14}^{*} & F_{13}^{*} & F_{12}^{*} & F_{11}^{*}
\end{array}\right], & f_{0}=-Tr(M_{l}+M_{g}),\end{array}\label{Fp}
\end{equation}
where $F_{11}=-\frac{i}{2}(A-A^{T})-i(B_{a}+\widetilde{B}_{a})$,
$F_{12}=-\frac{1}{2}(A+A^{T})-(B_{a}-\widetilde{B}_{a})$, $F_{22}=-\frac{i}{2}(A-A^{T})+i(B_{a}+\widetilde{B}_{a})$,
$F_{13}=p(M_{g}^{T}-M_{l}^{T}+M_{c}^{*}-M_{c}^{T})$, $F_{14}=ip(M_{g}^{T}+M_{l}^{T}+M_{c}^{*}+M_{c}^{T})$,
$F_{23}=ip(M_{g}^{T}+M_{l}^{T}-M_{c}^{*}-M_{c}^{T})$.

It is evident that $F_{p}$ is an antisymmetric matrix, i.e., $F_{p}^{T}=-F_{p}$.
If we have $F_{p}\left|\xi_{j,p}\right\rangle =\alpha_{j,p}\left|\xi_{j,p}\right\rangle $,
then we have $\left(\left|\xi_{j,p}\right\rangle \right){}^{T}F_{p}=\left(F_{p}^{T}\left|\xi_{j,p}\right\rangle \right)^{T}=-\alpha_{j,p}\left(\left|\xi_{j,p}\right\rangle \right){}^{T}$.
The eigenvalues of $F_{p}$ have been paired with $\alpha_{j,p}$
and $-\alpha_{j,p}$. We can label the set of eigenvalues of $F_{p}$
and the corresponding set of left and right eigenvectors as
\begin{equation}
\{\alpha_{1,p},\alpha_{2,p},\cdots,\alpha_{2L,p},-\alpha_{1,p},-\alpha_{2,p},\cdots,-\alpha_{2L,p}\},\label{eq:alpha}
\end{equation}
\[
\left\{ \left\langle \zeta_{1,p}\right|,\left\langle \zeta_{2,p}\right|,\cdots,\left\langle \zeta_{4L,p}\right|\right\} ,\ \ \left\{ \left|\xi_{1,p}\right\rangle ,\left|\xi_{2,p}\right\rangle ,\cdots,\left|\xi_{4L,p}\right\rangle \right\} ,
\]
then we have $\left\langle \zeta_{j,p}\right|=\left(\left|\xi_{2L+j,p}\right\rangle \right)^{T},\ \ \left\langle \zeta_{2L+j,p}\right|=\left(\left|\xi_{j,p}\right\rangle \right)^{T},\ \ j=1,2,\cdots,2L$.

If we specify the orthogonal conditions for left and right eigenvectors
as $\left\langle \zeta_{m,p}\right.\left|\xi_{n,p}\right\rangle =\delta_{mn}$,
then we have
\begin{equation}
\Omega_{p}^{T}\Omega_{p}=\left[\begin{array}{cc}
 & I_{2L}\\
I_{2L}
\end{array}\right]\label{eq:Psi}
\end{equation}
with $\Omega_{p}=\left[\left|\xi_{1,p}\right\rangle ,\left|\xi_{2,p}\right\rangle ,\cdots,\left|\xi_{4L,p}\right\rangle \right]$.
Defining $\Lambda_{p}=Diag(\alpha_{1,p},\alpha_{2,p},\cdots,\alpha_{2L,p})$,
we can write down the eigendecomposition of $F_{p}$
\begin{equation}
F_{p}=\Omega_{p}\left[\begin{array}{cc}
\Lambda_{p}\\
 & -\Lambda_{p}
\end{array}\right]\Omega_{p}^{-1}. \label{eq:F_eig}
\end{equation}
Combined with Eq.(\ref{eq:Psi}), we have
\begin{equation}
F_{p}=\Omega_{p}\left[\begin{array}{cc}
 & \Lambda_{p}\\
-\Lambda_{p}
\end{array}\right]\Omega_{p}^{T},
\end{equation}
\begin{equation}
\mathcal{\hat{L}}_{quad}=\sum_{p=1,-1}\mathbb{\hat{P}}_{p}\mathbf{\widehat{f}}\Omega_{p}\left[\begin{array}{cc}
 & \Lambda_{p}\\
-\Lambda_{p}
\end{array}\right]\Omega_{p}^{T}\widehat{\mathbf{f}}^{T}\mathbb{\hat{P}}_{p}+f_{0}.
\end{equation}
If we label $\left[\overbrace{\mathbf{\hat{g}}_{p},\mathbf{\hat{g}}_{p}^{\prime}}\right]=[\hat{g}_{1,p},\hat{g}_{2,p},\cdots,\hat{g}_{2L,p},\hat{g}_{1,p}^{\prime},\hat{g}_{2,p}^{\prime},\cdots,\hat{g}_{2L,p}^{\prime}]=\widehat{\mathbf{f}}\Omega_{p}$,
the superoperators $\hat{g}_{i,p}$ and $\hat{g}_{j,p}^{\prime}$
will satisfy the Fermion commutation relation
\begin{equation}
	\begin{array}{l}
		\begin{array}{cl}
			\{\hat{g}_{i,p},\hat{g}_{j,p}^{\prime}\} & =\stackrel[k=1]{2L}{\sum}(\Omega_{p})_{k,i}(\Omega_{p})_{k,2L+j}=(\Omega_{p}^{T}\Omega_{p})_{i,2L+j}=\delta_{ij},\end{array}\\
		\begin{array}{cl}
			\{\hat{g}_{i,p},\hat{g}_{j,p}\} & =\stackrel[k=1]{2L}{\sum}(\Omega_{p})_{k,i}(\Omega_{p})_{k,j}=(\Omega_{p}^{T}\Omega_{p})_{i,j}=0,\end{array}\\
		\begin{array}{cl}
			\{\hat{g}_{i,p}^{\prime},\hat{g}_{j,p}^{\prime}\} & =\stackrel[k=1]{2L}{\sum}(\Omega_{p})_{k,2L+i}(\Omega_{p})_{k,2L+j}=(\Omega_{p}^{T}\Omega_{p})_{2L+i,2L+j}\\
			& =0,
		\end{array}
	\end{array}
\end{equation}
the Liouvillian superoperator $\mathcal{\hat{L}}_{quad}$ will be
rewritten as
\begin{equation}
\begin{array}{cl}
\mathcal{\hat{L}}_{quad} & =\underset{p=\pm1}{\sum}\stackrel[j=1]{2L}{\sum}\alpha_{j,p}\mathbb{\hat{P}}_{p}(\hat{g}_{j,p}\hat{g}_{j,p}^{\prime}-\hat{g}_{j,p}^{\prime}\hat{g}_{j,p})\mathbb{\hat{P}}_{p}+f_{0}\\
 & =-2\underset{p=\pm1}{\sum}\stackrel[j=1]{2L}{\sum}\alpha_{j,p}\mathbb{\hat{P}}_{p}\hat{g}_{j,p}^{\prime}\hat{g}_{j,p}\mathbb{\hat{P}}_{p}+\stackrel[j=1]{2L}{\sum}\alpha_{j,p}+f_{0}\\
 & =-2\underset{p=\pm1}{\sum}\stackrel[j=1]{2L}{\sum}\alpha_{j}\mathbb{\hat{P}}_{p}\hat{g}_{j,p}^{\prime}\hat{g}_{j,p}\mathbb{\hat{P}}_{p},
\end{array},\label{eq:L_p}
\end{equation}
the last equal sign holds because $\stackrel[j=1]{2L}{\sum}\alpha_{j,p}=Tr(X)=Tr(M_{l}+M_{g})=-f_{0}$
and $\alpha_{j,1}=\alpha_{j,-1}=\alpha_{j}$, which can be obtained
by means of Eq.(\ref{eq:F_XY},\ref{eq:X define}).

It is easy to know the relationship between fermion operators $\hat{d}_{j},\ \ \hat{d}_{j}^{\dagger}$
and  $\hat{g}_{j,p},\ \ \hat{g}_{j,p}^{\prime}$ as follows
\begin{equation}
\begin{array}{cl}
\left[\overbrace{\mathbf{\hat{g}}_{p},\mathbf{\hat{g}}_{p}^{\prime}}\right] & =[\hat{g}_{1,p},\hat{g}_{2,p},\cdots,\hat{g}_{2L,p},\hat{g}_{1,p}^{\prime},\hat{g}_{2,p}^{\prime},\cdots,\hat{g}_{2L,p}^{\prime}]\\
 & =[\hat{d}_{1},\hat{d}_{2},\cdots,\hat{d}_{2L},\hat{d}_{1}^{\dagger},\hat{d}_{2}^{\dagger},\cdots,\hat{d}_{2L}^{\dagger}]S_{p}\\
 & =\left[\overbrace{\mathbf{\hat{d}},\mathbf{\hat{d}^{\dagger}}}\right]S_{p}
\end{array}\label{eq:d2g}
\end{equation}
with
\[
S_{p}=\frac{1}{\sqrt{2}}\left[\begin{array}{cc|cc}
I_{L} & iI_{L}\\
 &  & I_{L} & iI_{L}\\
\hline I_{L} & -iI_{L}\\
 &  & I_{L} & -iI_{L}
\end{array}\right]\Omega_{p}.
\]

We know that in the evolution process described by Eq.(\ref{eq:lindblad}),
the trace of the density matrix remains unchanged, that is, $Tr(\rho)=\llangle\mathbb{I}|\rho\rrangle=1$
with $\llangle\mathbb{I}|=\left(|\mathbb{I}\rrangle\right)^{T}$,
so $0=\frac{d}{dt}\llangle\mathbb{I}|\rho\rrangle=\llangle\mathbb{I}|\mathcal{\hat{L}}|\rho\rrangle$,
which indicates that the Liouvillian superoperator has a left eigenvector
$\llangle\mathbb{I}|$, corresponding to the eigenvalue 0. We
mark its corresponding right eigenvector as $|0_{\mathcal{L}}\rrangle$ and label $|\Theta_{R,\overrightarrow{\nu},p}\rrangle=\stackrel[s=1]{2L}{\overrightarrow{\prod}}\hat{g}_{s,p}^{\prime\nu_{s}}|0_{\mathcal{L}}\rrangle\coloneqq\hat{g}_{1,p}^{\prime\nu_{1}}\hat{g}_{2,p}^{\prime\nu_{2}}\cdots\hat{g}_{2L,p}^{\prime\nu_{2L}}|0_{\mathcal{L}}\rrangle$,
$\llangle\Theta_{L,\overrightarrow{\nu},p}|=\llangle\mathbb{I}|\stackrel[s=1]{2L}{\overleftarrow{\prod}}\hat{g}_{s,p}^{\nu_{s}}\coloneqq\llangle\mathbb{I}|\hat{g}_{2L,p}^{\nu_{2L}}\hat{g}_{2L-1,p}^{\nu_{2L-1}}\cdots\hat{g}_{1,p}^{\nu_{1}}$,
$\nu_{s}\in\{0,1\}$, it is easy to know $\llangle\Theta_{L,\overrightarrow{\nu},p}|\Theta_{R,\overrightarrow{\nu}^{\prime},p}\rrangle=\delta_{\overrightarrow{\nu},\overrightarrow{\nu}^{\prime}}$.
Using Eq.(\ref{eq:d2g}), it follows that $\{\mathcal{\hat{P}},\hat{g}_{j}\}=\{\mathcal{\hat{P}},\hat{g}_{j}^{\prime}\}=0$.
Considering that $\llangle\mathbb{I}|\mathcal{\hat{P}}=\llangle\mathbb{I}|$
(proof in the Appendix \ref{subsec:Proof IP_I}), we thus have $\llangle\Theta_{L,\overrightarrow{\nu},p}|\mathcal{\hat{P}}=p_{\overrightarrow{\nu}}\llangle\Theta_{L,\overrightarrow{\nu},p}|$,
$\mathcal{\hat{P}}|\Theta_{R,\overrightarrow{\nu},p}\rrangle=p_{\overrightarrow{\nu}}|\Theta_{R,\overrightarrow{\nu},p}\rrangle$
with $p_{\overrightarrow{\nu}}=(-1)^{\sum_{s=1}^{2L}\nu_{s}}$. From
this, we deduce that $\hat{\mathbb{P}}_{p}|\Theta_{R,\overrightarrow{\nu},p^{\prime}}\rrangle=p\delta_{p,p_{\overrightarrow{\nu}}}*|\Theta_{R,\overrightarrow{\nu},p^{\prime}}\rrangle$,
$\llangle\Theta_{L,\overrightarrow{\nu},p^{\prime}}|\hat{\mathbb{P}}_{p}=p\delta_{p,p_{\overrightarrow{\nu}}}*\llangle\Theta_{L,\overrightarrow{\nu},p^{\prime}}|$.
By inserting the identity superoperator $\hat{\mathbb{I}}=\underset{\overrightarrow{\nu}}{\sum}|\Theta_{R,\overrightarrow{\nu},p}\rrangle\llangle\Theta_{L,\overrightarrow{\nu},p}|$
into Eq.(\ref{eq:L_p}), we obtain the eigen decomposition of the
quadratic Liouvillian superoperator as
\begin{equation}
\begin{array}{rl}
\mathcal{\hat{L}}_{quad}= & -2\underset{p=\pm1}{\sum}\mathbb{\hat{P}}_{p}\left(\underset{\overrightarrow{\nu}}{\sum}|\Theta_{R,\overrightarrow{\nu},p}\rrangle\llangle\Theta_{L,\overrightarrow{\nu},p}|\right)\\
 & *\left(\stackrel[j=1]{2L}{\sum}\alpha_{j}\hat{g}_{j,p}^{\prime}\hat{g}_{j,p}\right)\left(\underset{\overrightarrow{\nu}}{\sum}|\Theta_{R,\overrightarrow{\nu},p}\rrangle\llangle\Theta_{L,\overrightarrow{\nu},p}|\right)\mathbb{\hat{P}}_{p}\\
= & \underset{\overrightarrow{\nu}}{\sum}\eta_{\overrightarrow{\nu}}|\Theta_{R,\overrightarrow{\nu},p_{\overrightarrow{\nu}}}\rrangle\llangle\Theta_{L,\overrightarrow{\nu},p_{\overrightarrow{\nu}}}|.
\end{array}
\end{equation}
Here, $\eta_{\overrightarrow{\nu}}=-2\sum_{s=1}^{2L}\nu_{s}\alpha_{s}$
with $\nu_{s}\in\{0,1\}$ represents the Liouvillian spectrum, $p_{\overrightarrow{\nu}}=(-1)^{\sum_{s=1}^{2L}\nu_{s}}=1$
corresponds to even parity and $p_{\overrightarrow{\nu}}=-1$ corresponds
to odd parity.

\subsection{Diagonalization of the matrix $F_{p}$}

In the above derivation, we have assumed that the eigen decomposition
of $F_{p}$ is given by Eq.(\ref{eq:F_eig}). Here, we will provide
a detailed explanation of this assumption. See Eq.(\ref{Fp}) for
the specific form of matrix $F_{p}$, and it is easy to know that
\begin{equation}
\varOmega_{1,p}^{-1}F_{p}\varOmega_{1,p}=\left[\begin{array}{cc}
X\\
-pY & -X^{T}
\end{array}\right]\label{eq:F_XY}
\end{equation}
with
\begin{equation}
\begin{array}{cc}
X=\left[\begin{array}{cc}
X_{1} & iX_{2}\\
-iX_{2}^{*} & X_{1}^{*}
\end{array}\right], & \left\{ \begin{array}{l}
X_{1}=\frac{1}{2}(M_{l}+M_{g}^{T}-ih),\\
X_{2}=\frac{1}{2}(M_{c}+M_{c}^{T})+i\Delta^{*},
\end{array}\right.\end{array}\label{eq:X define}
\end{equation}
\begin{equation}
\begin{array}{cc}
Y=\left[\begin{array}{cc}
-iM_{c}^{*} & M_{l}^{T}\\
M_{g}^{T} & iM_{c}^{T}
\end{array}\right], & \varOmega_{1,p}=\end{array}\left[\begin{array}{cc|cc}
0 & 0 & ip & p\\
0 & 0 & p & ip\\
\hline -p & ip & i & -1\\
ip & -p & -1 & i
\end{array}\right]\otimes I_{L}.\label{eq:Y define}
\end{equation}

Combined with Eq.(\ref{eq:alpha}), we can see that the eigenvalues
of $X$ are $\{\alpha_{1},\alpha_{2},\cdots,\alpha_{2L}\}$, which
have nothing to do with what $p$ is. According to Reference {[}\citep{T. Prosen 2008}{]},
we can refer to it as the \textquotedbl rapid spectrum\textquotedbl .
In the next section, we will demonstrate the existence of $Z$ such
that $ZX+X^{T}Z=Y$. Meanwhile, we assume that $X$ is diagonalizable
and provide its eigen-decomposition $X\varOmega=\varOmega\Lambda$,
$\Lambda=Diag(\alpha_{1},\alpha_{2},\cdots,\alpha_{2L})$. Defining
\begin{equation}
\begin{array}{cc}
\varOmega_{2,p}=\left[\begin{array}{cc}
I_{2L}\\
-pZ & I_{2L}
\end{array}\right], & \varOmega_{3}=\left[\begin{array}{cc}
\varOmega\\
 & (\varOmega^{T}){}^{-1}
\end{array}\right],\end{array}
\end{equation}
then we have $F_{p}(\varOmega_{1,p}\varOmega_{2,p}\varOmega_{3})=(\varOmega_{1,p}\varOmega_{2,p}\varOmega_{3})\left[\begin{array}{cc}
\Lambda\\
 & -\Lambda
\end{array}\right]$ and this means $\Omega_{p}=\varOmega_{1,p}\varOmega_{2,p}\varOmega_{3}$.

\section{\label{sec:Higher-Order-Correlation}High-Order Correlation Functions}

In the above section, we have presented the eigen decomposition of the Liouvillian superoperator.
Now, we explore the dynamical evolution of the system from a
different perspective, namely through the correlation functions. When
considering the dynamical behavior of the expectation value of any
operator $O$ with $\left\langle O\right\rangle _{t}=Tr[O\rho(t)]=\llangle\mathbb{I}|O\otimes\mathbb{I}|\rho(t)\rrangle$,
using $\llangle\mathbb{I}|\mathcal{\hat{L}}=0$, we deduce that
\begin{equation}
\begin{array}{cl}
\frac{d}{dt}\left\langle O\right\rangle _{t} & =\llangle\mathbb{I}|(O\otimes\mathbb{I})\mathcal{\hat{L}}|\rho(t)\rrangle\\
 & =\llangle\mathbb{I}|\left[(O\otimes\mathbb{I}),\mathcal{\hat{L}}\right]|\rho(t)\rrangle.
\end{array}\label{eq:DO_Dt}
\end{equation}
Following the previous section, we still keep $\mathcal{\hat{L}}$
as quadratic. By choosing $O=\stackrel[l=1]{n}{\overrightarrow{\prod}}\omega_{j_{l}}$,
$j_{l}\in\{1,2,\cdots,2L\}$ with $\omega_{s}=\frac{1}{\sqrt{2}}(a_{s}+a_{s}^{\dagger})$,
$\omega_{L+s}=\frac{i}{\sqrt{2}}(a_{s}-a_{s}^{\dagger})$, $s\in\{1,2,\cdots,L\}$,
we obtain the dynamical equation for the $n$-th order correlation
function $\left\langle \stackrel[l=1]{n}{\overrightarrow{\prod}}\omega_{j_{l}}\right\rangle _{t}=Tr\left[\left(\stackrel[l=1]{n}{\overrightarrow{\prod}}\omega_{j_{l}}\right)\rho(t)\right]$,
which satisfies $\frac{d}{dt}\left\langle \stackrel[l=1]{n}{\overrightarrow{\prod}}\omega_{j_{l}}\right\rangle _{t}=\llangle\mathbb{I}|\left[\stackrel[l=1]{n}{\overrightarrow{\prod}}\hat{f}_{j_{l}},\mathcal{\hat{L}}\right]|\rho(t)\rrangle$.
For simplicity in notation, we define $T_{j_{1},j_{2},\cdots,j_{n}}\coloneqq\llangle\mathbb{I}|\stackrel[l=1]{n}{\overrightarrow{\prod}}\hat{f}_{j_{l}}|\rho(t)\rrangle$,
clearly, when $j_{l}\in\{1,2,\cdots,2L\}$, it represents the $n$-th
order correlation function $\left\langle \stackrel[l=1]{n}{\overrightarrow{\prod}}\omega_{j_{l}}\right\rangle _{t}$.

Using Eq.(\ref{eq:L_f}) and the properties $\llangle\mathbb{I}|\mathcal{\hat{P}}=\llangle\mathbb{I}|$,
$\llangle\mathbb{I}|\left(\stackrel[l=1]{n}{\overrightarrow{\prod}}\hat{f}_{j_{l}}\right)\mathcal{\hat{P}}=(-1)^{n}\llangle\mathbb{I}|\left(\stackrel[l=1]{n}{\overrightarrow{\prod}}\hat{f}_{j_{l}}\right)$,
when $n$ is odd, we have
\[
\begin{array}{rl}
\frac{d}{dt}T_{j_{1},\cdots,j_{n}}= & \llangle\mathbb{I}|\left(\stackrel[l=1]{n}{\overrightarrow{\prod}}\hat{f}_{j_{l}}\right)\left(\mathbf{\widehat{f}}F_{-}\widehat{\mathbf{f}}^{T}\right)|\rho(t)\rrangle\\
 & -\llangle\mathbb{I}|\left(\mathbf{\widehat{f}}F_{+}\widehat{\mathbf{f}}^{T}\right)\left(\stackrel[l=1]{n}{\overrightarrow{\prod}}\hat{f}_{j_{l}}\right)|\rho(t)\rrangle.
\end{array}
\]
Here, we denote $F_{p}$ as $F_{+}$ when $p=1$ (and as $F_{-}$
when $p=-1$). It is obvious that the dynamic evolution of lower odd-order
correlation functions is associated with higher odd-order correlation
functions, so we can say odd-order correlation functions are generally
not closed unless appropriate Hamiltonian and dissipation operators
are selected and this is consistent with the prediction in Reference
{[}\citep{B. Zunkovic 2014}{]}. Similarly, when $n=2m$ is an even
number, we obtain
\begin{equation}
\frac{d}{dt}T_{j_{1},\cdots,j_{2m}}=\llangle\mathbb{I}|\left[\stackrel[l=1]{2m}{\overrightarrow{\prod}}\hat{f}_{j_{l}},\left(\mathbf{\widehat{f}}F_{+}\widehat{\mathbf{f}}^{T}\right)\right]|\rho(t)\rrangle.
\end{equation}
Since $\{\hat{f}_{i},\hat{f}_{j}\}=\delta_{i,j}$, $F_{+}^{T}=-F_{+}$,
we have
\[
\begin{array}{rl}
 & \left[\stackrel[l=1]{n}{\overrightarrow{\prod}}\hat{f}_{j_{l}},\mathbf{\widehat{f}}F_{+}\widehat{\mathbf{f}}^{T}\right]\\
= & 2\stackrel[k=1]{n}{\sum}\stackrel[s=1]{4L}{\sum}\left[\left(F_{+}\right)_{j_{k},s}(\stackrel[l=1]{k-1}{\overrightarrow{\prod}}\hat{f}_{j_{l}})\hat{f}_{s}(\stackrel[l=k+1]{n}{\overrightarrow{\prod}}\hat{f}_{j_{l}})\right],
\end{array}
\]
so we have
\begin{equation}
\frac{d}{dt}T_{j_{1},\cdots,j_{2m}}=2\sum_{k=1}^{2m}\sum_{s=1}^{4L}(F_{+})_{j_{k},s}T_{j_{1},\cdots,j_{k-1},s,j_{k+1},\cdots,j_{2m}}.
\end{equation}
It is evident that even-order correlation functions are closed, meaning
the evolution of the $2m$-th order correlation function only depends
on the $2m$-th, $(2m-2)$-th, ..., $0$-th order correlation functions.
It is easy to see that $\hat{f}_{j}$ has the following properties:
\begin{equation}
\begin{array}{ccc}
\llangle\mathbb{I}|\hat{f}_{2L+j}=i\llangle\mathbb{I}|\hat{f}_{L+j}, & \llangle\mathbb{I}|\hat{f}_{3L+j}=i\llangle\mathbb{I}|\hat{f}_{j}, & j=1,2,\cdots,L.\end{array}\label{eq:f2f}
\end{equation}
If we specify that $j_{l}\in\{1,2,\cdots,2L\}$, $s\in\{2L+1,2L+2,\cdots,4L\}$,
then we have
\begin{equation}
\begin{array}{rl}
 & T_{j_{1},\cdots,j_{k-1},s,j_{k+1},\cdots,j_{2m}}\\
= & (-1)^{k-1}T_{s,j_{1},\cdots,j_{k-1},j_{k+1},\cdots,j_{2m}}\\
= & (-1)^{k-1}iT_{s^{\prime},j_{1},\cdots,j_{k-1},j_{k+1},\cdots,j_{2m}}
\end{array}
\end{equation}
with $s^{\prime}=s-\frac{3}{2}L+(-1)^{\lfloor\frac{s}{L}\rfloor}*\frac{L}{2}$.
So 
\begin{equation}
\begin{array}{rl}
 & \frac{d}{dt}T_{j_{1},\cdots,j_{2m}}\\
= & 2\stackrel[k=1]{2m}{\sum}\left[\stackrel[s=1]{2L}{\sum}\left(F_{+}\right)_{j_{k},s}T_{j_{1},\cdots,j_{k-1},s,j_{k+1},\cdots,j_{2m}}\right.\\
 & \left.+\stackrel[s=2L+1]{4L}{\sum}\left(F_{+}\right)_{j_{k},s}T_{j_{1},\cdots,j_{k-1},s,j_{k+1},\cdots,j_{2m}}\right]\\
= & \stackrel[k=1]{2m}{\sum}\stackrel[s=1]{2L}{\sum}\left[\left(F_{A}\right)_{j_{k},s}T_{j_{1},\cdots,j_{k-1},s,j_{k+1},\cdots,j_{2m}}\right.\\
 & \left.+(-1)^{k-1}i\left(F_{B}\right)_{j_{k},s}T_{s,j_{1},\cdots,j_{k-1},j_{k+1},\cdots,j_{2m}}\right],
\end{array}\label{eq:dT_j}
\end{equation}
where
\begin{equation}
\begin{array}{cc}
F_{A}=\left[\begin{array}{cc}
F_{11} & F_{12}\\
-F_{12}^{T} & F_{22}
\end{array}\right], & F_{B}=\left.\left[\begin{array}{cc}
F_{14} & F_{13}\\
-F_{13}^{\dagger} & F_{23}
\end{array}\right]\right|_{p=1}\end{array}.
\end{equation}
Combining the definition of $X$, $Y$ by Eqs.(\ref{eq:X define},
\ref{eq:Y define}), we can get
\[
F_{A}+iF_{B}=-2Q^{-1}X^{T}Q,\ F_{B}=2iQ^{-1}Y\left[\begin{array}{cc}
 & I_{L}\\
I_{L}
\end{array}\right]Q
\]
with $Q=\frac{1}{\sqrt{2}}\left[\begin{array}{cc}
I_{L} & iI_{L}\\
iI_{L} & I_{L}
\end{array}\right]$.

We think of the sequence $(j_{1},j_{2},\cdots,j_{2m})$ as a $2m$-bit
$2L$-base number and arrange it in ascending order as a column vector
\begin{equation}
\boldsymbol{\left|T_{2m,t}\right\rangle }=\left[\begin{array}{c}
T_{1,\cdots,1,1}\\
\vdots\\
T_{1,\cdots,1,2L}\\
T_{1,\cdots,2,1}\\
\vdots\\
T_{1,\cdots,2,2L}\\
\vdots\\
T_{2L,\cdots,2L}
\end{array}\right],
\end{equation}
where each element represents a specific $2m$-th order correlation
function, for example, $T_{1,\cdots,1,2L}=\llangle\mathbb{I}|\hat{f}_{1}\cdots\hat{f}_{1}\hat{f}_{2L}|\rho(t)\rrangle=Tr\left[\omega_{1}\cdots\omega_{1}\omega_{2L}\rho(t)\right]$.
Given the $2L*1$-dimensional standard basis vector $\overrightarrow{e}_{j}$, which satisfies $(\overrightarrow{e}_{j})_{k}=\delta_{j,k}$, we
define the swap matrix $R=\sum_{j,k=1}^{2L}(\overrightarrow{e}_{j}\overrightarrow{e}_{k}^{T})\otimes(\overrightarrow{e}_{k}\overrightarrow{e}_{j}^{T})$.
Clearly, for any $2L*1$-dimensional column vectors $\left|\psi_{1}\right\rangle $,
$\left|\psi_{2}\right\rangle $ and any $2L$-dimensional matrices
$M_{A}$, $M_{B}$, it holds that
\begin{equation}
\left\{ \begin{array}{c}
R\left(\left|\psi_{1}\right\rangle \otimes\left|\psi_{2}\right\rangle \right)=\left|\psi_{2}\right\rangle \otimes\left|\psi_{1}\right\rangle ,\\
R(M_{A}\otimes M_{B})=(M_{B}\otimes M_{A})R.
\end{array}\right.\label{eq:R}
\end{equation}
Based on this, by defining $R_{k}=\stackrel[l=0]{k-2}{\overrightarrow{\prod}}\left(\mathbf{I}_{k-2-l}\otimes R\otimes\mathbf{I}_{l}\right)$
with $\mathbf{I}_{l}=I_{(2L)^{l}}$, we consequently obtain
\begin{equation}
\begin{array}{rl}
 & \boldsymbol{\left|T_{2m,t}\right\rangle }_{j_{k},j_{1},\cdots,j_{k-1},j_{k+1},\cdots,j_{2m}}\\
= & \left(\left[R_{k}\otimes\mathbf{I}_{2m-k}\right]\boldsymbol{\left|T_{2m,t}\right\rangle }\right)_{j_{1},\cdots,j_{2m}}.
\end{array}\label{eq:prop of Rk}
\end{equation}
Then Eq.(\ref{eq:dT_j}) can be rewritten as
\begin{equation}
\frac{d}{dt}\boldsymbol{\left|T_{2m,t}\right\rangle }=\mathbf{D}_{2m}\boldsymbol{\left|T_{2m,t}\right\rangle }
\end{equation}
with
\begin{equation}
\begin{array}{rl}
\mathbf{D}_{2m} & =\stackrel[k=1]{2m}{\sum}\left[\mathbf{I}_{k-1}\otimes F_{A}-(-1)^{k}i(\mathbf{I}_{k-1}\otimes F_{B})R_{k}\right]\otimes\mathbf{I}_{n-k}\\
 & =\mathbf{F}_{2m}+\mathbf{F}_{2m}^{(B)},
\end{array}
\end{equation}
\begin{equation}
	\left\{ \begin{array}{l}
		\mathbf{F}_{n}=\stackrel[k=1]{n}{\sum}\mathbf{I}_{k-1}\otimes(F_{A}+iF_{B})\otimes\mathbf{I}_{n-k},\\
		\mathbf{F}_{n}^{(B)}=-i\stackrel[k=1]{n}{\sum}\left\{ \left[\mathbf{I}_{k-1}\otimes F_{B}\right]\left[\mathbf{I}_{k}+(-1)^{k}R_{k}\right]\right\} \otimes\mathbf{I}_{n-k}.
	\end{array}\right.\label{eq:Fn}
\end{equation}

Considering $\{\hat{f}_{i},\hat{f}_{j}\}=\delta_{i,j}$, we can get
(see Appendix \ref{subsec:Derivation FT_DT} for detailed derivation)
\begin{equation}
\mathbf{F}_{2m}^{(B)}\boldsymbol{\left|T_{2m,t}\right\rangle }=\mathbf{G}_{2m}\boldsymbol{\left|T_{2m-2,t}\right\rangle }
\end{equation}
with
\begin{equation}
\left\{ \begin{array}{l}
\mathbf{G}_{2m}=i\mathbf{R}_{2m}\left(|F_{B}\rrangle\otimes\mathbf{I}_{2m-2}\right),\\
\mathbf{R}_{2m}=\stackrel[k=1]{2m}{\sum}\stackrel[s=1]{k-1}{\sum}(-1)^{k-s}\left[R_{s}\otimes\mathbf{I}_{2m-s}\right]\left[R_{k}\otimes\mathbf{I}_{2m-k}\right],
\end{array}\right.\label{eq:G2m}
\end{equation}
so we have
\begin{equation}
\frac{d}{dt}\boldsymbol{\left|T_{2m,t}\right\rangle }=\mathbf{F}_{2m}\boldsymbol{\left|T_{2m,t}\right\rangle }+\mathbf{G}_{2m}\boldsymbol{\left|T_{2m-2,t}\right\rangle }.\label{eq:dt_2m_2m-2}
\end{equation}

When $m=1$, it is easy to obtain
\[
\boldsymbol{T_{2,t}}=e^{\mathbf{F}_{1}t}\left(\boldsymbol{T_{2,0}}-\boldsymbol{T_{2,\infty}}\right)e^{\mathbf{F}_{1}^{T}t}+\boldsymbol{T_{2,\infty}},
\]
where $\left(\boldsymbol{T_{2,t}}\right)_{j_{1},j_{2}}=T_{j_{1},j_{2}}$
represents the two-order correlation function, $\boldsymbol{T_{2,0}}$
represents the two-order correlation function corresponding to the
initial state. And $\boldsymbol{T_{2,\infty}}$ is the solution of
$\mathbf{F}_{1}\boldsymbol{T_{2,\infty}}+\boldsymbol{T_{2,\infty}}\mathbf{F}_{1}^{T}-i\left(F_{B}\right)^{T}=0$,
with the existence of the solution provided in the Appendix \ref{subsec:Proof_T2inf}.
When the non-equilibrium steady state is unique (i.e., $\det(\mathbf{F}_{2})\neq0$),
$\boldsymbol{T_{2,\infty}}$ represents the two-order correlation
function corresponding to the non-equilibrium steady state. Comparing
with the previous equation $ZX+X^{T}Z=Y$, it is easy to deduce that
$Z=-iQ\boldsymbol{T_{2,\infty}}Q$, and thus $ZX+X^{T}Z=Y$ must also
have a solution $Z$.

For $m>1$, the solution to Eq.(\ref{eq:dt_2m_2m-2}) is (For the
proof and derivation process, see the Appendix \ref{sec:The solution T2m})\begin{widetext}
\begin{equation}
\boldsymbol{\left|T_{2m,t}\right\rangle }=\stackrel[l=0]{m}{\sum}\frac{(-1)^{m-l}}{(m-l)!}\mathbb{R}_{(2m,2l+2)}\left\{ \left[\stackrel[r=1]{m-l}{\bigotimes}\left(\boldsymbol{\left|T_{2,t}^{T}\right\rangle }-e^{\mathbf{F}_{2}t}\boldsymbol{\left|T_{2,0}^{T}\right\rangle }\right)\right]\otimes\left[e^{\mathbf{F}_{2l}t}\boldsymbol{\left|T_{2l,0}\right\rangle }\right]\right\} \label{eq:High Order}
\end{equation}
\end{widetext}with $\mathbb{R}_{(2m,2l+2)}=\stackrel[r=0]{m-l-1}{\overrightarrow{\prod}}\left(\mathbf{I}_{2r}\otimes\mathbf{R}_{2m-2r}\right)$.
It should be noted that in this formula, the tensor product involving
$\left(\boldsymbol{\left|T_{2,t}^{T}\right\rangle }-e^{\mathbf{F}_{2}t}\boldsymbol{\left|T_{2,0}^{T}\right\rangle }\right)$
clearly does not contain the index $r$, thus, $\stackrel[r=1]{m-l}{\bigotimes}\left(\boldsymbol{\left|T_{2,t}^{T}\right\rangle }-e^{\mathbf{F}_{2}t}\boldsymbol{\left|T_{2,0}^{T}\right\rangle }\right)$
represents the tensor product of $\left(\boldsymbol{\left|T_{2,t}^{T}\right\rangle }-e^{\mathbf{F}_{2}t}\boldsymbol{\left|T_{2,0}^{T}\right\rangle }\right)$
taken $m-l$ times. Eq.(\ref{eq:High Order}) gives the relationship
between the higher-order correlation function and the second-order
correlation functions, as well as the initial state, which can be
regarded as an extension of the Wick's theorem.

A property
of $\mathbb{R}_{(2m,2l+2)}$ is described by the following Lemma:

$Lemma\ 1:\forall\ 1\leqslant s\leqslant m-l,\ \mathbb{R}_{(2m,2l+2)}=\mathbb{R}_{(2m,2l+2)}\left(R_{2s}\otimes\mathbf{I}_{2m-2s}\right)^{2}$.

The proof of the Lemma can be found in
the Appendix \ref{subsec:Prove-of-Lemma1}. It plays an important role in solving Eq.(\ref{eq:dt_2m_2m-2}) and
in the subsequent derivations.

\subsection{\label{subsec:Gaussian-initial-state}The initial state is a Gaussian
state}

When the initial state is a Gaussian state, it satisfies the Wick's
theorem. To maintain consistency in the equations, we present the
matrix version of the Wick's theorem as follows (proof can be found
in the Appendix \ref{sec:Wick}):
\begin{equation}
\begin{array}{rl}
\bm{\left|T_{2l,0}\right\rangle } & =\frac{-1}{l}\mathbf{R}_{2l}\left(\bm{\left|T_{2,0}^{T}\right\rangle }\otimes\bm{\left|T_{2l-2,0}\right\rangle }\right)\\
 & =\frac{(-1)^{l}}{l!}\mathbb{R}_{(2l,2)}\left(\stackrel[r=1]{l}{\bigotimes}\bm{\left|T_{2,0}^{T}\right\rangle }\right).
\end{array}\label{eq:wick}
\end{equation}
We can see that $\mathbf{R}_{2l}$ acts as a bridge connecting the
$(2l-2)$-order correlation functions and the $2l$-order correlation
functions. Substituting this into Eq.(\ref{eq:High Order}), we get
\begin{equation}
\begin{array}{rl}
 & \bm{\left|T_{2m,t}\right\rangle }\\
= & \stackrel[l=0]{m}{\sum}\frac{(-1)^{m-l}}{(m-l)!}\mathbb{R}_{(2m,2l+2)}\left\{ \left[\stackrel[r=1]{m-l}{\bigotimes}\left(\boldsymbol{\left|T_{2,t}^{T}\right\rangle }-e^{\mathbf{F}_{2}t}\boldsymbol{\left|T_{2,0}^{T}\right\rangle }\right)\right]\right.\\
 & \left.\otimes\left[\frac{(-1)^{l}}{l!}e^{\mathbf{F}_{2l}t}\mathbb{R}_{(2l,2)}\left(\stackrel[r=1]{l}{\bigotimes}\bm{\left|T_{2,0}^{T}\right\rangle }\right)\right]\right\} \\
= & \frac{(-1)^{m}}{m!}\mathbb{R}_{(2m,2)}\stackrel[l=0]{m}{\sum}C_{m}^{l}\left\{ \left[\stackrel[r=1]{m-l}{\bigotimes}\left(\boldsymbol{\left|T_{2,t}^{T}\right\rangle }-e^{\mathbf{F}_{2}t}\boldsymbol{\left|T_{2,0}^{T}\right\rangle }\right)\right]\right.\\
 & \left.\otimes\left[\stackrel[r=1]{l}{\bigotimes}e^{\mathbf{F}_{2}t}\bm{\left|T_{2,0}^{T}\right\rangle }\right]\right\} \\
= & \frac{(-1)^{m}}{m!}\mathbb{R}_{(2m,2)}\left(\stackrel[r=1]{m}{\bigotimes}\bm{\left|T_{2,t}^{T}\right\rangle }\right).
\end{array}\label{eq:gaussian rho}
\end{equation}
Here, $C_{m}^{l}$ represents a binomial coefficient, the second equality
holds because $e^{\mathbf{F}_{2l}t}\mathbb{R}_{(2l,2)}=\mathbb{R}_{(2l,2)}e^{\mathbf{F}_{2l}t}$,
as discussed in Appendix B. The third equality uses Binomial theorem
and Lemma 1, which states that after left-multiplying by $\mathbb{R}_{(2m,2)}$,
the order of $\left(\boldsymbol{\left|T_{2,t}^{T}\right\rangle }-e^{\mathbf{F}_{2}t}\boldsymbol{\left|T_{2,0}^{T}\right\rangle }\right)$
and $e^{\mathbf{F}_{2}t}\bm{\left|T_{2,0}^{T}\right\rangle }$ in
the direct product is no longer important and they can be exchanged
arbitrarily. Evidently, Eq.(\ref{eq:gaussian rho}) indicates that
the evolution process still satisfies the Wick's theorem. Since we
assume the initial state to be a Gaussian state, we can denote 
$\rho(0)=e^{\stackrel[j_{1},j_{2}=1]{2L}{\sum}\varrho_{j_{1},j_{2}}\omega_{j_{1}}\omega_{j_{2}}}$,
so $|\rho(0)\rrangle=e^{\stackrel[j_{1},j_{2}=1]{2L}{\sum}\varrho_{j_{1},j_{2}}\hat{f}_{j_{1}}\hat{f}_{j_{2}}}|\mathbb{I}\rrangle$.
It is evident that the initial state is an even-parity state, therefore,
$\rho(t)$ remains an even-parity state. Thus, we know that
$\rho(t)$ remains a Gaussian state, and the system
evolution process described by Eqs.(\ref{eq:lindblad},\ref{eq:H},\ref{eq:L})
is a Gaussian channel {[}\citep{D. Speed 2023,G. Lindblad Gauss 1976,O. Brodier 2010,YKZhang,C. Weedbrook 2012, A. Serafini 2017}{]}, that is, if $\rho(0)$ is a Gaussian state,
$\rho(t)$ remains a Gaussian state. Using the properties of
the Gaussian state {[}\citep{M. Fagotti 2013,S. Groha 2018}{]}, we
deduce that
\begin{equation}
\rho(t)=\frac{1}{\sqrt{\det\left[\mathbf{I}_{1}+e^{2\varrho_{t}}\right]}}e^{\stackrel[j_{1},j_{2}=1]{2L}{\sum}\left(\varrho_{t}\right)_{j_{1},j_{2}}\omega_{j_{1}}\omega_{j_{2}}}
\end{equation}
with $e^{2\varrho_{t}}=\left(\boldsymbol{T_{2,t}}\right)^{-1}-\mathbf{I}_{1}$,
combined with Klich's formula {[}\citep{I. Klich 2014}{]}, we know
that
\begin{eqnarray}
 Tr\left[\left(\stackrel[r=1]{n}{\overrightarrow{\prod}}e^{\stackrel[j_{1},j_{2}=1]{2L}{\sum}\left(W_{r}\right)_{j_{1},j_{2}}\omega_{j_{1}}\omega_{j_{2}}}\right)\rho(t)\right] \nonumber
\\
=  \sqrt{\det\left[\boldsymbol{T_{2,t}}+\left(\mathbf{I}_{1}-\boldsymbol{T_{2,t}}\right)\left(\stackrel[r=1]{n}{\overrightarrow{\prod}}e^{2W_{r}}\right)\right]},
\label{eq:full counting}
\end{eqnarray}
where the chosen $W_{r}$ must satisfy $W_{r}^{T}=-W_{r}$.

\begin{figure}
\includegraphics[scale=0.11]{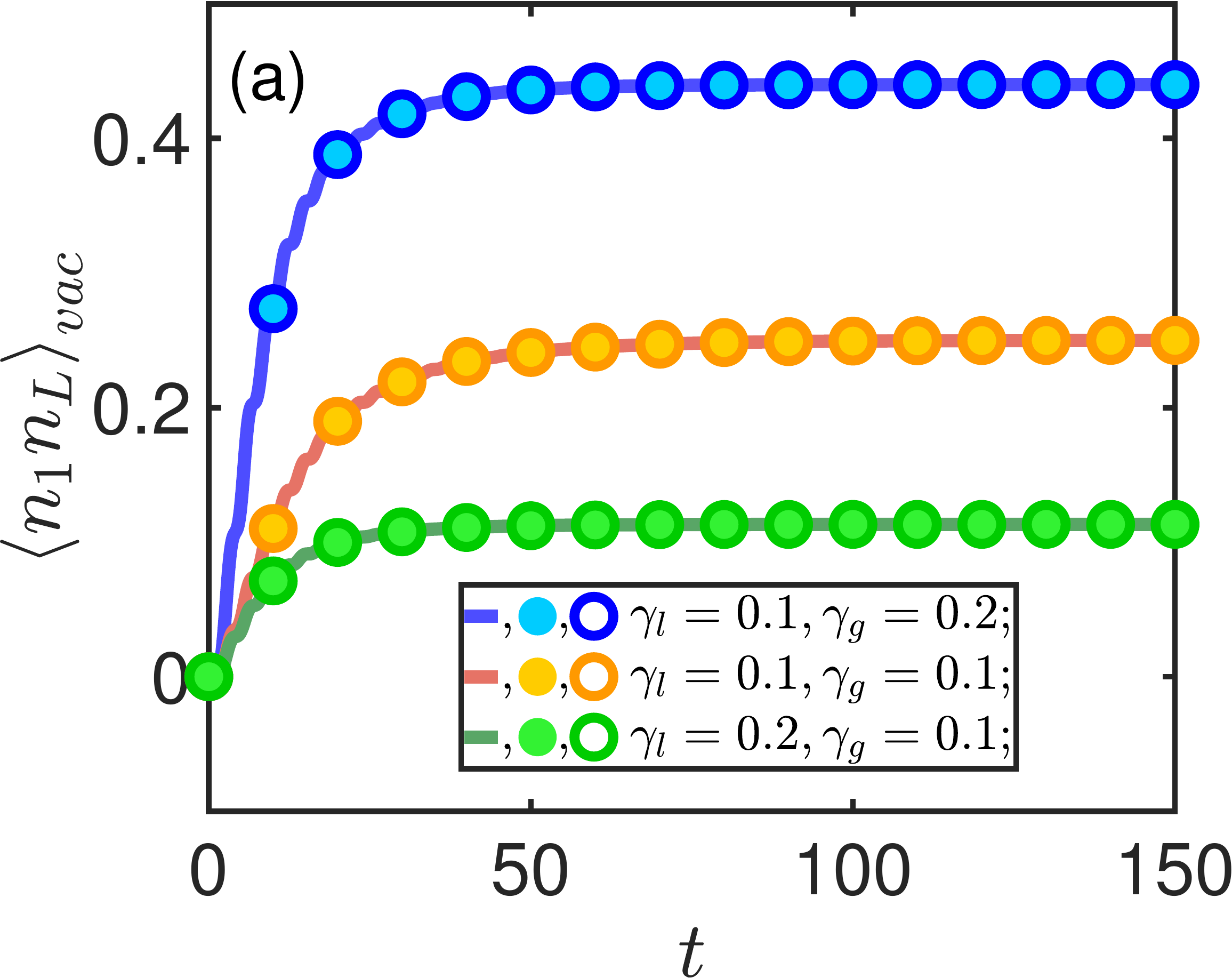}\includegraphics[scale=0.11]{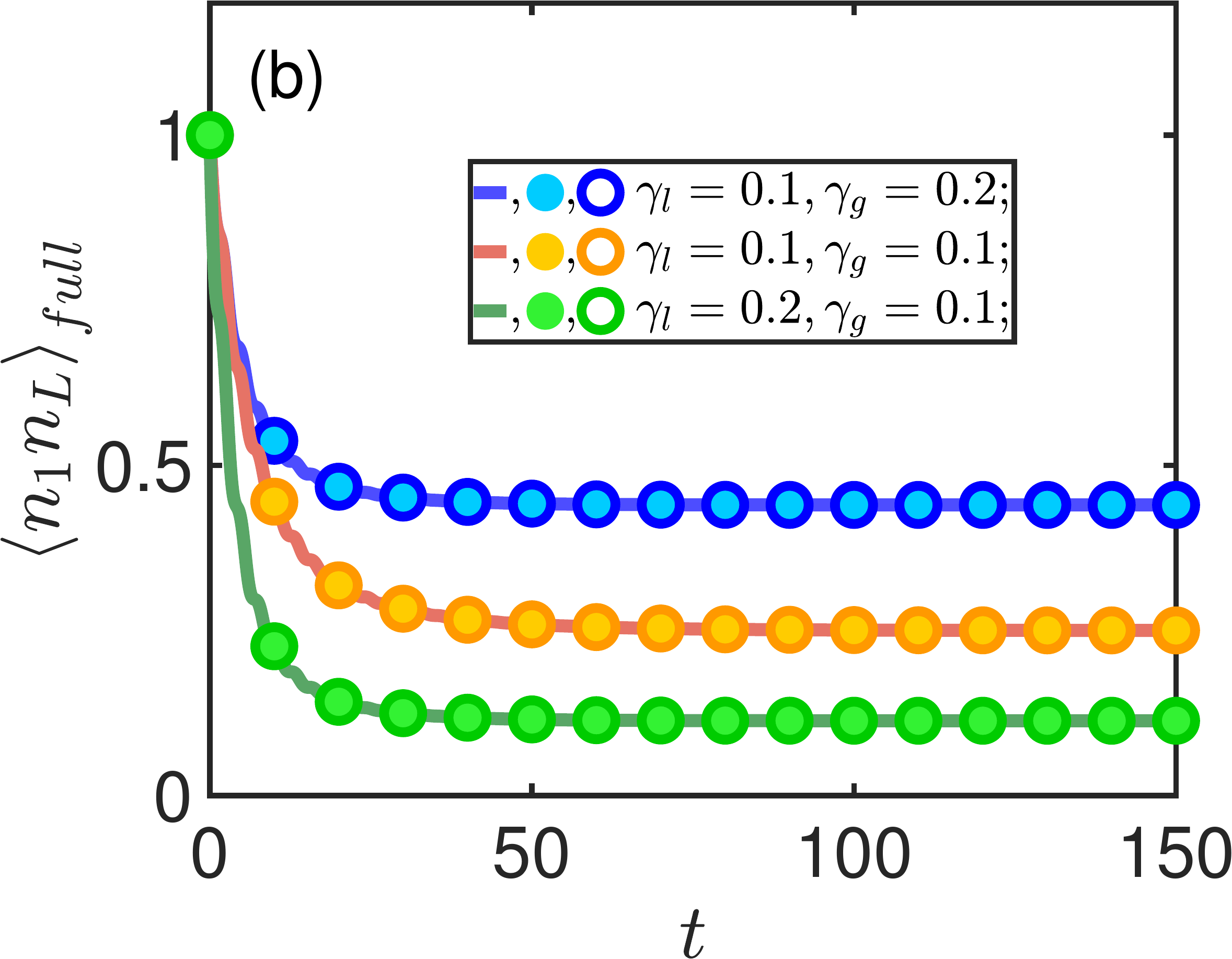}

\caption{\label{fig:n1nL_two}Under different conditions, the corresponding
$\left\langle n_{1}n_{L}\right\rangle $ are shown. We fix $L=4$
and $J=1$. In subplot (a), the initial state is chosen as the fully
empty state, with the dissipation strengths set to $\gamma_{l}=0.1$,
$\gamma_{g}=0.2$ and $\gamma_{l}=0.1$, $\gamma_{g}=0.1$ and $\gamma_{l}=0.2$,
$\gamma_{g}=0.1$, respectively, each represented by different colors
as shown in the legend. In the figure, solid lines denote results
calculated from Eq.(\ref{eq:High Order}), filled dots represent results
obtained by exact diagonalization, and open circles indicate results
obtained via the Wick\textquoteright s theorem. Subplot (b) is identical
to subplot (a) except that the initial state is set as the fully occupied
state.}

\end{figure}

\begin{figure}
\includegraphics[scale=0.11]{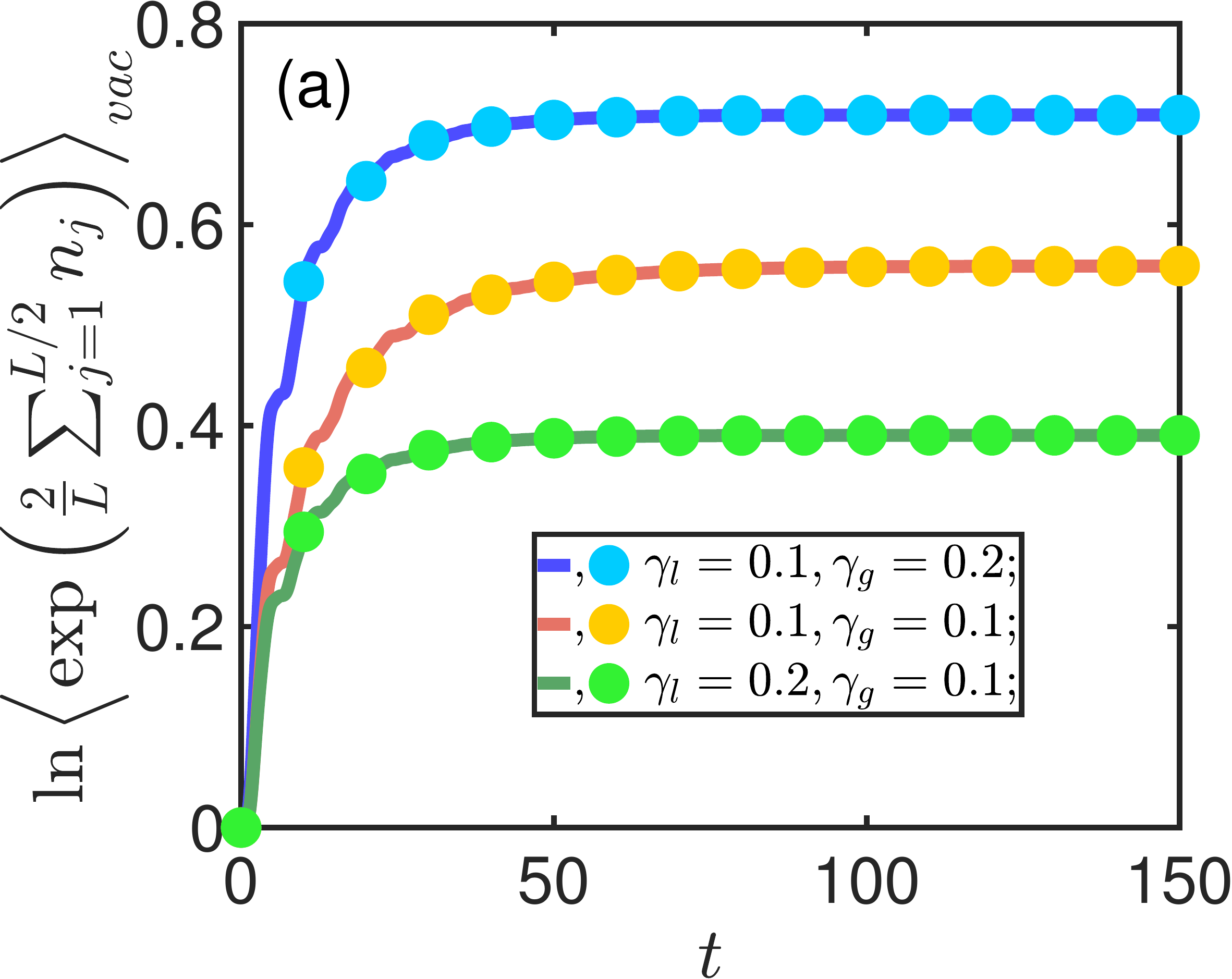}\includegraphics[scale=0.11]{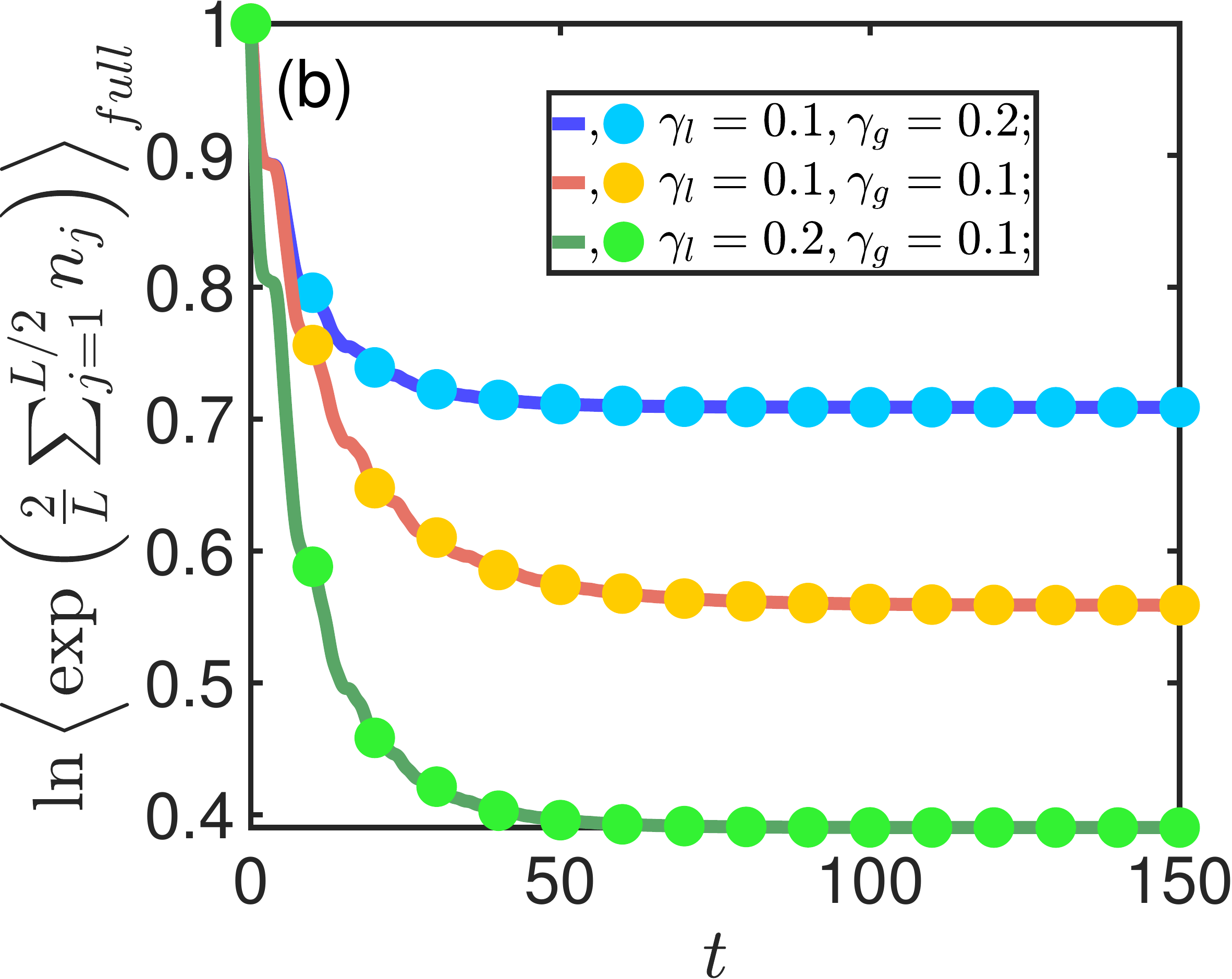}

\caption{\label{fig:expn_two}The $\ln\left\langle \exp\left(\frac{2}{L}\sum_{j=1}^{L/2}n_{j}\right)\right\rangle $
calculated under different conditions. We also fix $L=4$ and $J=1$,
and the dissipation strengths are chosen in the same way as in Figure
1, with different colors representing each case, as shown in the legend.
Here, solid lines represent results calculated from Eq.(\ref{eq:full counting}),
and filled dots represent results obtained by exact diagonalization.
Subplot (a) and (b) correspond to the initial state being the fully
empty state and the fully occupied state, respectively.}
\end{figure}

Let us choose a simple model to verify the above conclusions by taking
\begin{equation}
H=J\stackrel[j=1]{L-1}{\sum}\left(a_{j}^{\dagger}a_{j+1}+a_{j+1}^{\dagger}a_{j}\right) \label{model}
\end{equation}
and
\begin{equation}
L_{l}=\sqrt{\gamma_{l}}a_{1}, ~~~ L_{g}=\sqrt{\gamma_{g}}a_{L}^{\dagger}. \label{model-L}
\end{equation}
Here, we define $n_{j}=a_{j}^{\dagger}a_{j}$, the initial states
are chosen as the fully empty state and the fully filled state, with
the results denoted by the subscripts \textquotedbl vac\textquotedbl{}
and \textquotedbl full\textquotedbl , respectively. The numerical
results are shown in Fig.\ref{fig:n1nL_two} and Fig.\ref{fig:expn_two},
where the points represent the results from exact diagonalization,
the solid lines represent $\left\langle n_{1}n_{L}\right\rangle $
calculated using Eq.(\ref{eq:High Order}) and $\ln\left\langle \exp\left(\frac{2}{L}\sum_{j=1}^{L/2}n_{j}\right)\right\rangle $
calculated using Eq.(\ref{eq:full counting}), while the circles represent
results derived using the Wick's theorem. Three different cases with $\gamma_{l}=0.1$, $\gamma_{g}=0.2$; $\gamma_{l}=0.1$, $\gamma_{g}=0.1$ and $\gamma_{l}=0.2$, $\gamma_{g}=0.1$ are considered,
with the results displayed in different colors. We observe that in
all cases, the results obtained through exact diagonalization, as
well as those from Eqs.(\ref{eq:High Order},\ref{eq:full counting})
and the Wick's theorem, perfectly match, which verifies the correctness
of our formulas and conclusions.

\subsection{The initial state is not a Gaussian state}

When the initial state is non-Gaussian, the Wick's theorem no longer
holds during the dynamical evolution. However, if we are interested in the nonequilibrium
steady state and when the steady state is unique, i.e., $\det\left[\mathbf{F}_{2}\right]\neq0$,
we have $\bm{\left|T_{2m,\infty}\right\rangle }=\frac{(-1)^{m}}{m!}\mathbb{R}_{(2m,2)}\left(\stackrel[r=1]{m}{\bigotimes}\bm{\left|T_{2,\infty}^{T}\right\rangle }\right)$,
which satisfies the Wick's theorem. Since this unique non-equilibrium
steady state must be an even parity state, we know that it must be
a Gaussian state, which has nothing to do with the initial state.
The corresponding density matrix and the full counting statistics
can refer to the discussion with Gaussian state above.

\begin{figure}
\includegraphics[scale=0.11]{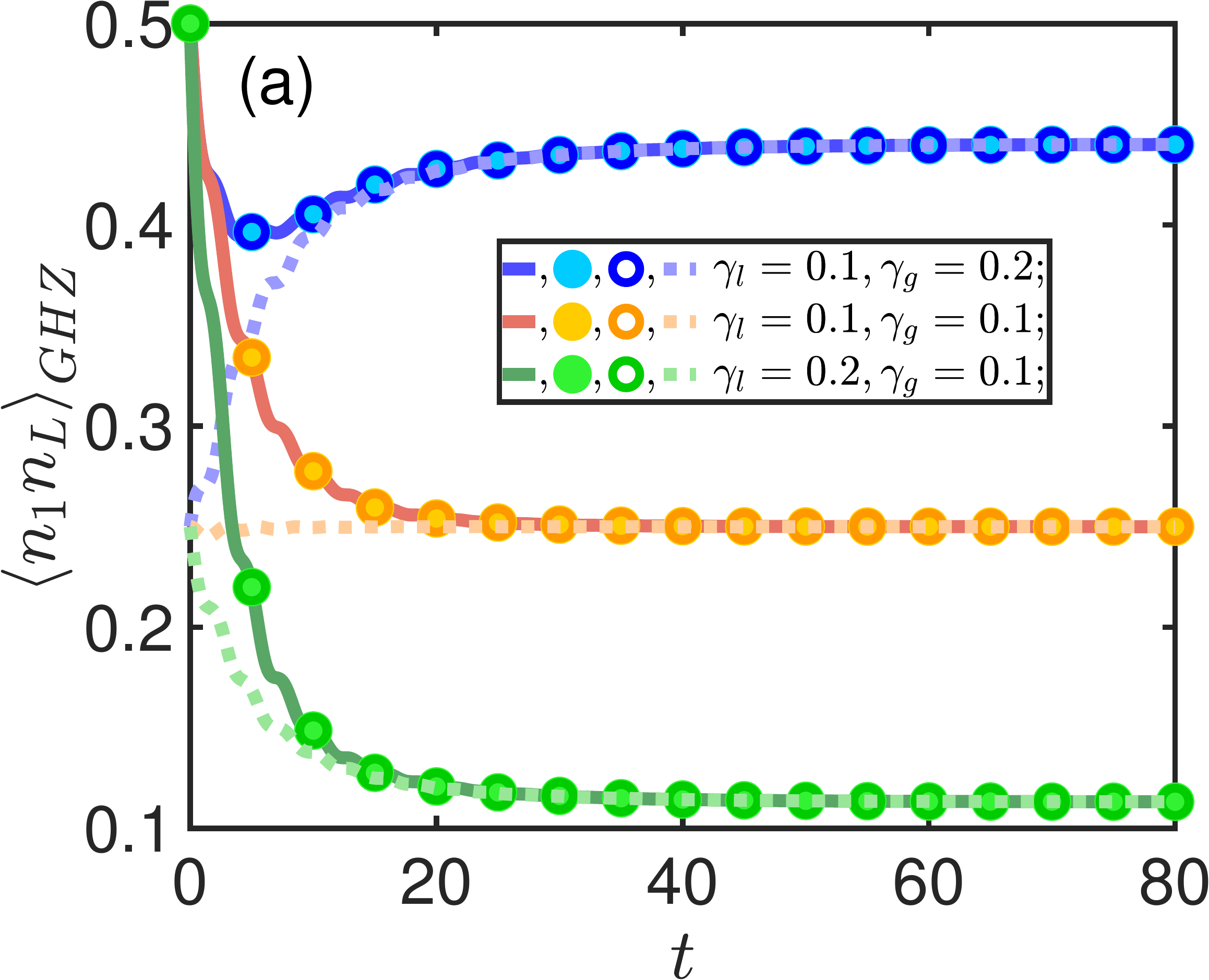}\includegraphics[scale=0.11]{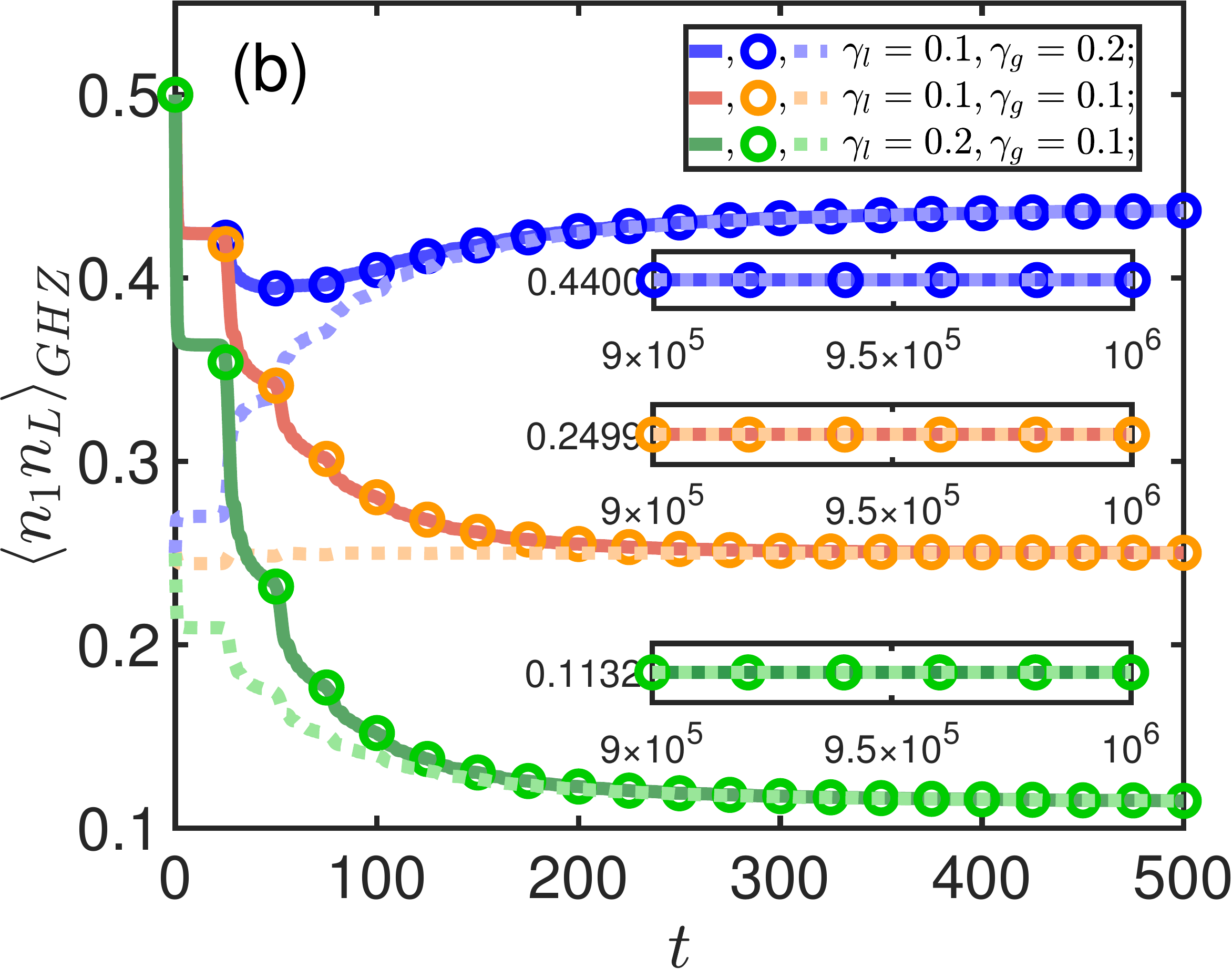}\caption{\label{fig:GHZ_n1nL_two}The initial state is chosen as the GHZ state,
and the calculation results of $\left\langle n_{1}n_{L}\right\rangle $
for different conditions are shown with $J=1$ fixed. Subplot (a)
corresponds to $L=4$, and subplot (b) corresponds to $L=50$. The
dissipation strengths are selected in the same way as in Figure 1
and represented by different colors. In the figure, solid lines represent
results calculated from Eq.(\ref{eq:High Order}), dashed lines represent
results obtained via the Wick's theorem, filled dots represent results
from exact diagonalization, and open circles represent results calculated
from Equation $\left\langle n_{1}n_{L}\right\rangle _{GHZ}=\frac{1}{2}\left(\left\langle n_{1}n_{L}\right\rangle _{vac}+\left\langle n_{1}n_{L}\right\rangle _{full}\right)$.
The inset in subplot (b) shows the results of the non-equilibrium
steady state when the evolution time reaches $10^{6}$,  the numerical results corresponding to different methods are still indistinguishable even at the highest precision of the software.}
\end{figure}

Using the same model as above, now we consider the case with  a non-Gaussian initial state by choosing the initial state
as a GHZ state, which is an equal-weight superposition of the fully
empty and fully occupied states.
The calculation results of $\left\langle n_{1}n_{L}\right\rangle $
are shown in Fig.\ref{fig:GHZ_n1nL_two}. Similar to the above, we represent
the results obtained from Eq.(\ref{eq:High Order}) with solid lines, and
the results from exact diagonalization with points. For comparison, we use
dashed lines to denote the results derived from the Wick's theorem.
The open circles represent results calculated using the formula $\left\langle n_{1}n_{L}\right\rangle _{GHZ}=\frac{1}{2}\left(\left\langle n_{1}n_{L}\right\rangle _{vac}+\left\langle n_{1}n_{L}\right\rangle _{full}\right)$,
specifically the average of the results from the vacuum state and
the fully occupied state as initial states. We also use different
colors to indicate the different values of $\gamma_{l}$ and $\gamma_{g}$,
with the values chosen the same as in Fig.1. Clearly,
since the initial state is non-Gaussian, the Wick's theorem no longer
holds during the dynamical evolution. However, when the evolution time is sufficiently
long for the system to approach the nonequilibrium steady state, we
observe that the Wick's theorem aligns very well (See the inset in
panel (b) of Fig.\ref{fig:GHZ_n1nL_two}), which is consistent with
our expectations. In all cases, we observe that the open circles consistently
align very well with numerical results obtained from Eq.(\ref{eq:High Order}). Below, we provide an explanation for this observation.

We label the fully empty state, the fully occupied state, and the
GHZ state as $\left|vac\right\rangle $, $\left|full\right\rangle $,
and $\left|GHZ\right\rangle =\frac{1}{\sqrt{2}}\left(\left|vac\right\rangle +\left|full\right\rangle \right)$,
respectively, and that implies $\rho_{GHZ}=\frac{1}{2}\left(\rho_{vac}+\rho_{full}+\left|vac\right\rangle \left\langle full\right|+\left|full\right\rangle \left\langle vac\right|\right)$.
It is easy to see that for any $j_{1},j_{2},j_{3},j_{4}\in\{1,2,\cdots,2L\}$,
when $L>4$, we have $Tr\left[\omega_{j_{1}}\omega_{j_{2}}\left|vac\right\rangle \left\langle full\right|\right]=0=Tr\left[\omega_{j_{1}}\omega_{j_{2}}\omega_{j_{3}}\omega_{j_{4}}\left|vac\right\rangle \left\langle full\right|\right]$.
Therefore, when $L>4$, we have $\bm{\left|T_{2,0}\right\rangle }_{GHZ}=\frac{1}{2}\left(\bm{\left|T_{2,0}\right\rangle }_{vac}+\bm{\left|T_{2,0}\right\rangle }_{full}\right)$,
$\bm{\left|T_{4,0}\right\rangle }_{GHZ}=\frac{1}{2}\left(\bm{\left|T_{4,0}\right\rangle }_{vac}+\bm{\left|T_{4,0}\right\rangle }_{full}\right)$.
By setting $m=2$ in Eq.(\ref{eq:High Order}), we have
\[
\begin{array}{rl}
	\boldsymbol{\left|T_{4,t}\right\rangle }= & \frac{1}{2}\mathbb{R}_{(4,2)}\left[\stackrel[r=1]{2}{\bigotimes}\left(\boldsymbol{\left|T_{2,t}^{T}\right\rangle }-e^{\mathbf{F}_{2}t}\boldsymbol{\left|T_{2,0}^{T}\right\rangle }\right)\right]\\
	& -\mathbf{R}_{4}\left[\left(\boldsymbol{\left|T_{2,t}^{T}\right\rangle }-e^{\mathbf{F}_{2}t}\boldsymbol{\left|T_{2,0}^{T}\right\rangle }\right)\otimes\left(e^{\mathbf{F}_{2}t}\boldsymbol{\left|T_{2,0}\right\rangle }\right)\right]\\
	& +e^{\mathbf{F}_{4}t}\boldsymbol{\left|T_{4,0}\right\rangle }.
\end{array}
\]
As mentioned earlier, the required $\boldsymbol{T_{2,\infty}}$ is
guaranteed to exist, so $\left(\boldsymbol{\left|T_{2,t}^{T}\right\rangle }-e^{\mathbf{F}_{2}t}\boldsymbol{\left|T_{2,0}^{T}\right\rangle }\right)=\left(\boldsymbol{\left|T_{2,\infty}^{T}\right\rangle }-e^{\mathbf{F}_{2}t}\boldsymbol{\left|T_{2,\infty}^{T}\right\rangle }\right)$
is independent of the initial state. It follows naturally that when
$L>4$, we have $\boldsymbol{\left|T_{4,t}\right\rangle }_{GHZ}=\frac{1}{2}\left(\bm{\left|T_{4,t}\right\rangle }_{vac}+\bm{\left|T_{4,t}\right\rangle }_{full}\right)$.
Similarly, using Eq.(\ref{eq:High Order}), we can further generalize
this conclusion to
\begin{equation}
\boldsymbol{\left|T_{2m,t}\right\rangle }_{GHZ}=\frac{1}{2}\left(\bm{\left|T_{2m,t}\right\rangle }_{vac}+\bm{\left|T_{2m,t}\right\rangle }_{full}\right),\ if\ L>2m.\label{GHZ_vac_full}
\end{equation}
Thus, we explain the perfect agreement of the open-circle-marked data when
$L=50$. As for the agreement of the open-circle-marked data even when $L=4$,
this is due to the simplicity of the chosen Hamiltonian and the specific
nature of the computed quantity $\left\langle n_{1}n_{L}\right\rangle $.
As a counterexample, if we were to compute $\left\langle a_{1}^{\dagger}a_{2}^{\dagger}a_{3}^{\dagger}a_{4}^{\dagger}\right\rangle $,
it is clear that at the initial moment for $L=4$, $\left\langle a_{1}^{\dagger}a_{2}^{\dagger}a_{3}^{\dagger}a_{4}^{\dagger}\right\rangle _{GHZ}=\frac{1}{2}\left(\left\langle a_{1}^{\dagger}a_{2}^{\dagger}a_{3}^{\dagger}a_{4}^{\dagger}\right\rangle _{vac}+\left\langle a_{1}^{\dagger}a_{2}^{\dagger}a_{3}^{\dagger}a_{4}^{\dagger}\right\rangle _{full}\right)$
would not hold. Previously, we observed that, when the initial states
are the fully empty and fully occupied states, $\left\langle n_{1}n_{L}\right\rangle $
exhibits monotonic growth and decay, respectively, over time. Using
Eq.(\ref{GHZ_vac_full}), we can easily explain the non-monotonic
behavior seen in Fig.\ref{fig:GHZ_n1nL_two} when the initial state
is the GHZ state and $\gamma_{g}>\gamma_{l}$. It should be noted
that even if we choose a specific non-Gaussian initial state, making
it impossible to find a convenient formula like Eq.(\ref{GHZ_vac_full}),
we can still calculate for relatively large lattice points using Eq.(\ref{eq:High Order}).

\subsection{\label{subsec:Constructing-the-Liouvillian-spectrum}Constructing the
Liouvillian spectrum using the dynamical Eq.(\ref{eq:dt_2m_2m-2})}

It is easy to see that the $\boldsymbol{\left|T_{2m,t}\right\rangle }$
we defined has significant redundancy. For example, when $m=2$, there
exists $\boldsymbol{\left|T_{4,t}\right\rangle }_{j_{1},j_{2},j_{1},j_{4}}=Tr\left(\omega_{j_{1}}\omega_{j_{2}}\omega_{j_{1}}\omega_{j_{4}}\rho(t)\right)=\delta_{j_{1},j_{2}}\boldsymbol{\left|T_{2,t}\right\rangle }_{j_{1},j_{4}}-\frac{1}{2}\boldsymbol{\left|T_{2,t}\right\rangle }_{j_{2},j_{4}}$
within $\boldsymbol{\left|T_{4,t}\right\rangle }$, which can clearly
be obtained from $\boldsymbol{\left|T_{2,t}\right\rangle }$. Given
$\{\omega_{j_{1}},\omega_{j_{2}}\}=\delta_{j_{1},j_{2}}$, we will
show below how this redundancy can be eliminated using a fully antisymmetric
projection $\mathbb{P}_{FA}$.

It can be observed that the redundancy allowing for reduction in order,
as mentioned above, arises from repeated indices in the subscript
sequence, such as the $j_{1}$ in the sequence $(j_{1},j_{2},j_{1},j_{4})$.
Clearly, replacing $\boldsymbol{\left|T_{4,t}\right\rangle }_{j_{1},j_{2},j_{3},j_{4}}$
with $\frac{1}{2}\left(\boldsymbol{\left|T_{4,t}\right\rangle }_{j_{1},j_{2},j_{3},j_{4}}-\boldsymbol{\left|T_{4,t}\right\rangle }_{j_{3},j_{2},j_{1},j_{4}}\right)$
enables $\boldsymbol{\left|T_{4,t}\right\rangle }_{j_{1},j_{2},j_{3},j_{4}}$
with $j_{1}=j_{3}$ to become zero, while terms satisfying $j_{1}\neq j_{3}$
remain unaffected (Here, we focus on the indices $j_{1}$ and $j_{3}$,
asserting that $j_{1}$, $j_{3}$ are fundamentally different from
other indices.). This effectively applies an antisymmetric projection
to indices $j_{1}$ and $j_{3}$. Since indices at any positions may
be identical, it is necessary to perform an antisymmetric projection
across all pairs of positions, thus requiring a fully antisymmetric
projection on the whole structure. Using the properties of $R_{s}$
as expressed in Eq.(\ref{eq:prop of Rk}), we can construct the required
fully antisymmetric projection as
\begin{equation}
\begin{array}{rl}
\mathbb{P}_{FA,n} & =\frac{1}{n!}\stackrel[l=0]{n}{\overrightarrow{\prod}}\left[\stackrel[s=1]{n-l}{\sum}(-1)^{s+1}\mathbf{I}_{l}\otimes R_{s}\otimes\mathbf{I}_{n-l-s}\right]\\
 & =\frac{1}{n!}\stackrel[l=0]{n}{\overleftarrow{\prod}}\left[\stackrel[s=1]{n-l}{\sum}(-1)^{s+1}\mathbf{I}_{l}\otimes R_{s}^{T}\otimes\mathbf{I}_{n-l-s}\right].
\end{array}\label{eq:P_FA}
\end{equation}
As a result, for any index sequence $(j_{1},j_{2},\cdots,j_{2m})$
containing identical elements, $\left(\mathbb{P}_{FA,2m}\boldsymbol{\left|T_{2m,t}\right\rangle }\right)_{j_{1},j_{2},\cdots,j_{2m}}=0$,
while for sequences without identical elements, $\left(\mathbb{P}_{FA,2m}\boldsymbol{\left|T_{2m,t}\right\rangle }\right)_{j_{1},j_{2},\cdots,j_{2m}}=\boldsymbol{\left|T_{2m,t}\right\rangle }_{j_{1},j_{2},\cdots,j_{2m}}$.

We observe that the $2m$ positions involved in the direct product
of $\mathbf{F}_{2m}$ are equivalent, as reflected by $\left[\mathbf{F}_{2m},\mathbf{I}_{l}\otimes R\otimes\mathbf{I}_{2m-l-2}\right]=0$.
Therefore, $\left[\mathbf{F}_{2m},\mathbb{P}_{FA,2m}\right]=0$. Considering
$\left(\mathbb{P}_{FA,2m}\right)^{2}=\mathbb{P}_{FA,2m}$ and $\mathbb{P}_{FA,2m}\left(\mathbf{I}_{l}\otimes R\otimes\mathbf{I}_{2m-l-2}\right)=\left(\mathbf{I}_{l}\otimes R\otimes\mathbf{I}_{2m-l-2}\right)\mathbb{P}_{FA,2m}=-\mathbb{P}_{FA,2m}$,
we can rewrite Eq.(\ref{eq:dt_2m_2m-2}) as
\[
\begin{array}{rl}
 & \frac{d}{dt}\boldsymbol{\left|\widetilde{T}_{2m,t}\right\rangle }\\
= & \frac{d}{dt}\left(\mathbb{P}_{FA,2m}\boldsymbol{\left|T_{2m,t}\right\rangle }\right)\\
= & \left(\mathbb{P}_{FA,2m}\mathbf{F}_{2m}\mathbb{P}_{FA,2m}\right)\left(\mathbb{P}_{FA,2m}\boldsymbol{\left|T_{2m,t}\right\rangle }\right)\\
 & +\mathbb{P}_{FA,2m}\mathbf{G}_{2m}\boldsymbol{\left|T_{2m-2,t}\right\rangle }\\
= & \left(\mathbb{P}_{FA,2m}\mathbf{F}_{2m}\mathbb{P}_{FA,2m}\right)\left(\mathbb{P}_{FA,2m}\boldsymbol{\left|T_{2m,t}\right\rangle }\right)+\\
 & iC_{2m}^{2}\left[\mathbb{P}_{FA,2m}\left(|F_{B}\rrangle\otimes\mathbf{I}_{2m-2}\right)\right]\left(\mathbb{P}_{FA,2m-2}\boldsymbol{\left|T_{2m-2,t}\right\rangle }\right)\\
= & \widetilde{\mathbf{F}}_{2m}\boldsymbol{\left|\widetilde{T}_{2m,t}\right\rangle }+\widetilde{\mathbf{G}}_{2m}\boldsymbol{\left|\widetilde{T}_{2m-2,t}\right\rangle }
\end{array}
\]
with $\boldsymbol{\left|\widetilde{T}_{2m,t}\right\rangle }=\mathbb{P}_{FA,2m}\boldsymbol{\left|T_{2m,t}\right\rangle }$,
$\widetilde{\mathbf{F}}_{2m}=\mathbb{P}_{FA,2m}\mathbf{F}_{2m}\mathbb{P}_{FA,2m}$,
$\widetilde{\mathbf{G}}_{2m}=iC_{2m}^{2}\left[\mathbb{P}_{FA,2m}\left(|F_{B}\rrangle\otimes\mathbf{I}_{2m-2}\right)\right]$.
We can express it integrally as
\[
\frac{d}{dt}\left[\begin{array}{c}
\boldsymbol{\left|\widetilde{T}_{2L,t}\right\rangle }\\
\vdots\\
\boldsymbol{\left|\widetilde{T}_{0,t}\right\rangle }
\end{array}\right]=\left[\begin{array}{cccc}
\widetilde{\mathbf{F}}_{2L} & \widetilde{\mathbf{G}}_{2L}\\
 & \ddots & \ddots\\
 &  & \ddots & \widetilde{\mathbf{G}}_{2}\\
 &  &  & \widetilde{\mathbf{F}}_{0}
\end{array}\right]\left[\begin{array}{c}
\boldsymbol{\left|\widetilde{T}_{2L,t}\right\rangle }\\
\vdots\\
\boldsymbol{\left|\widetilde{T}_{0,t}\right\rangle }
\end{array}\right].
\]
Clearly, the even-parity portion of the Liouvillian spectrum, which
describes the decay exponent of the dynamics, is composed of the eigenvalues
of $\widetilde{\mathbf{F}}_{2m}$ for $m=0,1,\ldots,L$. Although
$\mathbb{P}_{FA,2m}\boldsymbol{\left|T_{2m,t}\right\rangle }$ still
retains redundancies like $j_{1}\neq j_{2},\ \boldsymbol{\left|T_{2m,t}\right\rangle }_{j_{1},j_{2},\cdots,j_{2m}}=-\boldsymbol{\left|T_{2m,t}\right\rangle }_{j_{2},j_{1},j_{3,}\cdots,j_{2m}}$,
caused by sequence ordering, which leads to a high degeneracy in the
eigenvalues of $\widetilde{\mathbf{F}}_{2m}$, this does not affect
our discussion of the Liouvillian spectrum here. We will address such
ordering-related redundancies in the next section when considering
more complex cases. If we denote $\mathbf{F}_{1}\left|\psi_{j}\right\rangle =\lambda_{j}\left|\psi_{j}\right\rangle $,
it is clear that ${\mathbf{F}}_{2m}\left(\stackrel[l=1]{2m}{\overrightarrow{\bigotimes}}\left|\psi_{j_{l}}\right\rangle \right)=\left(\stackrel[l=1]{2m}{\sum}\lambda_{j_{l}}\right)\left(\stackrel[l=1]{2m}{\overrightarrow{\bigotimes}}\left|\psi_{j_{l}}\right\rangle \right)$.
When considering the eigenvalues of $\widetilde{\mathbf{F}}_{2m}$,
the previous discussion reminds us to select the fully symmetric part
of the corresponding eigenvectors, that is, to require $\mathbb{P}_{FA,2m}\boldsymbol{\left|\psi_{2m,\overrightarrow{j}}\right\rangle }=\boldsymbol{\left|\psi_{2m,\overrightarrow{j}}\right\rangle }$
with $\widetilde{\mathbf{F}}_{2m}\boldsymbol{\left|\psi_{2m,\overrightarrow{j}}\right\rangle }=\lambda_{2m,\overrightarrow{j}}\boldsymbol{\left|\psi_{2m,\overrightarrow{j}}\right\rangle }$.
Thus, we have $\boldsymbol{\left|\psi_{2m,\overrightarrow{j}}\right\rangle }=\mathbb{P}_{FA,2m}\left(\stackrel[l=1]{2m}{\overrightarrow{\bigotimes}}\left|\psi_{j_{l}}\right\rangle \right)$,
$\lambda_{2m,\overrightarrow{j}}=\stackrel[l=1]{2m}{\sum}\lambda_{j_{l}}$,
and this means that there should be no identical elements in sequence
$(j_{1},j_{2},\cdots,j_{2m})$. Thus, we conclude that the Liouvillian
even-parity spectrum is $\sum_{j=1}^{2L}\nu_{j}\lambda_{j}$ with
$(-1)^{\sum_{j=1}^{2L}\nu_{j}}=1$. With $F_{A}+iF_{B}=-2Q^{-1}X^{T}Q$
, we obtain $\lambda_{j}=-2\alpha_{j}$, leading us to the same conclusion
as in Section \ref{subsec:Eigen-decomposition}. Since the odd parity
is not closed, we do not derive the Liouvillian spectrum corresponding
to the odd parity here. However, as seen in Section \ref{subsec:Eigen-decomposition},
the only difference between the even and odd parity Liouvillian spectra
is the requirement changing to $(-1)^{\sum_{j=1}^{2L}\nu_{j}}=-1$.
Here, it corresponds to taking the eigenvalues of $\mathbb{P}_{FA,n}\mathbf{F}_{n}\mathbb{P}_{FA,n}$
with $n=1,3,\cdots,2L-1$. Similarly, only the fully anti-symmetric
part is selected.

\section{\label{sec:Quartic-Liouvillian-Superoperator}Quartic Liouvillian Superoperator:
Generalization and Spectrum Construction}

Here, we consider extending the previously discussed quadratic Liouvillian
superoperator to a quartic form, allowing the Hamiltonian to include
interaction terms and adding quadratic dissipation in the form
\begin{equation}
L_{t,\mu}=\stackrel[j_{1},j_{2}=1]{2L}{\sum}\left(U_{L,\mu}\right)_{j_{1},j_{2}}\omega_{j_{1}}\omega_{j_{2}}.\label{eq:Lt_mu}
\end{equation}
 At this point, it is clear that we can no longer provide an eigen-decomposition
for $\hat{\mathcal{L}}_{quar}$. Instead, as in Section \ref{sec:Higher-Order-Correlation},
we attempt to derive the dynamical equation for higher-order correlation
functions and use it to construct the Liouvillian spectrum. Similarly,
we focus solely on even-order correlation functions and require them
to be closed. Through intricate derivations (see Appendix for details),
we can establish the condition for the closure of second-order correlation
functions, which requires the Hamiltonian to remain quadratic, i.e.,
without interaction terms, and the quadratic dissipation operator
must satisfy
\begin{equation}
\Im\left(\underset{\mu}{\sum}U_{a,\mu}\otimes U_{a,\mu}^{*}\right)=0,\label{eq:closure}
\end{equation}
where $U_{a,\mu}=U_{L,\mu}-U_{L,\mu}^{T}$ and the symbol $\Im$ represents
taking the imaginary part. Under this condition, higher-order correlation
functions are also closed, allowing us to obtain the dynamical equation
satisfied by $2m$-order correlation functions $\bm{\left|T_{2m,t}\right\rangle }$
as
\begin{equation}
\frac{d}{dt}\bm{\left|T_{2m,t}\right\rangle }=\mathbf{F}_{M,2m}\bm{\left|T_{2m,t}\right\rangle }+\mathbf{G}_{M,2m}\bm{\left|T_{2m-2,t}\right\rangle }\label{eq:dTdt-many}
\end{equation}
with\begin{widetext}
\[
\left\{ \begin{array}{l}
\mathbf{F}_{M,n}=\mathbf{F}_{n}-\underset{\mu}{\sum}\mathbf{U_{a,\mu}^{(n)}}\left(\mathbf{U_{a,\mu}^{(n)}}\right)^{\dagger}+\Re\left[\underset{\mu}{\sum}\left(Tr(U_{L,\mu}^{*})*\mathbf{U_{a,\mu}^{(n)}}\right)\right],\\
\mathbf{G}_{M,2m}=\mathbf{G}_{2m}+\stackrel[k=1]{m}{\sum}\left[R_{2k}\otimes\mathbf{I}_{2m-2k}\right]^{2}\left(|F_{\gamma}\rrangle\otimes\mathbf{I}_{2m-2}\right)
\end{array}\right.
\]
and
\[
\left\{ \begin{array}{l}
\mathbf{U_{a,\mu}^{(n)}}=\stackrel[k=1]{n}{\sum}\mathbf{I}_{k-1}\otimes U_{a,\mu}\otimes\mathbf{I}_{n-k},\\
\left(F_{\gamma}\right)_{j_{1},j_{2}}=\Re\left\{ \underset{\mu}{\sum}\left[\left(U_{a,\mu}\right)_{j_{1},j_{2}}\left(\frac{1}{2}(U_{L,\mu}^{*})_{j_{1},j_{1}}+\frac{1}{2}(U_{L,\mu}^{*})_{j_{2},j_{2}}-Tr(U_{L,\mu}^{*})\right)\right]\right\} .
\end{array}\right.
\]
\end{widetext}Here, $\Re$ represents taking the real part, the definitions
of $\mathbf{F}_{2m}$ and $\mathbf{G}_{2m}$ refer to Eq.(\ref{eq:Fn})
and (\ref{eq:G2m}) mentioned earlier, we add the subscript ''M''
to represent the results of the quartic Liouvillian superoperator. We
can formally present the solution of Eq.(\ref{eq:dTdt-many}) as
\begin{equation}
\begin{array}{rl}
\bm{\left|T_{2m,t}\right\rangle }= & e^{\mathbf{F}_{M,2m}t}\bm{\left|T_{2m,0}\right\rangle }+\\
 & e^{\mathbf{F}_{M,2m}t}\int_{0}^{t}\left(e^{-\mathbf{F}_{M,2m}\tau}\mathbf{G}_{M,2m}\bm{\left|T_{2m-2,\tau}\right\rangle }\right)d\tau.
\end{array}\label{eq:T2m_M}
\end{equation}
When selecting a special case of the closure condition in Eq.(\ref{eq:closure})
as $U_{L,\mu}=U_{L,\mu}^{\dagger}=-U_{L,\mu}^{T}$ and setting $m=1$
in Eq.(\ref{eq:dTdt-many}), we obtain
\begin{equation}
\begin{array}{rl}
\frac{d}{dt}\boldsymbol{T_{2,t}}= & M_{quar}\boldsymbol{T_{2,t}}+\boldsymbol{T_{2,t}}M_{quar}^{T}\\
 & -8\sum_{\mu}U_{L,\mu}\boldsymbol{T_{2,t}}U_{L,\mu}^{T}-iF_{B}^{T}
\end{array}
\end{equation}
with $M_{quar}=F_{A}+iF_{B}-4\sum_{\mu}U_{L,\mu}^{2}$. This is precisely
the result given in Reference {[}\citep{YKZhang}{]}.

\subsection{The removal of redundancy in $\bm{\left|T_{2m,t}\right\rangle }$
and the construction of the Liouvillian spectrum}

As mentioned before, $\bm{\left|T_{2m,t}\right\rangle }$ has a high
degree of redundancy, which limits the lattice size that Eq.(\ref{eq:T2m_M})
can compute. Therefore, we will seek to eliminate this redundancy.
First, by using the fully anti-symmetric projection $\mathbb{P}_{FA,2m}$,
we can eliminate redundancies caused by identical elements in the
subscripts, rewriting Eq.(\ref{eq:dTdt-many}) as
\begin{equation}
\begin{array}{rl}
 & \frac{d}{dt}\left(\mathbb{P}_{FA,2m}\boldsymbol{\left|T_{2m,t}\right\rangle }\right)\\
= & \left(\mathbb{P}_{FA,2m}\mathbf{F}_{M,2m}\mathbb{P}_{FA,2m}\right)\left(\mathbb{P}_{FA,2m}\boldsymbol{\left|T_{2m,t}\right\rangle }\right)\\
 & +\mathbb{P}_{FA,2m}\mathbf{G}_{M,2m}\boldsymbol{\left|T_{2m-2,t}\right\rangle }\\
= & \left(\mathbb{P}_{FA,2m}\mathbf{F}_{M,2m}\mathbb{P}_{FA,2m}\right)\left(\mathbb{P}_{FA,2m}\boldsymbol{\left|T_{2m,t}\right\rangle }\right)\\
 & +\mathbb{P}_{FA,2m}\left(|mF_{\gamma}+iC_{2m}^{2}F_{B}\rrangle\otimes\mathbf{I}_{2m-2}\right)\\
 & *\left(\mathbb{P}_{FA,2m-2}\boldsymbol{\left|T_{2m-2,t}\right\rangle }\right).
\end{array}\label{eq:dT_PM}
\end{equation}
In the derivation process, we used $\left[\mathbf{F}_{M,2m},\left(\mathbf{I}_{l}\otimes R\otimes\mathbf{I}_{2m-l-2}\right)\right]=0$
and $\mathbb{P}_{FA,2m}\left(\mathbf{I}_{l}\otimes R\otimes\mathbf{I}_{2m-l-2}\right)=\left(\mathbf{I}_{l}\otimes R\otimes\mathbf{I}_{2m-l-2}\right)\mathbb{P}_{FA,2m}=-\mathbb{P}_{FA,2m}$.
At this point, $\mathbb{P}_{FA,2m}\boldsymbol{\left|T_{2m,t}\right\rangle }$
still retains redundancies due to different permutations of the indices.
To eliminate this redundancy, we establish a fixed ordering of the
indices, which we will explain next.

For simplicity, we denote $\widetilde{\mathbf{F}}_{M,2m}=\mathbb{P}_{FA,2m}\mathbf{F}_{M,2m}\mathbb{P}_{FA,2m}$,
$\boldsymbol{\left|\widetilde{T}_{2m,t}\right\rangle }=\mathbb{P}_{FA,2m}\boldsymbol{\left|T_{2m,t}\right\rangle }$,
$\widetilde{\mathbf{G}}_{M,2m}=\mathbb{P}_{FA,2m}\left(|mF_{\gamma}+iC_{2m}^{2}F_{B}\rrangle\otimes\mathbf{I}_{2m-2}\right)$,
$\boldsymbol{\left|\vec{j}_{2m}\right\rangle }=\stackrel[r=1]{2m}{\overrightarrow{\bigotimes}}\overrightarrow{e}_{j_{r}}$,
$\boldsymbol{\left\langle \vec{j}_{2m}\right|}=\stackrel[r=1]{2m}{\overrightarrow{\bigotimes}}\overrightarrow{e}_{j_{r}}^{T}$.
Setting $j_{1}<j_{2}<\cdots<j_{2m}$ and using $\boldsymbol{\left|T_{2m,t}\right\rangle }_{j_{1},\cdots,j_{2m}}=\boldsymbol{\left\langle \vec{j}_{2m}\right|}\boldsymbol{\left.\widetilde{T}_{2m,t}\right\rangle }$,
left-multiplying Eq.(\ref{eq:dT_PM}) by $\boldsymbol{\left\langle \vec{j}_{2m}\right|}$
yields
\[
\begin{array}{rl}
 & \frac{d}{dt}\boldsymbol{\left\langle \vec{j}_{2m}\right|}\boldsymbol{\left.\widetilde{T}_{2m,t}\right\rangle }\\
= & \boldsymbol{\left\langle \vec{j}_{2m}\right|}\widetilde{\mathbf{F}}_{M,2m}\left(\underset{s_{1},\cdots,s_{2m}}{\sum}\boldsymbol{\left|\vec{s}_{2m}\right\rangle }\boldsymbol{\left\langle \vec{s}_{2m}\right|}\right)\boldsymbol{\left|\widetilde{T}_{2m,t}\right\rangle }\\
 & +\boldsymbol{\left\langle \vec{j}_{2m}\right|}\widetilde{\mathbf{G}}_{M,2m}\\
 & *\left(\underset{s_{1},\cdots,s_{2m-2}}{\sum}\boldsymbol{\left|\vec{s}_{2m-2}\right\rangle }\boldsymbol{\left\langle \vec{s}_{2m-2}\right|}\right)\boldsymbol{\left|\widetilde{T}_{2m-2,t}\right\rangle }\\
= & (2m)!\underset{s_{1}<\cdots<s_{2m}}{\sum}\boldsymbol{\left\langle \vec{j}_{2m}\right|}\widetilde{\mathbf{F}}_{M,2m}\boldsymbol{\left|\vec{s}_{2m}\right\rangle }\boldsymbol{\left\langle \vec{s}_{2m}\right|}\boldsymbol{\left.\widetilde{T}_{2m,t}\right\rangle }\\
 & +(2m-2)!\underset{s_{1}<\cdots<s_{2m-2}}{\sum}\boldsymbol{\left\langle \vec{j}_{2m}\right|}\widetilde{\mathbf{G}}_{M,2m}\boldsymbol{\left|\vec{s}_{2m-2}\right\rangle }\\
 & *\boldsymbol{\left\langle \vec{s}_{2m-2}\right|}\boldsymbol{\left.\widetilde{T}_{2m-2,t}\right\rangle }.
\end{array}
\]
Thus, we can arrange all $\boldsymbol{\left|T_{2m,t}\right\rangle }_{j_{1},\cdots,j_{2m}}$
satisfying $j_{1}<j_{2}<\cdots<j_{2m}$ in ascending order by the
sequence $(j_{1},j_{2},\cdots,j_{2m})$ into a column vector $\boldsymbol{\left|\bar{T}_{2m,t}\right\rangle }$
of dimension $C_{2L}^{2m}*1$. Simultaneously, we define a $C_{2L}^{2m}$-dimensional
square matrix $\bar{\mathbf{F}}_{M,2m}$ and an $C_{2L}^{2m}$-row,
$C_{2L}^{2m-2}$-column matrix $\bar{\mathbf{G}}_{M,2m}$, which satisfy
$(\bar{\mathbf{F}}_{M,2m})_{\vec{j}_{2m},\vec{s}_{2m}}=(2m)!\boldsymbol{\left\langle \vec{j}_{2m}\right|}\widetilde{\mathbf{F}}_{M,2m}\boldsymbol{\left|\vec{s}_{2m}\right\rangle }$
and $(\bar{\mathbf{G}}_{M,2m})_{\vec{j}_{2m},\vec{s}_{2m-2}}=(2m-2)!\boldsymbol{\left\langle \vec{j}_{2m}\right|}\widetilde{\mathbf{G}}_{M,2m}\boldsymbol{\left|\vec{s}_{2m-2}\right\rangle }$.
Consequently, we obtain the high-order correlation function dynamical
equation with redundancy removed as
\begin{equation}
\frac{d}{dt}\boldsymbol{\left|\bar{T}_{2m,t}\right\rangle }=\bar{\mathbf{F}}_{M,2m}\boldsymbol{\left|\bar{T}_{2m,t}\right\rangle }+\bar{\mathbf{G}}_{M,2m}\boldsymbol{\left|\bar{T}_{2m-2,t}\right\rangle }.\label{eq:dTdt_bar}
\end{equation}
Similarly, its solution is given by
\begin{equation}
\begin{array}{rl}
\boldsymbol{\left|\bar{T}_{2m,t}\right\rangle }= & e^{\bar{\mathbf{F}}_{M,2m}t}\boldsymbol{\left|\bar{T}_{2m,0}\right\rangle }+\\
 & e^{\bar{\mathbf{F}}_{M,2m}t}\int_{0}^{t}\left(e^{-\bar{\mathbf{F}}_{M,2m}\tau}\bar{\mathbf{G}}_{M,2m}\boldsymbol{\left|\bar{T}_{2m-2,\tau}\right\rangle }\right)d\tau.
\end{array}\label{eq:T2m_bar}
\end{equation}

From Eq.(\ref{eq:dTdt_bar}), we see that the even-parity Liouvillian
spectrum consists of the eigenvalues of $\bar{\mathbf{F}}_{M,2m}$
for $m=0,1,\cdots,L$. We denote the eigenvalues of matrix $\bar{\mathbf{F}}_{M,2m}$
as $eig(\bar{\mathbf{F}}_{M,2m})$. Clearly, $eig(\bar{\mathbf{F}}_{M,2m})\subset eig(\mathbf{F}_{M,2m})$.
Considering that $\mathbf{F}_{M,2m}=(\mathbf{F}_{M,2m})^{*}$, we
then have $\lambda_{M}^{*}\in eig(\bar{\mathbf{F}}_{M,2m})$ if $\lambda_{M}\in eig(\bar{\mathbf{F}}_{M,2m})$,
meaning that $eig(\bar{\mathbf{F}}_{M,2m})$ appears in the form of
conjugate pairs. By referring to the derivation of $\Re(\alpha_{s})\geqslant0$
in Appendix A.2, it is easy to deduce that $\mathbf{F}_{M,2m}+(\mathbf{F}_{M,2m})^{T}\preceq0$.
Further, using $\mathbf{F}_{M,2m}=(\mathbf{F}_{M,2m})^{*}$, we can
conclude that $\Re\left[eig(\bar{\mathbf{F}}_{M,2m})\right]\leqslant0$.
Since $\mathbf{F}_{M,0}=0$, this implies $eig(\bar{\mathbf{F}}_{M,0})=0$.
These three properties correspond to the preservation of Hermiticity,
semi-positive definiteness, and trace in the evolution process of
Eq.(\ref{eq:lindblad}). In Section \ref{subsec:Constructing-the-Liouvillian-spectrum},
we observed that the spectrum corresponding to odd parity can be obtained
by simply replacing $\mathbf{F}_{2m}$ with $\mathbf{F}_{n}$ and
taking $n=1,3,\cdots,2L-1$. Thus, we have reason to conjecture that
$eig(\bar{\mathbf{F}}_{M,n})$ represents the spectrum of odd parity
when $(\bar{\mathbf{F}}_{M,n})_{\vec{j}_{n},\vec{s}_{n}}=(n)!\boldsymbol{\left\langle \vec{j}_{n}\right|}\mathbb{P}_{FA,n}\mathbf{F}_{M,n}\mathbb{P}_{FA,n}\boldsymbol{\left|\vec{s}_{n}\right\rangle }$
and $n=1,3,\cdots,2L-1$.

\subsection{An illustrative example}

\begin{figure}
\includegraphics[scale=0.2]{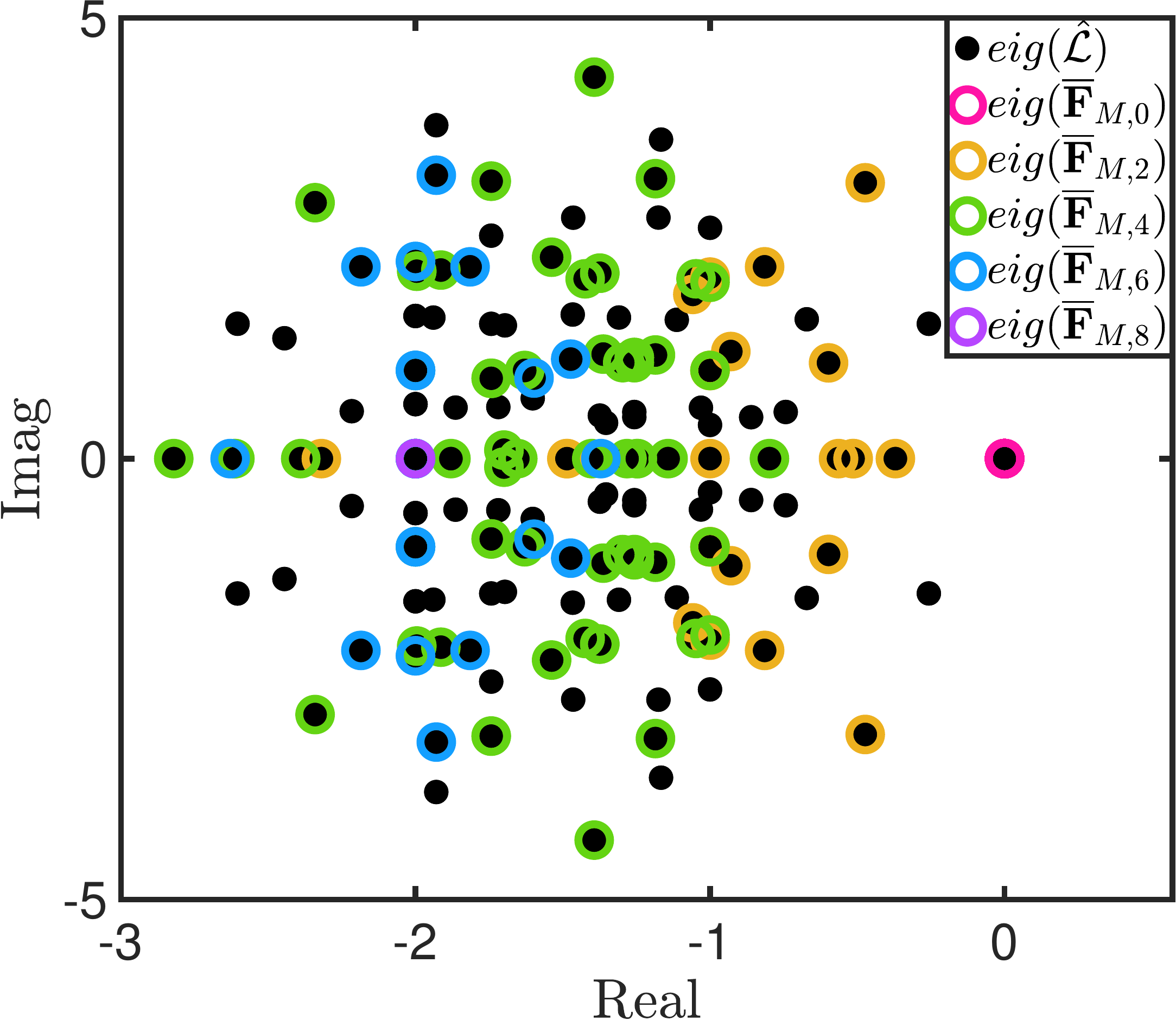}\caption{\label{fig:Spec_even}The correspondence between the Liouvillian spectrum
and the even parity spectrum. Here, we set $L=4$, $J=1$, $\gamma_{l}=\gamma_{g}=\gamma_{t}=0.5$.
In the figure, filled dots represent the spectrum of the Liouvillian
superoperator obtained via exact diagonalization, which corresponds
to the full Liouvillian spectrum. Open circles of different colors represent
the eigenvalues of the matrix $\bar{\mathbf{F}}_{M,2m}$ corresponding
to different $m$, which correspond to the even parity Liouvillian spectrum.}
\end{figure}

\begin{figure}
\includegraphics[scale=0.2]{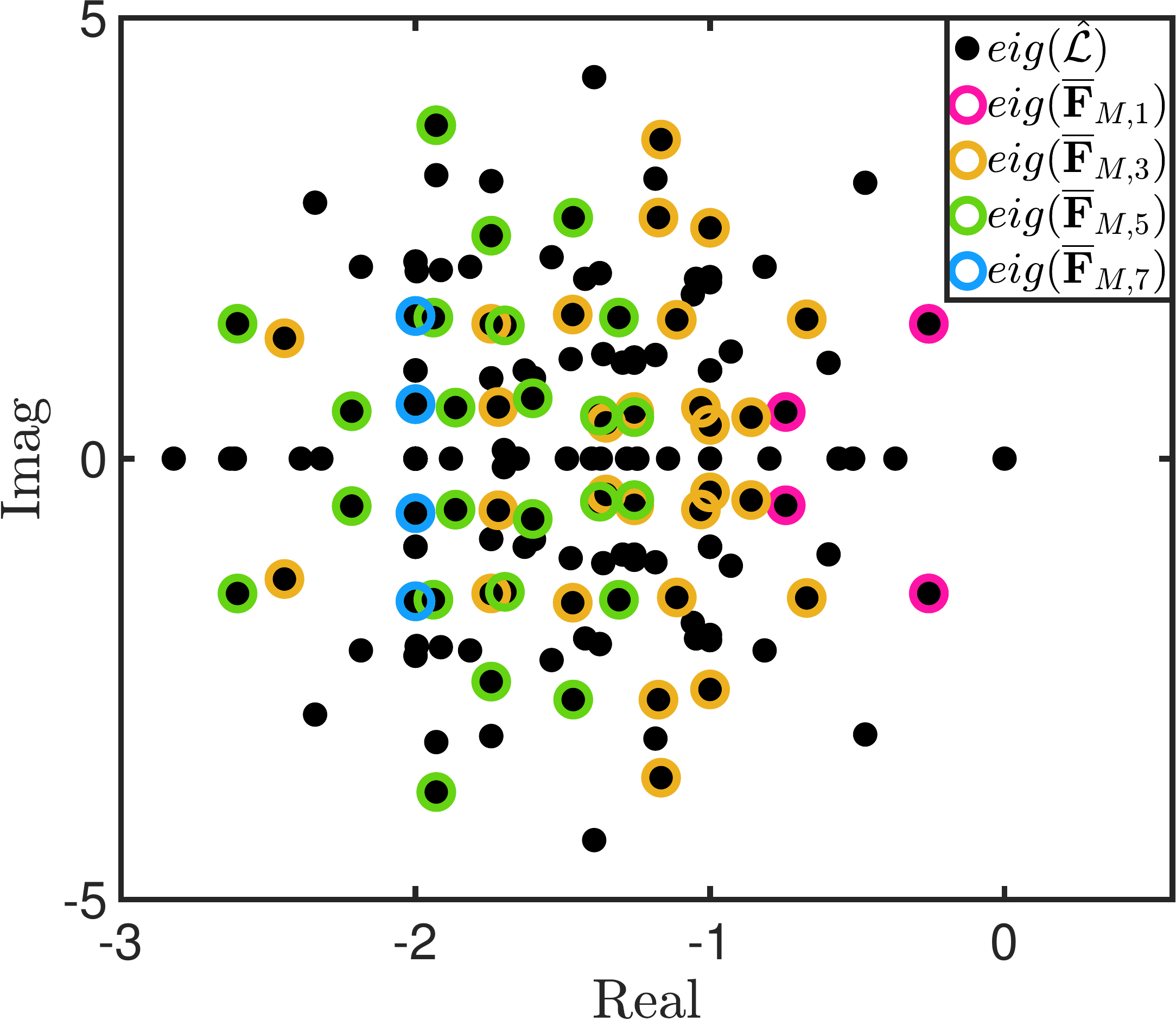}\caption{\label{fig:Spec_odd}The correspondence between the Liouvillian spectrum
and the conjectured odd-parity spectrum. We similarly set $L=4$,
$J=1$, $\gamma_{l}=\gamma_{g}=\gamma_{t}=0.5$. In the figure, filled
dots represent the full Liouvillian spectrum obtained via exact diagonalization,
while open circles of different colors represent the eigenvalues of
the matrix $\bar{\mathbf{F}}_{M,n}$ corresponding to different $n$,
which we conjecture to be the odd-parity Liouvillian spectrum.}
\end{figure}

\begin{figure}
\includegraphics[scale=0.2]{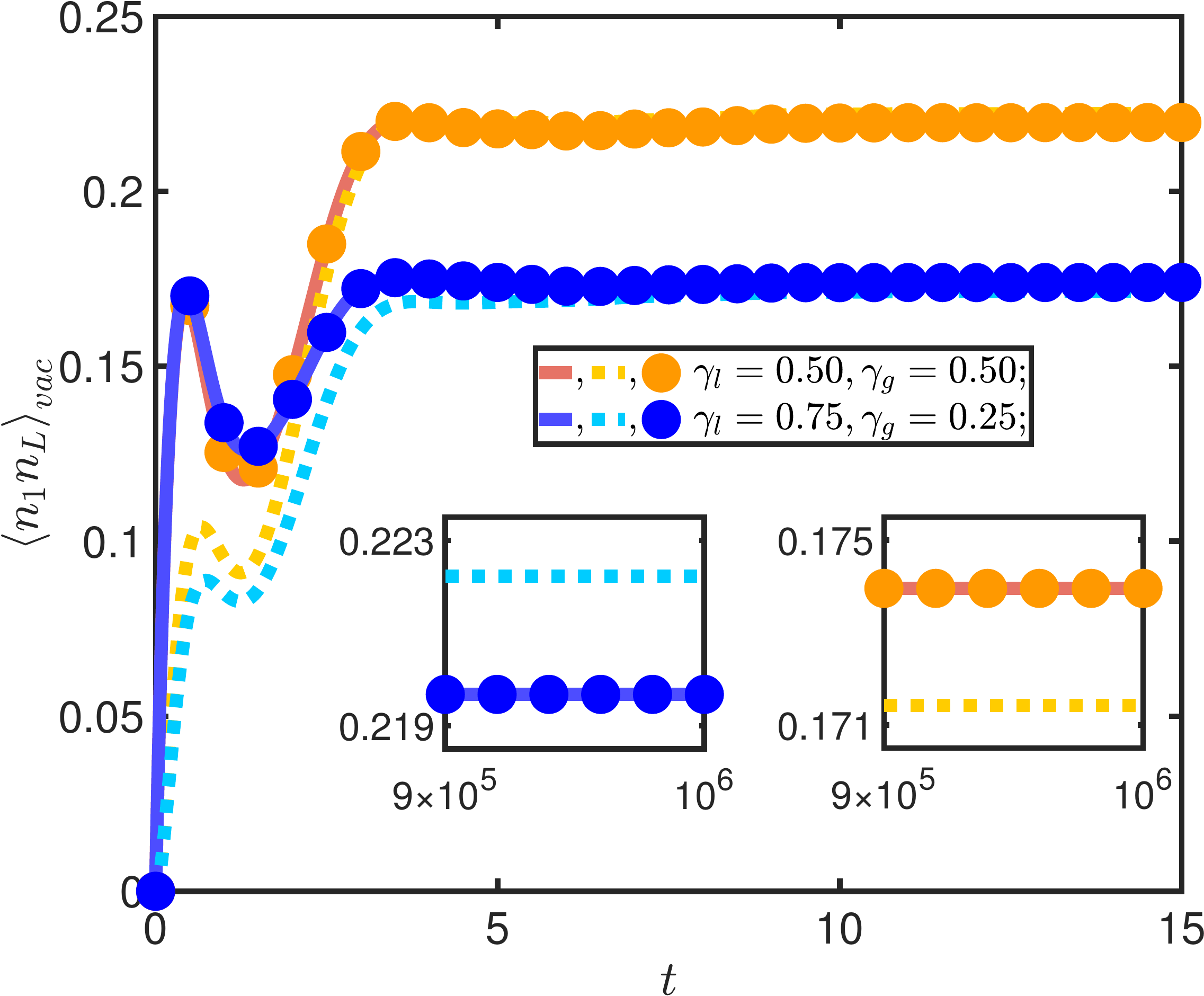}\caption{\label{fig:n1nL_four}The calculation results of $\left\langle n_{1}n_{L}\right\rangle $
under different conditions after adding quadratic dissipation. We
fix $L=4$, $J=1$, and $\gamma_{t}=0.5$, and different colors represent
two different choices of dissipative strength for the linear
dissipation, $\gamma_{l}=\gamma_{g}=0.5$ and $\gamma_{l}=0.75$,
$\gamma_{g}=0.25$, the initial state is chosen as the vacuum state.
In the figure, solid lines represent results obtained from Eq.(\ref{eq:T2_bar},\ref{eq:T4_bar}),
dashed lines represent results from the Wick's theorem, and filled dots
represent results from exact diagonalization. The inset shows the
results of the non-equilibrium steady state when the evolution time
reaches $10^{6}$.}
\end{figure}

\begin{figure}
\includegraphics[scale=0.2]{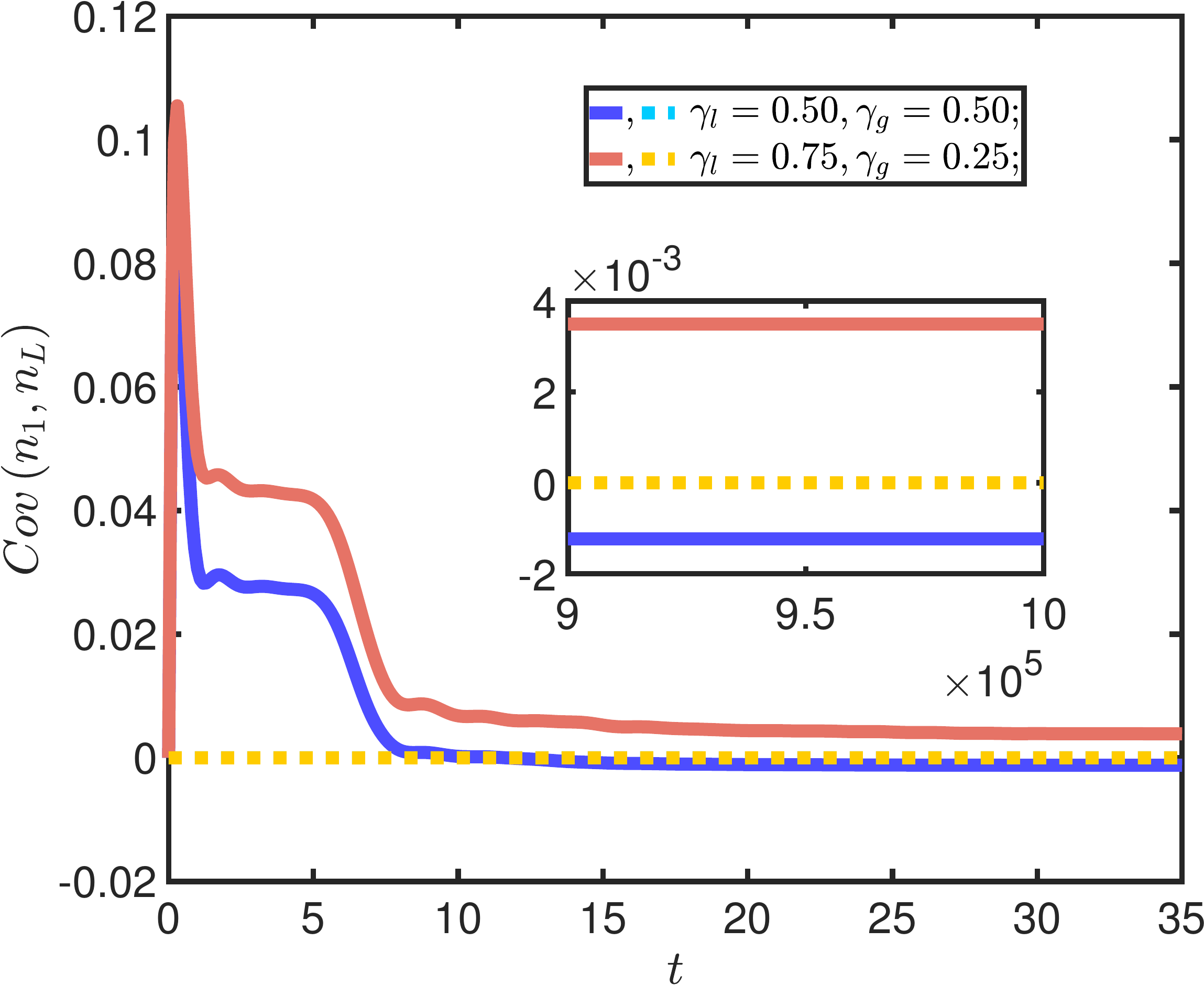}\caption{\label{fig:Cov_four}The particle number covariance $Cov\left(n_{1},n_{L}\right)$
at the left and right boundaries under different conditions. Here,
we set $L=12$, $J=1$, $\gamma_{t}=0.5$, with different colors representing
the cases $\gamma_{l}=\gamma_{g}=0.5$ and $\gamma_{l}=0.75$, $\gamma_{g}=0.25$,
the initial state is chosen as the vacuum state. The solid lines represent
the results calculated using Eq.(\ref{eq:T2_bar},\ref{eq:T4_bar}),
while the dashed lines represent the results calculated using
the Wick's theorem. In this case, the results from the two choices of $\gamma_{l},\ \gamma_{g}$
using the Wick's theorem overlap. The inset shows the results for the
non-equilibrium steady state when the evolution time reaches $10^{6}$.}
\end{figure}

The dimension of $\bar{\mathbf{F}}_{M,n}$ is $C_{2L}^{n}$, considering
that $2^{2L}=\sum_{s=0}^{2L}C_{2L}^{s}$, this indicates that Eq.(\ref{eq:dTdt_bar})
reflects the idea of dividing the entire Fock space into a series
of subspaces, each with dimension $C_{2L}^{s}$, through the closure
of correlation functions. This clearly enables us to calculate larger
lattice sizes. If we assume that $\det\left(\bar{\mathbf{F}}_{M,2}\right)\neq0$,
$\det\left(\bar{\mathbf{F}}_{M,4}\right)\neq0$, then from Eq.(\ref{eq:T2m_bar})
we obtain
\begin{equation}
\begin{array}{rl}
\boldsymbol{\left|\bar{T}_{2,t}\right\rangle }= & e^{\bar{\mathbf{F}}_{M,2}t}\left(\boldsymbol{\left|\bar{T}_{2,0}\right\rangle }+(\bar{\mathbf{F}}_{M,2})^{-1}\bar{\mathbf{G}}_{M,2}\right)\\
 & -(\bar{\mathbf{F}}_{M,2})^{-1}\bar{\mathbf{G}}_{M,2},
\end{array}\label{eq:T2_bar}
\end{equation}
\begin{equation}
	\begin{array}{rl}
		\boldsymbol{\left|\bar{T}_{4,t}\right\rangle }= & (\bar{\mathbf{F}}_{M,4})^{-1}\bar{\mathbf{G}}_{M,4}(\bar{\mathbf{F}}_{M,2})^{-1}\bar{\mathbf{G}}_{M,2}+e^{\bar{\mathbf{F}}_{M,4}t}\left\{ \boldsymbol{\left|\bar{T}_{4,0}\right\rangle }\right.\\
		& -(\bar{\mathbf{F}}_{M,4})^{-1}\bar{\mathbf{G}}_{M,4}(\bar{\mathbf{F}}_{M,2})^{-1}\bar{\mathbf{G}}_{M,2}\\
		& +\left(e^{-\bar{\mathbf{F}}_{M,4}t}K-Ke^{\bar{\mathbf{F}}_{M,2}t}\right)\\
		& \left.*\left(\boldsymbol{\left|\bar{T}_{2,t}\right\rangle }+(\bar{\mathbf{F}}_{M,2})^{-1}\bar{\mathbf{G}}_{M,2}\right)\right\} ,
	\end{array}\label{eq:T4_bar}
\end{equation}
where $K$ satisfies $\bar{\mathbf{F}}_{M,4}K-K\bar{\mathbf{F}}_{M,2}+\bar{\mathbf{G}}_{M,4}=0$.
Considering $\mathbb{P}_{FA,2m}\left(\mathbf{I}_{l}\otimes R\otimes\mathbf{I}_{2m-l-2}\right)$$=\left(\mathbf{I}_{l}\otimes R\otimes\mathbf{I}_{2m-l-2}\right)\mathbb{P}_{FA,2m}=-\mathbb{P}_{FA,2m}$,
we know that $\mathbb{P}_{FA,n}\mathbf{F}_{M,n}\mathbb{P}_{FA,n}=\mathbb{P}_{FA,n}\left(\mathbf{I}_{n-2}\otimes M_{F,n}\right)\mathbb{P}_{FA,n}$
with $M_{F,n}=n\mathbf{I}_{1}\otimes\left\{ \mathbf{F}_{1}-\underset{\mu}{\sum}U_{a,\mu}U_{a,\mu}^{\dagger}+\Re\left[\underset{\mu}{\sum}\left(Tr(U_{L,\mu}^{*})*U_{a,\mu}\right)\right]\right\} -2C_{n}^{2}\underset{\mu}{\sum}\Re\left(U_{a,\mu}\otimes U_{a,\mu}^{\dagger}\right)$.
Noting the two representations of $\mathbb{P}_{FA,n}$ given in Eq.(\ref{eq:P_FA}),
it becomes clear that $\bar{\mathbf{F}}_{M,n}$ and $\bar{\mathbf{G}}_{M,2m}$
can be constructed straightforwardly through an iterative method.

To validate the above conclusion, we use the model from Section \ref{subsec:Gaussian-initial-state} described by Eq.(\ref{model}) and Eq.(\ref{model-L})
with additionally introduced two types of dissipation:
\begin{equation}
L_{t,1}=\sqrt{\gamma_{t}}a_{1}a_{L},~~~ L_{t,2}=\sqrt{\gamma_{t}}a_{1}^{\dagger}a_{L}^{\dagger}.\label{eq:Lt_1_2}
\end{equation}
At this point, the condition in Eq.(\ref{eq:closure}) is clearly
satisfied. Choosing the parameters $L=4$, $J=1$, $\gamma_{l}=\gamma_{g}=\gamma_{t}=0.5$,
we present the correspondence between the full Liouvillian spectrum
and the even (odd) parity spectra, as shown in Fig.\ref{fig:Spec_even} (Fig.\ref{fig:Spec_odd}).
The even parity spectrum indeed corresponds to the eigenvalues of
$\bar{\mathbf{F}}_{M,2m}$ for $m=0,1,\cdots,L$, as derived, while
the odd parity spectrum confirms our hypothesis, corresponding to
the eigenvalues of $\bar{\mathbf{F}}_{M,n}$ for $n=1,3,\cdots,2L-1$.

Keeping the model unchanged, with the parameters $L=4$, $J=1$, $\gamma_{t}=0.5$,
we plot the dynamical evolution of $\left\langle n_{1}n_{L}\right\rangle $
for the cases corresponding to $\gamma_{l}=\gamma_{g}=0.5$ and $\gamma_{l}=0.75$,
$\gamma_{g}=0.25$ with the initial state being the empty state, as
shown in Fig.\ref{fig:n1nL_four}. Solid lines represent the results
obtained from Eqs.(\ref{eq:T2_bar},\ref{eq:T4_bar}), dashed lines
represent the results calculated using the Wick's theorem, and dots
represent the exact diagonalization results. We observe that, despite
the initial state being Gaussian, the Wick\textquoteright s theorem
does not hold during the dynamics. This indicates that the dynamics
described by Eqs.(\ref{eq:lindblad},\ref{eq:H},\ref{eq:L},\ref{eq:Lt_mu})
do not preserve Gaussianity. As the evolution time becomes sufficiently
long, we observe that the difference between the actual results and
those obtained using the Wick's theorem does not diminish, indicating
that the non-equilibrium steady state is no longer a Gaussian state.
Although the Wick's theorem does not agree with the actual results, the evolution behaviors display similar behaviors as the actual one. This is due to
the term $\boldsymbol{\left|T_{4,t}\right\rangle }_{1,L,L+1,2L}$
is relatively small compared to the other terms, when we calculate $\left\langle n_{1}n_{L}\right\rangle =0.25-0.5i\left(\boldsymbol{\left|T_{2,t}\right\rangle }_{1,L+1}+\boldsymbol{\left|T_{2,t}\right\rangle }_{L,2L}\right)+\boldsymbol{\left|T_{4,t}\right\rangle }_{1,L,L+1,2L}$.
By eliminating the influence of lower-order terms and focusing on the true fourth-order correlation effects, such as the particle number covariance at the left and right
edges $Cov\left(n_{1},n_{L}\right)=\left\langle n_{1}n_{L}\right\rangle -\left\langle n_{1}\right\rangle \left\langle n_{L}\right\rangle $,
we plot the dynamics of $Cov\left(n_{1},n_{L}\right)$ for the cases
$\gamma_{l}=\gamma_{g}=0.5$ and $\gamma_{l}=0.75$, $\gamma_{g}=0.25$
in Fig.\ref{fig:Cov_four}, where we take $L=12$, $J=1$ and $\gamma_{t}=0.5$.
Here, we can see that the Wick's theorem fails to capture  the information of covariance.
As shown in the inset, when the evolution
time is sufficiently large, the covariance corresponding to the non-equilibrium
steady state is not zero, and there are both positive and negative
values. This indicates that in the non-equilibrium steady state, there
are effective attractive or repulsive interactions  between particles on the left and right boundaries,
depending on $\gamma_{l},\ \gamma_{g}\ ,\gamma_{t}$. This clearly
cannot be captured by the Wick's theorem. Below, we will provide a rough
explanation for this phenomenon.
\begin{figure}
\includegraphics[scale=0.2]{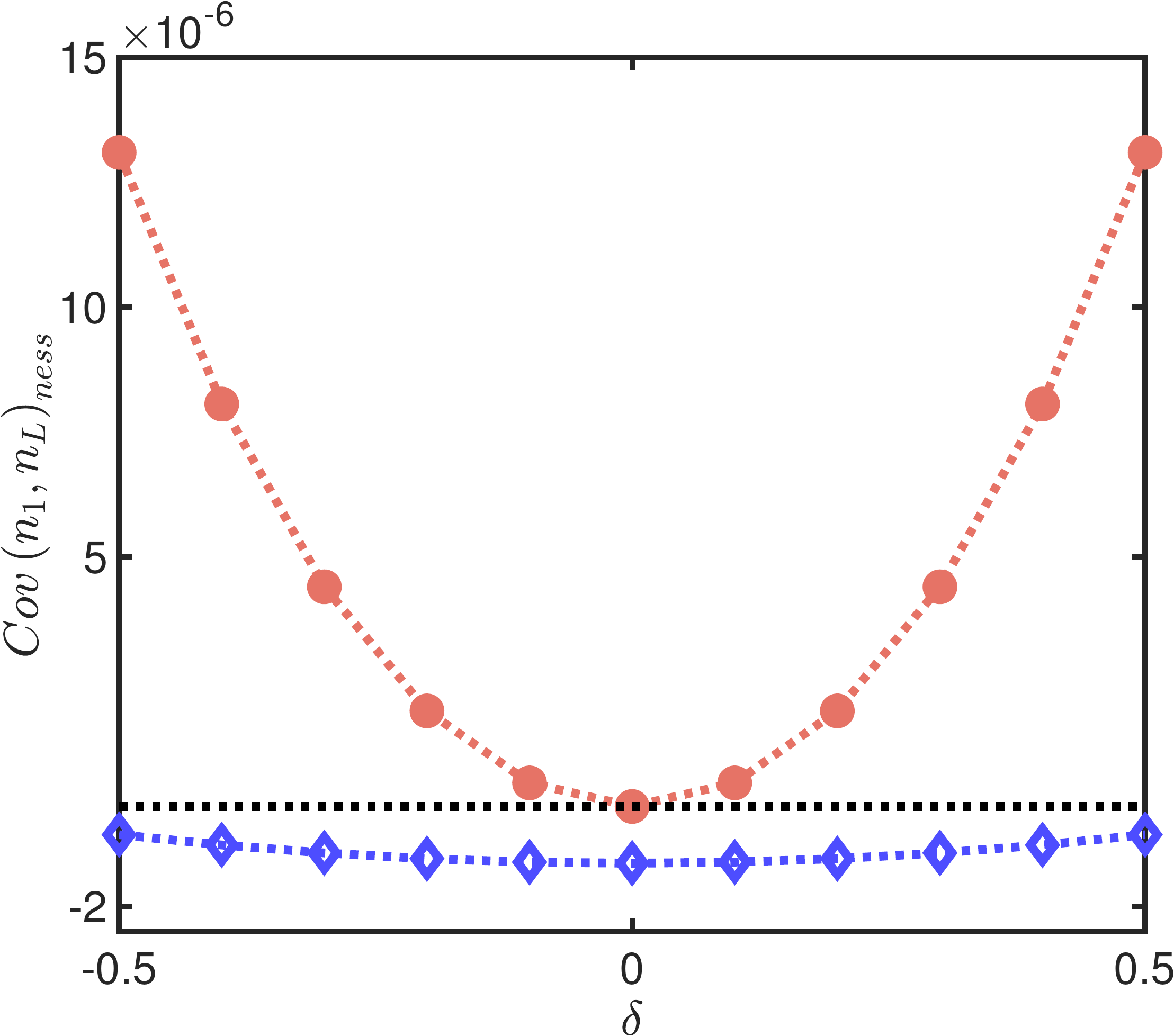}
\caption{\label{fig:small_gamma}The curve of the particle number covariance
at the left and right boundaries in the non-equilibrium steady state
as a function of $\delta$. Here, we fix $L=12$ and $J=1$. First,
we select $\gamma_{l}=10^{-4}(1+\delta)$, $\gamma_{g}=10^{-4}(1-\delta)$,
and $\gamma_{t}=0.5$, the $Cov\left(n_{1},n_{L}\right)_{ness}$ calculated
using Eq.(\ref{eq:ness_bar}) is represented by diamond-shaped points.
Then, we select $\gamma_{l}=0.5(1+\delta)$, $\gamma_{g}=0.5(1-\delta)$,
and $\gamma_{t}=10^{-4}$, and the computed $Cov\left(n_{1},n_{L}\right)_{ness}$
is represented by solid points. The black dashed line represents $Cov\left(n_{1},n_{L}\right)_{ness}=0$.}
\end{figure}

\begin{figure}
\includegraphics[scale=0.19]{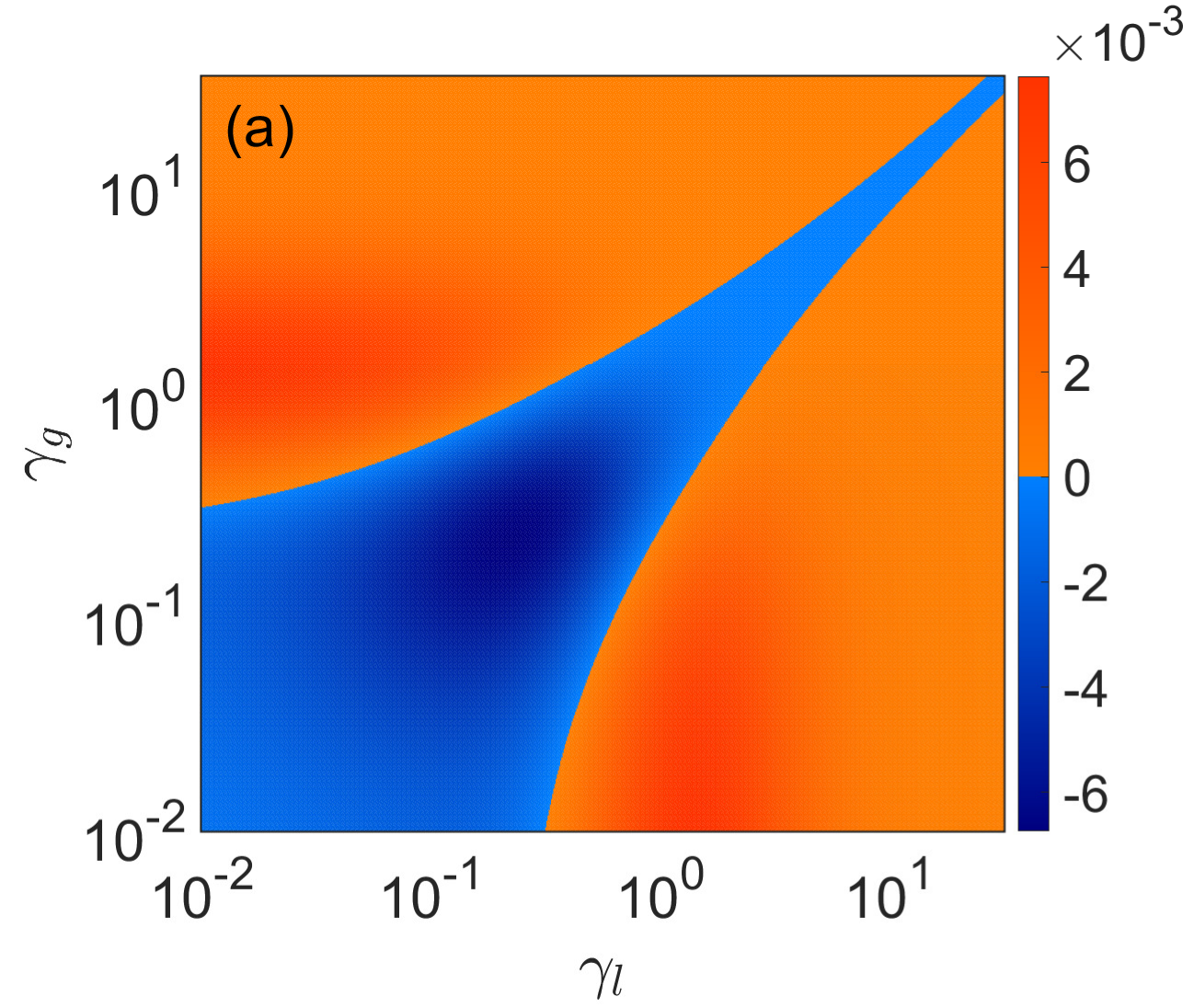}\includegraphics[scale=0.19]{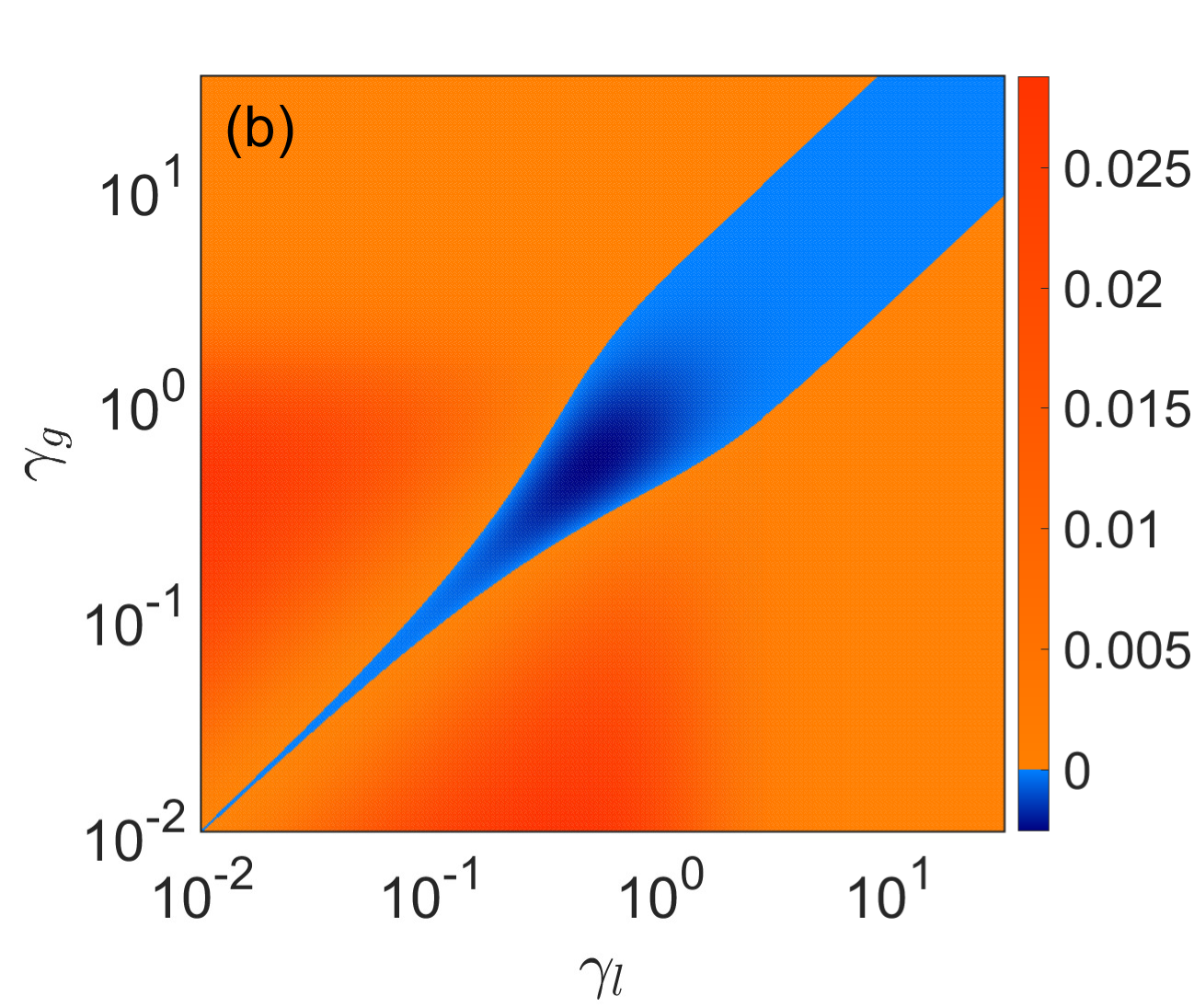}\caption{\label{fig:L_4_Cov_ness}Plot of the variation of $Cov\left(a_{1}^{\dagger}a_{1},a_{L}^{\dagger}a_{L}\right)_{ness}$
with parameters $\gamma_{l}$ and $\gamma_{g}$. Here we select $L=4$
and $J=1$, with $\gamma_{t}=1$ in subfigure (a) and $\gamma_{t}=\frac{1}{2}(\gamma_{l}+\gamma_{g})$
in subfigure (b).}
\end{figure}

For the unique non-equilibrium steady state, we have
\begin{equation}
\bar{\mathbf{F}}_{M,2m}\boldsymbol{\left|\bar{T}_{2m,\infty}\right\rangle }+\bar{\mathbf{G}}_{M,2m}\boldsymbol{\left|\bar{T}_{2m-2,\infty}\right\rangle }=0.\label{eq:ness_bar}
\end{equation}
Since we have already eliminated the redundant terms in $\boldsymbol{\left|\bar{T}_{2m,\infty}\right\rangle }$,
using Eq.(\ref{eq:ness_bar}), we can provide analytical expressions
for all even-order correlation functions in the non-equilibrium steady
state for smaller system sizes. For the model of interest here, with
$L=4$, we have
\begin{equation}
\left\langle n_{1}\right\rangle _{ness}=\frac{J^{2}(\gamma_{g}+2\gamma_{t})}{\Gamma},\label{eq:n1_ness}
\end{equation}
\begin{equation}
\left\langle n_{L}\right\rangle _{ness}=1-\frac{J^{2}(\gamma_{l}+2\gamma_{t})}{\Gamma},\label{eq:nL_ness}
\end{equation}
where $\Gamma=(\gamma_{l}+\gamma_{g})(J^{2}+\gamma_{l}\gamma_{g}+2\gamma_{t}^{2})+\gamma_{t}(4J^{2}+\gamma_{l}^{2}+\gamma_{g}^{2}+4\gamma_{l}\gamma_{g})$.
The expression for $Cov\left(n_{1},n_{L}\right)_{ness}$ is overly lengthy and cumbersome, and thus is omitted here. However, from its expression, we can deduce that
$Cov\left(n_{1},n_{L}\right)_{ness}=0$ when one of the conditions: $J=0$, $\gamma_{l}=\gamma_{g}=0$, and $\gamma_{t}=0$,
is fulfilled. In other words, the absence of any one of the Hamiltonian,
linear dissipation, or quadratic dissipation terms will cause the disappearance of effective  interactive between the particles on the boundaries. This conclusion is not
limited to $L=4$. Specifically, in the absence of quadratic dissipation,
according to Reference {[}\citep{M. Znidaric 2010b}{]},
the non-equilibrium steady state $\rho_{ness}$ can be written as
a matrix product state. At this point we find $\left\langle n_{1}n_{L}\right\rangle _{ness}=\frac{x_{1}x_{L}+x_{1}+x_{L}+1}{4}$,
$\left\langle n_{1}\right\rangle _{ness}=\frac{x_{1}+1}{2}$, $\left\langle n_{L}\right\rangle _{ness}=\frac{x_{L}+1}{2}$
with $x_{1}=\frac{J^{2}(\gamma_{g}-\gamma_{l})-\gamma_{l}\gamma_{g}(\gamma_{l}+\gamma_{g})}{(J^{2}+\gamma_{l}\gamma_{g})(\gamma_{l}+\gamma_{g})}$,
$x_{L}=\frac{J^{2}(\gamma_{g}-\gamma_{l})+\gamma_{l}\gamma_{g}(\gamma_{l}+\gamma_{g})}{(J^{2}+\gamma_{l}\gamma_{g})(\gamma_{l}+\gamma_{g})}$,
which are independent of system size, and it is easy to see that $Cov\left(n_{1},n_{L}\right)_{ness}=0$.
In the absence of linear dissipation, the nonequilibrium steady state
is $\rho_{ness}=\frac{1}{2^{L}}\mathbb{I}$, and clearly, in this
case, $Cov\left(n_{1},n_{L}\right)_{ness}=0$. When $J=0$, the system
effectively reduces to  only two lattice sites, which is identical to the case of $L=4$ with $J=0$,
naturally resulting
in $Cov\left(n_{1},n_{L}\right)_{ness}=0$.

To clarify the reasons for the emergence of boundary attraction and
repulsion interactions, we consider treating linear dissipation and
quadratic dissipation as perturbations for analysis. First, consider
the quadratic dissipation as a perturbation, i.e., $\gamma_{t}\ll\gamma_{l},\ \gamma_{g}$.
We select $L=12$, $J=1$, $\gamma_{l}=0.5(1+\delta)$, $\gamma_{g}=0.5(1-\delta)$,
$\gamma_{t}=10^{-4}$, and the actual results are shown as solid points
in Fig.\ref{fig:small_gamma}. Except for the special case with $\delta=0$,
we have $Cov\left(n_{1},n_{L}\right)_{ness}>0$. When only linear dissipation is present, if $x_{1}<0$,
$x_{L}>0$, then $\left\langle n_{1}\right\rangle _{ness,1}<0.5<\left\langle n_{L}\right\rangle _{ness,1}$.
The left boundary tends to occupy no particles and the right boundary
tends to occupy particles. Here, the subscript $1$ indicates the
inclusion of only linear dissipation, while the subscript $2$ in
subsequent formulas represents the inclusion of only quadratic dissipation.
When only quadratic dissipation is present, $\left\langle n_{1}\right\rangle _{ness,2}=\left\langle n_{L}\right\rangle _{ness,2}=0.5$,
the left and right boundaries are equally populated. Thus, we might
naturally think that after adding quadratic dissipation as a perturbation,
the two boundaries would tend towards opposite occupation configurations,
leading to the incorrect conclusion that the boundaries exhibit repulsive
effects, i.e., $Cov\left(n_{1},n_{L}\right)_{ness}<0$, which is exactly
the opposite of the result given by the data. This reasoning overlooks
the conclusion that when only linear dissipation is added, $Cov\left(n_{1},n_{L}\right)_{ness,1}=0$.
Because $Cov\left(n_{1},n_{L}\right)_{ness,1}=0$ exists, the main
effect on $Cov\left(n_{1},n_{L}\right)_{ness}$ after adding quadratic
dissipation as a perturbation comes from the perturbation term itself.
That is, the most important factor is the transition from opposite
occupation configurations at the two boundaries to the same occupation
configuration due to the perturbation. Since it tends towards the
same occupation configuration, it naturally manifests as an attractive
interaction: $Cov\left(n_{1},n_{L}\right)_{ness}>0$. Next, consider
taking the linear dissipation as a perturbation. We select $L=12$,
$J=1$, $\gamma_{t}=0.5$, $\gamma_{l}=10^{-4}(1+\delta)$, $\gamma_{g}=10^{-4}(1-\delta)$,
and the actual results are shown as diamond-shaped points in Fig.\ref{fig:small_gamma},
which indicate $Cov\left(n_{1},n_{L}\right)_{ness}<0$. We can explain
this for the same reason, since $Cov\left(n_{1},n_{L}\right)_{ness,2}=0$,
when linear dissipation is taken as a perturbation, the main effect
still comes from the transition tendency from the same occupation
configuration to the opposite occupation configurations. Since it
tends towards opposite occupation configurations, it manifests as
a repulsive interaction: $Cov\left(n_{1},n_{L}\right)_{ness}<0$.
This conclusion is quite counterintuitive.

From this, we see that although $Cov\left(n_{1},n_{L}\right)_{ness}=0$
when only one type of dissipation is added, different perturbation
cases lead to different signs of $Cov\left(n_{1},n_{L}\right)_{ness}$,
indicating that the interaction between boundary particle numbers
arises from the competition between linear and quadratic dissipation.
Here the Hamiltonian provides a platform for their competition.
Using Eq.(\ref{eq:ness_bar}),
we have plotted a more comprehensive covariance image in Fig.\ref{fig:L_4_Cov_ness}.
In subplot (a), we set $J=1$ and $\gamma_{t}=1$, which illustrates
the boundary interactions under the competition of two dissipation
types with different strengths. In subplot (b), we set $J=1$ and
$\gamma_{t}=\frac{1}{2}(\gamma_{l}+\gamma_{g})$, keeping the strengths
of both types of dissipation roughly the same. As shown in Fig.\ref{fig:L_4_Cov_ness},
the covariance $Cov\left(n_{1},n_{L}\right)_{ness}$ is symmetric
with respect to $\gamma_{l}$ and $\gamma_{g}$. This is because after
performing the transformation $a_{j}\rightarrow(-1)^{j}a_{L+1-j}^{\dagger}$
and swapping $\gamma_{l}$ and $\gamma_{g}$, Eq.(\ref{eq:lindblad})
remains unchanged. From this symmetry, we can deduce that $\left\langle n_{1}\right\rangle _{\gamma_{l}\leftrightarrow\gamma_{g}}=1-\left\langle n_{L}\right\rangle $,
$\left\langle n_{1}n_{L}\right\rangle _{\gamma_{l}\leftrightarrow\gamma_{g}}=1-\left\langle n_{1}\right\rangle -\left\langle n_{L}\right\rangle +\left\langle n_{1}n_{L}\right\rangle $,
$Cov\left(n_{1},n_{L}\right)_{\gamma_{l}\leftrightarrow\gamma_{g}}=Cov\left(n_{1},n_{L}\right)$.
Here, the subscript $\gamma_{l}\leftrightarrow\gamma_{g}$ represents
the exchange of parameters $\gamma_{l}$ and $\gamma_{g}$.

\section{Summary and outlook}

In summary, we developed a systematical framework for solving all even-order correlation functions via the study of a series of closed dynamical recursion equations for general quadratic Liouvillian systems and some quartic Liouvillian systems, and
demonstrated how to utilize these dynamical equations to construct the Liouvillian spectrum.
We first considered the quadratic Liouvillian superoperator,
presenting its eigen-decomposition and dynamical expressions for high-order
correlation functions. This represents a generalization of the Wick's
theorem. On this basis, we demonstrated the preservation of Gaussianity
during the system's dynamical evolution. For non-Gaussian initial
states, we provided a convenient computational method based on the
dynamical expressions of high-order correlation functions and showed
that when the nonequilibrium steady state is unique, it must be a
Gaussian state. Additionally, we demonstrated how to construct the
Liouvillian spectrum through these equations. We then extended the results
to the quartic Liouvillian superoperator, establishing the necessary
and sufficient conditions for the closure of correlation functions
and deriving the dynamical equations for high-order correlation
functions under these conditions. By eliminating redundancies in $\bm{\left|T_{2m,t}\right\rangle }$,
we showed how the closure of correlation functions enables the partitioning
of the entire Fock space into subspaces related to correlation functions,
and upon which we constructed the Liouvillian spectrum. We verified
all conclusions with a simple model, illustrating that this approach
allows the computation of high-order correlation functions and the
Liouvillian spectrum for larger lattice sizes. On this basis, we further
discussed the emergence of attractive or repulsive interactions between
boundary particle numbers, driven by the competition between linear
and quadratic dissipation.

Our work provides expressions for high-order correlation functions,
offering a key tool for gaining a deeper understanding of the complexity
of quantum systems. This aids in uncovering deeper dynamics and statistical
correlations within the system, as well as capturing the rich interactions
and coherence characteristics among particles. Via the calculation of concrete examples, we observed
that the odd-parity Liouvillian spectrum can be inferred by analogy with the even-parity Liouvillian spectrum. The underlying principle remains unclear and
is worth exploring. In this work, we only provided results for the fermionic
version, and it should be straightforward to apply this process to
obtain the bosonic version. This will be a task in our future work.
\begin{acknowledgments}
We thank C. G. Liang, Y. Zhang, Z.-Y. Zheng and X. D. Liu for helpful
discussions.
This work is supported by Key Research and Development Program
of China (Grant No. 2023YFA1406704) and the NSFC under
Grants No. 12174436, No.12474287 and No. T2121001.
\end{acknowledgments}

\appendix

\section{Proofs of some conclusions in the text}

\subsection{\label{subsec:Proof IP_I}Proof of $\llangle\mathbb{I}|\mathcal{\hat{P}}=\llangle\mathbb{I}|$}

We denote the vacuum state corresponding to fermion $a_{j}$ as $\left|0\right\rangle _{a}$,
meaning $a_{j}\left|0\right\rangle _{a}=0$. Thus, we can provide
the expression for $\mathbb{I}$ as
\[
\mathbb{I}=\underset{\overrightarrow{\nu}}{\sum}\left[\left(\stackrel[s=1]{L}{\overrightarrow{\prod}}\left(a_{s}^{\dagger}\right)^{\nu_{s}}\right)\left|0\right\rangle _{a}\tensor[_{a}]{\left\langle 0\right|}{}\left(\stackrel[s=1]{L}{\overleftarrow{\prod}}\left(a_{s}\right)^{\nu_{s}}\right)\right]
\]
where $\stackrel[s=1]{L}{\overrightarrow{\prod}}$ and $\stackrel[s=1]{L}{\overleftarrow{\prod}}$
represent successive rightward and leftward cumulative products as
$s$ increases, respectively, i.e., $\stackrel[s=1]{L}{\overrightarrow{\prod}}\left(a_{s}^{\dagger}\right)^{\nu_{s}}\coloneqq\left(a_{1}^{\dagger}\right)^{\nu_{1}}\left(a_{2}^{\dagger}\right)^{\nu_{2}}\cdots\left(a_{L}^{\dagger}\right)^{\nu_{L}}$
and $\stackrel[s=1]{L}{\overleftarrow{\prod}}\left(a_{s}\right)^{\nu_{s}}\coloneqq\left(a_{L}\right)^{\nu_{L}}\cdots\left(a_{2}\right)^{\nu_{2}}\left(a_{1}\right)^{\nu_{1}}$.
Then, we have
\[
\begin{array}{rl}
\llangle\mathbb{I}|= & \underset{\overrightarrow{\nu}}{\sum}\left(\tensor[_{a}]{\left\langle 0\right|}{}\otimes\tensor[_{a}]{\left\langle 0\right|}{}\right)\left[\left(\stackrel[s=1]{L}{\overleftarrow{\prod}}\left(a_{s}\right)^{\nu_{s}}\right)\otimes\left(\stackrel[s=1]{L}{\overleftarrow{\prod}}\left(a_{s}\right)^{\nu_{s}}\right)\right]\\
= & \underset{\overrightarrow{\nu}}{\sum}\tensor[_{d}]{\llangle0|}{}\left[\left(\stackrel[s=1]{L}{\overleftarrow{\prod}}\left(\hat{b}_{s}\right)^{\nu_{s}}\right)\left(\stackrel[s=1]{L}{\overleftarrow{\prod}}\left(\hat{c}_{s}\right)^{\nu_{s}}\right)\right]\\
= & \underset{\overrightarrow{\nu}}{\sum}\tensor[_{d}]{\llangle0|}{}r_{\overrightarrow{\nu}}\left[\left(\stackrel[s=1]{L}{\overleftarrow{\prod}}\left(\hat{d}_{s}\right)^{\nu_{s}}\right)\left(\stackrel[s=1]{L}{\overleftarrow{\prod}}\left(\hat{d}_{L+s}\right)^{\nu_{s}}\right)\right],
\end{array}
\]
where $\tensor[_{d}]{\llangle0|}{}=\tensor[_{a}]{\left\langle 0\right|}{}\otimes\tensor[_{a}]{\left\langle 0\right|}{}$
is the vacuum corresponding to the superoperator $\hat{d}_{s}^{\dagger}$,
satisfying $\tensor[_{d}]{\llangle0|}{}\hat{d}_{s}^{\dagger}=0$.
$r_{\overrightarrow{\nu}}$ is a number related to sequence $\overrightarrow{\nu}=(\nu_{1},\nu_{1},\cdots,\nu_{L})$,
introduced by the parity superoperator $\mathcal{\hat{P}}$, with
values of $\pm1$. Thus, it is easy to see that $\llangle\mathbb{I}|\mathcal{\hat{P}}=\llangle\mathbb{I}|$.

\subsection{\label{subsec:Proof_T2inf}Proof of the existence of solutions to
the equation $\mathbf{F}_{1}\boldsymbol{T_{2,\infty}}+\boldsymbol{T_{2,\infty}}\mathbf{F}_{1}^{T}-i\left(F_{B}\right)^{T}=0$}

We will now prove that there must exist a $\boldsymbol{T_{2,\infty}}$
such that $\mathbf{F}_{1}\boldsymbol{T_{2,\infty}}+\boldsymbol{T_{2,\infty}}\mathbf{F}_{1}^{T}-i\left(F_{B}\right)^{T}=0$.
Vectorizing this implies $\mathbf{F}_{2}|\boldsymbol{T_{2,\infty}}\rrangle=i|F_{B}^{T}\rrangle$,
so we only need to prove that $rank\left[\mathbf{F}_{2}\right]=rank\left[\begin{array}{cc}
\mathbf{F}_{2}, & i|F_{B}^{T}\rrangle\end{array}\right]$, this is equivalent to showing that for any $\boldsymbol{\left\langle \phi\right|}$
satisfying $\boldsymbol{\left\langle \phi\right|}\mathbf{F}_{2}=0$,
the equation $\boldsymbol{\left\langle \phi\right|}F_{B}^{T}\rrangle=0$
holds. Given that $\mathbf{F}_{2}=\mathbf{I}_{1}\otimes(F_{A}+iF_{B})+(F_{A}+iF_{B})\otimes\mathbf{I}_{1}$,
we can set $\boldsymbol{\left\langle \phi\right|}=\left\langle \phi_{1}\right|\otimes\left\langle \phi_{2}\right|$
with $\left\langle \phi_{1}\right|(F_{A}+iF_{B})=\lambda_{1}\left\langle \phi_{1}\right|$
and $\left\langle \phi_{2}\right|(F_{A}+iF_{B})=\lambda_{2}\left\langle \phi_{2}\right|$.
Therefore, we only need to prove that when $\lambda_{1}+\lambda_{2}=0$,
it must hold that $\boldsymbol{\left\langle \phi\right|}F_{B}^{T}\rrangle=\left\langle \phi_{1}\right|F_{B}^{T}\left(\left\langle \phi_{2}\right|\right)^{T}=0$.
Next, we will provide a proof for this point.

It is easy to know that $\mathbf{F}_{1}=F_{A}+iF_{B}=\mathbf{F}_{1}^{*}$,
$\mathbf{F}_{1}+\mathbf{F}_{1}^{T}=i(F_{B}+F_{B}^{T})=-2\underset{\mu}{\sum}\left[\left(Q^{\dagger}\left|\boldsymbol{l}_{\mu}\right\rangle \right)\left(Q^{\dagger}\left|\boldsymbol{l}_{\mu}\right\rangle \right)^{\dagger}+\left(Q\left|\boldsymbol{l}_{\mu}^{*}\right\rangle \right)\left(Q\left|\boldsymbol{l}_{\mu}^{*}\right\rangle \right)^{\dagger}\right]$
with $\left(\left|\boldsymbol{l}_{\mu}\right\rangle \right)^{T}=\left[\begin{array}{cccccc}
l_{\mu,1}, & \cdots, & l_{\mu,L}, & ig_{\mu,1}, & \cdots, & ig_{\mu,L}\end{array}\right]$, if $\left|\psi_{s}\right\rangle $ satisfies $\mathbf{F}_{1}\left|\psi_{s}\right\rangle =\lambda_{s}\left|\psi_{s}\right\rangle $,
we have $0\geqslant\left(\left|\psi_{s}\right\rangle \right)^{\dagger}\left(\mathbf{F}_{1}+\mathbf{F}_{1}^{T}\right)\left|\psi_{s}\right\rangle =2\Re(\lambda_{s})\left(\left|\psi_{s}\right\rangle \right)^{\dagger}\left|\psi_{s}\right\rangle $,
this means $\Re(\lambda_{s})=-2\Re(\alpha_{s})\leqslant0$. So when
$\lambda_{1}+\lambda_{2}=0$, we have $\lambda_{1}=\lambda_{2}^{*}$,
$\left\langle \phi_{1}^{*}\right|\left(F_{A}+iF_{B}\right)=\lambda_{2}\left\langle \phi_{1}^{*}\right|$,
$\left\langle \phi_{2}^{*}\right|\left(F_{A}+iF_{B}\right)=\lambda_{1}\left\langle \phi_{2}^{*}\right|$,
combined with $F_{A}^{T}=-F_{A}$, we have
\begin{equation}
\left\langle \phi_{1}\right|F_{B}\left(\left\langle \phi_{2}\right|\right)^{T}+\left\langle \phi_{1}\right|F_{B}^{T}\left(\left\langle \phi_{2}\right|\right)^{T}=0,\label{eq:F_B+F_BT}
\end{equation}
\begin{equation}
\left(\left\langle \phi_{1}\right|+\left\langle \phi_{2}\right|^{*}\right)\left(F_{B}+F_{B}^{T}\right)\left(\left\langle \phi_{1}\right|+\left\langle \phi_{2}\right|^{*}\right)^{\dagger}=0,\label{eq:phi_1_phi_2}
\end{equation}
\begin{equation}
\left(\left\langle \phi_{1}\right|+i\left\langle \phi_{2}^{*}\right|\right)\left(F_{B}+F_{B}^{T}\right)\left(\left\langle \phi_{1}\right|+i\left\langle \phi_{2}^{*}\right|\right)^{\dagger}=0.\label{eq:phi_1_phi_2_i}
\end{equation}
Considering that $F_{B}=2i\sum_{\mu}\left|\varphi_{\mu}\right\rangle \left(\left|\varphi_{\mu}\right\rangle \right)^{\dagger}$
with $\left|\varphi_{\mu}\right\rangle =Q^{\dagger}\left|\boldsymbol{l}_{\mu}\right\rangle $,
we conclude that $-iF_{B}$ is a positive semidefinite matrix, i.e.,
\begin{equation}
-iF_{B}\succeq0.\label{eq:F_B}
\end{equation}
By combining Eq.(\ref{eq:F_B},\ref{eq:phi_1_phi_2}), we get
\begin{equation}
\begin{array}{rl}
0= & \left(\left\langle \phi_{1}\right|+\left\langle \phi_{2}^{*}\right|\right)F_{B}\left(\left\langle \phi_{1}\right|+\left\langle \phi_{2}^{*}\right|\right)^{\dagger}\\
= & 2i\sum_{\mu}\left(\left|\left\langle \phi_{1}\right|\left.\varphi_{\mu}\right\rangle \right|^{2}+\left|\left\langle \phi_{2}^{*}\right|\left.\varphi_{\mu}\right\rangle \right|^{2}\right)\\
 & +2i\Im\left[\left\langle \phi_{1}\right|F_{B}\left(\left\langle \phi_{2}\right|\right)^{T}\right],
\end{array}
\end{equation}

\begin{equation}
\begin{array}{rl}
0= & \left(\left\langle \phi_{1}\right|+\left\langle \phi_{2}^{*}\right|\right)F_{B}^{T}\left(\left\langle \phi_{1}\right|+\left\langle \phi_{2}^{*}\right|\right)^{\dagger}\\
= & 2i\sum_{\mu}\left(\left|\left\langle \phi_{1}\right|\left.\varphi_{\mu}^{*}\right\rangle \right|^{2}+\left|\left\langle \phi_{2}^{*}\right|\left.\varphi_{\mu}^{*}\right\rangle \right|^{2}\right)\\
 & +2i\Im\left[\left\langle \phi_{1}\right|F_{B}^{T}\left(\left\langle \phi_{2}\right|\right)^{T}\right],
\end{array}
\end{equation}
this implies that $\Im\left[\left\langle \phi_{1}\right|F_{B}\left(\left\langle \phi_{2}\right|\right)^{T}\right]\leqslant0$,
$\Im\left[\left\langle \phi_{1}\right|F_{B}^{T}\left(\left\langle \phi_{2}\right|\right)^{T}\right]\leqslant0$,
where the symbol $\Im$ denotes taking the imaginary part. Combining
this with Eq(\ref{eq:F_B+F_BT}), we find that $\Im\left[\left\langle \phi_{1}\right|F_{B}^{T}\left(\left\langle \phi_{2}\right|\right)^{T}\right]=0$.
In the same way, combined with Eq.(\ref{eq:F_B},\ref{eq:F_B+F_BT},\ref{eq:phi_1_phi_2_i}),
we get $\Re\left[\left\langle \phi_{1}\right|F_{B}^{T}\left(\left\langle \phi_{2}\right|\right)^{T}\right]=0$,
where the symbol $\Re$ denotes taking the real part, so we have $\boldsymbol{\left\langle \phi\right|}F_{B}^{T}\rrangle=\left\langle \phi_{1}\right|F_{B}^{T}\left(\left\langle \phi_{2}\right|\right)^{T}=0$.
Thus, the conclusion is proven.

\begin{widetext}Since the equations in the following derivations
are relatively long, for the sake of readability, we will adopt a
single-column format starting from this point.

\subsection{\label{subsec:Derivation FT_DT}Derivation of formula $\mathbf{F}_{2m}^{(B)}\bm{\left|T_{2m}\right\rangle }=\mathbf{G}_{2m}\bm{\left|T_{2m-2}\right\rangle }$}

Since $\{\hat{f}_{i},\hat{f}_{j}\}=\delta_{i,j}$, we have

\[
\begin{array}{rl}
 & \left(\left[\mathbf{I}_{2m}+(-1)^{k}R_{k}\otimes\mathbf{I}_{2m-k}\right]\bm{\left|T_{2m}\right\rangle }\right)_{j_{1},\cdots,j_{2m}}\\
= & \bm{\left|T_{2m}\right\rangle }_{j_{1},\cdots,j_{2m}}+(-1)^{k}\bm{\left|T_{2m}\right\rangle }_{j_{k},j_{1},\cdots,j_{k-1},j_{k+1},\cdots,j_{2m}}\\
= & \left(\bm{\left|T_{2m}\right\rangle }_{j_{1},\cdots,j_{2m}}+\bm{\left|T_{2m}\right\rangle }_{j_{1},\cdots,j_{k-2},j_{k},j_{k-1},j_{k+1},\cdots,j_{2m}}\right)\\
 & -\left(\bm{\left|T_{2m}\right\rangle }_{j_{1},\cdots,j_{k-2},j_{k},j_{k-1},j_{k+1},\cdots,j_{2m}}+\bm{\left|T_{2m}\right\rangle }_{j_{1},\cdots,j_{k-3},j_{k},j_{k-2},j_{k-1},j_{k+1},\cdots,j_{2m}}\right)\\
 & +\cdots\\
 & +(-1)^{k}\left(\bm{\left|T_{2m}\right\rangle }_{j_{1},j_{k},j_{2},\cdots,j_{k-1},j_{k+1},\cdots,j_{2m}}+\bm{\left|T_{2m}\right\rangle }_{j_{k},j_{1},\cdots,j_{k-1},j_{k+1},\cdots,j_{2m}}\right)\\
= & \stackrel[s=1]{k-1}{\sum}(-1)^{k-s+1}\llangle\mathbb{I}|\left(\stackrel[l=1]{s-1}{\overrightarrow{\prod}}\hat{f}_{j_{l}}\right)\{\hat{f}_{j_{s}},\hat{f}_{j_{k}}\}\left(\stackrel[l=s+1]{k-1}{\overrightarrow{\prod}}\hat{f}_{j_{l}}\right)\left(\stackrel[l=k+1]{2m}{\overrightarrow{\prod}}\hat{f}_{j_{l}}\right)|\rho(t)\rrangle\\
= & \stackrel[s=1]{k-1}{\sum}(-1)^{k-s+1}\delta_{j_{k},j_{s}}\bm{\left|T_{2m-2}\right\rangle }_{j_{1},\cdots,j_{s-1},j_{s+1},\cdots,j_{k-1},j_{k+1},\cdots,j_{2m}}\\
= & \stackrel[s=1]{k-1}{\sum}(-1)^{k-s+1}\left(\left[R_{s}\otimes\mathbf{I}_{2m-s}\right]\left[R_{k}\otimes\mathbf{I}_{2m-k}\right]\left[\stackrel[i=1]{2L}{\sum}\overrightarrow{e}_{i}\otimes\overrightarrow{e}_{i}\otimes\bm{\left|T_{2m-2}\right\rangle }\right]\right)_{j_{1},\cdots,j_{2m}},
\end{array}
\]
considering Eq.(\ref{eq:R}), we have
\[
\begin{array}{rl}
 & \mathbf{F}_{2m}^{(B)}\bm{\left|T_{2m}\right\rangle }\\
= & -i\stackrel[k=1]{2m}{\sum}\left\{ \left[\mathbf{I}_{k-1}\otimes F_{B}\otimes\mathbf{I}_{2m-k}\right]\left(\stackrel[s=1]{k-1}{\sum}(-1)^{k-s+1}\left[R_{s}\otimes\mathbf{I}_{2m-s}\right]\left[R_{k}\otimes\mathbf{I}_{2m-k}\right]\right)\left[\stackrel[i=1]{2L}{\sum}\overrightarrow{e}_{i}\otimes\overrightarrow{e}_{i}\otimes\bm{\left|T_{2m-2}\right\rangle }\right]\right\} \\
= & i\stackrel[k=1]{2m}{\sum}\stackrel[s=1]{k-1}{\sum}(-1)^{k-s}\left[R_{s}\otimes\mathbf{I}_{2m-s}\right]\left[\mathbf{I}_{k-1}\otimes F_{B}\otimes\mathbf{I}_{2m-k}\right]\left[R_{k}\otimes\mathbf{I}_{2m-k}\right]\left[\stackrel[i=1]{2L}{\sum}\overrightarrow{e}_{i}\otimes\overrightarrow{e}_{i}\otimes\bm{\left|T_{2m-2}\right\rangle }\right]\\
= & i\stackrel[k=1]{2m}{\sum}\stackrel[s=1]{k-1}{\sum}(-1)^{k-s}\left[R_{s}\otimes\mathbf{I}_{2m-s}\right]\left[R_{k}\otimes\mathbf{I}_{2m-k}\right]\left[F_{B}\otimes\mathbf{I}_{2m-1}\right]\left[\stackrel[i=1]{2L}{\sum}\overrightarrow{e}_{i}\otimes\overrightarrow{e}_{i}\otimes\bm{\left|T_{2m-2}\right\rangle }\right]\\
= & i\left[\mathbf{R}_{2m}\left(\stackrel[i=1]{2L}{\sum}(F_{B}\overrightarrow{e}_{i})\otimes\overrightarrow{e}_{i}\otimes\mathbf{I}_{2m-2}\right)\right]\bm{\left|T_{2m-2}\right\rangle }\\
= & \mathbf{G}_{2m}\bm{\left|T_{2m-2}\right\rangle }
\end{array}
\]
with
\begin{equation}
\left\{ \begin{array}{l}
\mathbf{G}_{2m}=i\mathbf{R}_{2m}\left(|F_{B}\rrangle\otimes\mathbf{I}_{2m-2}\right),\\
\mathbf{R}_{2m}=\stackrel[k=1]{2m}{\sum}\stackrel[s=1]{k-1}{\sum}(-1)^{k-s}\left[R_{s}\otimes\mathbf{I}_{2m-s}\right]\left[R_{k}\otimes\mathbf{I}_{2m-k}\right].
\end{array}\right.
\end{equation}

\subsection{\label{subsec:Prove-of-Lemma1}Prove of Lemma 1}

Previously, we have seen the importance of Lemma 1, here, we will
prove it by dividing the proof into two parts. First, we will prove
$\mathbb{R}_{(2m,2m-2)}=\mathbb{R}_{(2m,2m-2)}\left(R_{4}\otimes\mathbf{I}_{2m-4}\right)^{2}$,
and then we will extend it to $\forall\ 1\leqslant s\leqslant m-l$,
$\mathbb{R}_{(2m,2l+2)}=\mathbb{R}_{(2m,2l+2)}\left(R_{2s}\otimes\mathbf{I}_{2m-2s}\right)^{2}$.
To prove $\mathbb{R}_{(2m,2m-2)}=\mathbb{R}_{(2m,2m-2)}\left(R_{4}\otimes\mathbf{I}_{2m-4}\right)^{2}$,
it's necessary to demonstrate that for any state vector $\stackrel[l=1]{2m}{\bigotimes}\left|\xi_{l}\right\rangle $,
we have
\[
\mathbb{R}_{(2m,2m-2)}\left(\stackrel[l=1]{2m}{\bigotimes}\left|\xi_{l}\right\rangle \right)=\mathbb{R}_{(2m,2m-2)}\left(R_{4}\otimes\mathbf{I}_{2m-4}\right)^{2}\left(\stackrel[l=1]{2m}{\bigotimes}\left|\xi_{l}\right\rangle \right),
\]
expanding the left side, we get $left=\stackrel[n=1]{6}{\sum}\left|\overrightarrow{\xi}_{n}\right\rangle $
with
\[
\begin{array}{rl}
\left|\overrightarrow{\xi}_{1}\right\rangle = & \underset{1\leqslant r<l<s+2<j+2\leqslant2m}{\sum}(-1)^{l-r+j-s}\left(\stackrel[i=5]{r+3}{\bigotimes}\left|\xi_{i}\right\rangle \right)\otimes\left|\xi_{2}\right\rangle \otimes\left(\stackrel[i=r+4]{l+2}{\bigotimes}\left|\xi_{i}\right\rangle \right)\otimes\left|\xi_{1}\right\rangle \\
 & \otimes\left(\stackrel[i=l+3]{s+3}{\bigotimes}\left|\xi_{i}\right\rangle \right)\otimes\left|\xi_{4}\right\rangle \otimes\left(\stackrel[i=s+4]{j+2}{\bigotimes}\left|\xi_{i}\right\rangle \right)\otimes\left|\xi_{3}\right\rangle \otimes\left(\stackrel[i=j+3]{2m}{\bigotimes}\left|\xi_{i}\right\rangle \right),
\end{array}
\]
\[
\begin{array}{rl}
\left|\overrightarrow{\xi}_{2}\right\rangle = & \underset{1\leqslant r<s+1<l<j+2\leqslant2m}{\sum}(-1)^{l-r+j-s}\left(\stackrel[i=5]{r+3}{\bigotimes}\left|\xi_{i}\right\rangle \right)\otimes\left|\xi_{2}\right\rangle \otimes\left(\stackrel[i=r+4]{s+3}{\bigotimes}\left|\xi_{i}\right\rangle \right)\otimes\left|\xi_{4}\right\rangle \\
 & \otimes\left(\stackrel[i=s+4]{l+1}{\bigotimes}\left|\xi_{i}\right\rangle \right)\otimes\left|\xi_{1}\right\rangle \otimes\left(\stackrel[i=l+2]{j+2}{\bigotimes}\left|\xi_{i}\right\rangle \right)\otimes\left|\xi_{3}\right\rangle \otimes\left(\stackrel[i=j+3]{2m}{\bigotimes}\left|\xi_{i}\right\rangle \right),
\end{array}
\]
\[
\begin{array}{rl}
\left|\overrightarrow{\xi}_{3}\right\rangle = & \underset{1\leqslant s<r<l<j+2\leqslant2m}{\sum}(-1)^{l-r+j-s}\left(\stackrel[i=5]{s+3}{\bigotimes}\left|\xi_{i}\right\rangle \right)\otimes\left|\xi_{4}\right\rangle \otimes\left(\stackrel[i=s+4]{r+2}{\bigotimes}\left|\xi_{i}\right\rangle \right)\otimes\left|\xi_{2}\right\rangle \\
 & \otimes\left(\stackrel[i=r+3]{l+1}{\bigotimes}\left|\xi_{i}\right\rangle \right)\otimes\left|\xi_{1}\right\rangle \otimes\left(\stackrel[i=l+2]{j+2}{\bigotimes}\left|\xi_{i}\right\rangle \right)\otimes\left|\xi_{3}\right\rangle \otimes\left(\stackrel[i=j+3]{2m}{\bigotimes}\left|\xi_{i}\right\rangle \right),
\end{array}
\]
\[
\begin{array}{rl}
\left|\overrightarrow{\xi}_{4}\right\rangle = & \underset{1\leqslant r<s+1<j+1<l\leqslant2m}{\sum}(-1)^{l-r+j-s}\left(\stackrel[i=5]{r+3}{\bigotimes}\left|\xi_{i}\right\rangle \right)\otimes\left|\xi_{2}\right\rangle \otimes\left(\stackrel[i=r+4]{s+3}{\bigotimes}\left|\xi_{i}\right\rangle \right)\otimes\left|\xi_{4}\right\rangle \\
 & \otimes\left(\stackrel[i=s+4]{j+2}{\bigotimes}\left|\xi_{i}\right\rangle \right)\otimes\left|\xi_{3}\right\rangle \otimes\left(\stackrel[i=j+3]{l}{\bigotimes}\left|\xi_{i}\right\rangle \right)\otimes\left|\xi_{1}\right\rangle \otimes\left(\stackrel[i=l+1]{2m}{\bigotimes}\left|\xi_{i}\right\rangle \right),
\end{array}
\]
\[
\begin{array}{rl}
\left|\overrightarrow{\xi}_{5}\right\rangle = & \underset{1\leqslant s<r<j+1<l\leqslant2m}{\sum}(-1)^{l-r+j-s}\left(\stackrel[i=5]{s+3}{\bigotimes}\left|\xi_{i}\right\rangle \right)\otimes\left|\xi_{4}\right\rangle \otimes\left(\stackrel[i=s+4]{r+2}{\bigotimes}\left|\xi_{i}\right\rangle \right)\otimes\left|\xi_{2}\right\rangle \\
 & \otimes\left(\stackrel[i=r+3]{j+2}{\bigotimes}\left|\xi_{i}\right\rangle \right)\otimes\left|\xi_{3}\right\rangle \otimes\left(\stackrel[i=j+3]{l}{\bigotimes}\left|\xi_{i}\right\rangle \right)\otimes\left|\xi_{1}\right\rangle \otimes\left(\stackrel[i=l+1]{2m}{\bigotimes}\left|\xi_{i}\right\rangle \right),
\end{array}
\]
\[
\begin{array}{rl}
\left|\overrightarrow{\xi}_{6}\right\rangle = & \underset{1\leqslant s<j<r<l\leqslant2m}{\sum}(-1)^{l-r+j-s}\left(\stackrel[i=5]{s+3}{\bigotimes}\left|\xi_{i}\right\rangle \right)\otimes\left|\xi_{4}\right\rangle \otimes\left(\stackrel[i=s+4]{j+2}{\bigotimes}\left|\xi_{i}\right\rangle \right)\otimes\left|\xi_{3}\right\rangle \\
 & \otimes\left(\stackrel[i=j+3]{r+1}{\bigotimes}\left|\xi_{i}\right\rangle \right)\otimes\left|\xi_{2}\right\rangle \otimes\left(\stackrel[i=r+2]{l}{\bigotimes}\left|\xi_{i}\right\rangle \right)\otimes\left|\xi_{1}\right\rangle \otimes\left(\stackrel[i=l+1]{2m}{\bigotimes}\left|\xi_{i}\right\rangle \right).
\end{array}
\]

Through simple variable substitution, we know that $\left|\overrightarrow{\xi}_{1}\right\rangle =\left|\overrightarrow{\xi}_{6}\right\rangle _{\left|\xi_{1}\right\rangle \leftrightarrow\left|\xi_{3}\right\rangle ,\left|\xi_{2}\right\rangle \leftrightarrow\left|\xi_{4}\right\rangle }$,
$\left|\overrightarrow{\xi}_{2}\right\rangle =\left|\overrightarrow{\xi}_{5}\right\rangle _{\left|\xi_{1}\right\rangle \leftrightarrow\left|\xi_{3}\right\rangle ,\left|\xi_{2}\right\rangle \leftrightarrow\left|\xi_{4}\right\rangle }$,
$\left|\overrightarrow{\xi}_{3}\right\rangle =\left|\overrightarrow{\xi}_{4}\right\rangle _{\left|\xi_{1}\right\rangle \leftrightarrow\left|\xi_{3}\right\rangle ,\left|\xi_{2}\right\rangle \leftrightarrow\left|\xi_{4}\right\rangle }$,
here, the subscript $\left|\xi_{1}\right\rangle \leftrightarrow\left|\xi_{3}\right\rangle ,\left|\xi_{2}\right\rangle \leftrightarrow\left|\xi_{4}\right\rangle $
represents the exchange of $\left|\xi_{1}\right\rangle $ and $\left|\xi_{3}\right\rangle $,
and the exchange of $\left|\xi_{2}\right\rangle $ and $\left|\xi_{4}\right\rangle $
in the direct product, so we know $left=\left(left\right)_{\left|\xi_{1}\right\rangle \leftrightarrow\left|\xi_{3}\right\rangle ,\left|\xi_{2}\right\rangle \leftrightarrow\left|\xi_{4}\right\rangle }=right$,
the conclusion $\mathbb{R}_{(2m,2m-2)}=\mathbb{R}_{(2m,2m-2)}\left(R_{4}\otimes\mathbf{I}_{2m-4}\right)^{2}$
is thus proven. Noting that $\left(R_{2s}\otimes\mathbf{I}_{2m-2s}\right)^{2}=\left(R_{4}\otimes\mathbf{I}_{2m-4}\right)^{2}\left(\mathbf{I}_{2}\otimes R_{2s-2}\otimes\mathbf{I}_{2m-2s}\right)^{2}$,
we can deduce that $\forall\ 1\leqslant s\leqslant m-l$,
\[
\begin{array}{rl}
 & \mathbb{R}_{(2m,2l+2)}\left(R_{2s}\otimes\mathbf{I}_{2m-2s}\right)^{2}\\
= & \mathbb{R}_{(2m,2m-2)}\left(R_{4}\otimes\mathbf{I}_{2m-4}\right)^{2}\left(\mathbf{I}_{4}\otimes\mathbb{R}_{(2m-4,2l+2)}\right)\left(\mathbf{I}_{2}\otimes R_{2s-2}\otimes\mathbf{I}_{2m-2s}\right)^{2}\\
= & \mathbb{R}_{(2m,2l+2)}\left(\mathbf{I}_{2}\otimes R_{2s-2}\otimes\mathbf{I}_{2m-2s}\right)^{2}\\
= & \mathbb{R}_{(2m,2m)}\left\{ \mathbf{I}_{2}\otimes\left[\mathbb{R}_{(2m-2,2l+2)}\left(R_{2s-2}\otimes\mathbf{I}_{2m-2s}\right)^{2}\right]\right\} \\
= & \cdots\\
= & \mathbb{R}_{(2m,2m-2s+6)}\left\{ \mathbf{I}_{2s-4}\otimes\left[\mathbb{R}_{(2m-2s+4,2l+2)}\left(R_{4}\otimes\mathbf{I}_{2m-2s}\right)^{2}\right]\right\} \\
= & \mathbb{R}_{(2m,2l+2)},
\end{array}
\]
which is the result of Lemma 1.

\section{\label{sec:The solution T2m}The solution of the equation $\frac{d}{dt}\bm{\left|T_{2m,t}\right\rangle }=\mathbf{F}_{2m}\bm{\left|T_{2m,t}\right\rangle }+\mathbf{G}_{2m}\bm{\left|T_{2m-2,t}\right\rangle }$}

Here, we prove that Eq.(\ref{eq:High Order}) is the solution to Eq.(\ref{eq:dt_2m_2m-2}).
Later, we will demonstrate the derivation of the solution under specific
conditions.

Considering the case where $m>1$, we directly differentiate $t$
in Eq.(\ref{eq:High Order}), yielding
\begin{equation}
\begin{array}{rl}
 & \frac{d}{dt}\left\{ \left[\stackrel[r=1]{m-l}{\bigotimes}\left(\bm{\left|T_{2,t}^{T}\right\rangle }-e^{\mathbf{F}_{2}t}\bm{\left|T_{2,0}^{T}\right\rangle }\right)\right]\otimes\left[e^{\mathbf{F}_{2l}t}\bm{\left|T_{2l,0}\right\rangle }\right]\right\} \\
= & \left(\stackrel[k=1]{m-l}{\sum}\mathbf{I}_{2k-2}\otimes\mathbf{F}_{2}\otimes\mathbf{I}_{2m-2k}\right)\left\{ \left[\stackrel[r=1]{m-l}{\bigotimes}\left(\bm{\left|T_{2,t}^{T}\right\rangle }-e^{\mathbf{F}_{2}t}\bm{\left|T_{2,0}^{T}\right\rangle }\right)\right]\otimes\left[e^{\mathbf{F}_{2l}t}\bm{\left|T_{2l,0}\right\rangle }\right]\right\} \\
 & +\left(\mathbf{I}_{2m-2l}\otimes\mathbf{F}_{2l}\right)\left\{ \left[\stackrel[r=1]{m-l}{\bigotimes}\left(\bm{\left|T_{2,t}^{T}\right\rangle }-e^{\mathbf{F}_{2}t}\bm{\left|T_{2,0}^{T}\right\rangle }\right)\right]\otimes\left[e^{\mathbf{F}_{2l}t}\bm{\left|T_{2l,0}\right\rangle }\right]\right\} \\
 & +\stackrel[k=1]{m-l}{\sum}\left\{ \left[\stackrel[r=1]{k-1}{\bigotimes}\left(\bm{\left|T_{2,t}^{T}\right\rangle }-e^{\mathbf{F}_{2}t}\bm{\left|T_{2,0}^{T}\right\rangle }\right)\right]\otimes\left(-i|F_{B}\rrangle\right)\otimes\left[\stackrel[r=k+1]{m-l}{\bigotimes}\left(\bm{\left|T_{2,t}^{T}\right\rangle }-e^{\mathbf{F}_{2}t}\bm{\left|T_{2,0}^{T}\right\rangle }\right)\right]\otimes\left[e^{\mathbf{F}_{2l}t}\bm{\left|T_{2l,0}\right\rangle }\right]\right\}.
\end{array}\label{eq:d-dt-no R}
\end{equation}
By utilizing Lemma 1, we know that after left-multiplying by $\mathbb{R}_{(2m,2l+2)}$,
$\left(\bm{\left|T_{2,t}^{T}\right\rangle }-e^{\mathbf{F}_{2}t}\bm{\left|T_{2,0}^{T}\right\rangle }\right)$
and $-i|F_{B}\rrangle$ can be interchanged in the last term of Eq.(\ref{eq:d-dt-no R}).
Thus, we have
\begin{equation}
	\begin{array}{rl}
		\frac{d}{dt}\bm{\left|T_{2m,t}\right\rangle }= & \stackrel[l=0]{m}{\sum}\frac{(-1)^{m-l}}{(m-l)!}\mathbb{R}_{(2m,2l+2)}\mathbf{F}_{2m}\left\{ \left[\stackrel[r=1]{m-l}{\bigotimes}\left(\bm{\left|T_{2,t}^{T}\right\rangle }-e^{\mathbf{F}_{2}t}\bm{\left|T_{2,0}^{T}\right\rangle }\right)\right]\otimes\left[e^{\mathbf{F}_{2l}t}\bm{\left|T_{2l,0}\right\rangle }\right]\right\} \\
		& +\stackrel[l=0]{m}{\sum}\frac{(-1)^{m-l}}{(m-l-1)!}\mathbb{R}_{(2m,2l+2)}\left\{ \left(-i|F_{B}\rrangle\right)\otimes\left[\stackrel[r=1]{m-l-1}{\bigotimes}\left(\bm{\left|T_{2,t}^{T}\right\rangle }-e^{\mathbf{F}_{2}t}\bm{\left|T_{2,0}^{T}\right\rangle }\right)\right]\otimes\left[e^{\mathbf{F}_{2l}t}\bm{\left|T_{2l,0}\right\rangle }\right]\right\} \\
		= & \mathbf{F}_{2m}\bm{\left|T_{2m,t}\right\rangle }+\mathbf{G}_{2m}\bm{\left|T_{2m-2,t}\right\rangle }.
	\end{array}\label{eq:dtT2m=00003DF2m...}
\end{equation}
Given $\mathbf{F}_{2m}=\sum_{k=1}^{2m}\mathbf{I}_{k-1}\otimes(F_{A}+iF_{B})\otimes\mathbf{I}_{2m-k}$
and combining with Eq.(\ref{eq:R}), it is easy to know that when
$s\leqslant m$, $\left(\mathbf{I}_{2m-2s}\otimes\mathbf{R}_{2s}\right)\mathbf{F}_{2m}=\mathbf{F}_{2m}\left(\mathbf{I}_{2m-2s}\otimes\mathbf{R}_{2s}\right)$.
Considering the definition of $\mathbb{R}_{(2m,2l+2)}$, it is clear
that $\mathbb{R}_{(2m,2l+2)}$ and $\mathbf{F}_{2m}$ can be interchanged.
The final equality in the above formula holds precisely because of
this. Eq.(\ref{eq:dtT2m=00003DF2m...}) demonstrates that Eq.(\ref{eq:High Order})
is a solution to Eq.(\ref{eq:dt_2m_2m-2}). Since the evolution of
Eq.(\ref{eq:lindblad}) is deterministic, its solution is unique.
Therefore, Eq.(\ref{eq:High Order}) is the unique solution to Eq.(\ref{eq:dt_2m_2m-2}).

We assume that $\mathbf{F}_{2}$ can be diagonalized and $\det(\mathbf{F}_{2})\neq0$.
Using this as an example, let us demonstrate the derivation process
of Eq.(\ref{eq:High Order}). It is important to note that Eq.(\ref{eq:High Order})
is generally valid and does not depend on these assumptions, as the
proof above has already shown this. The solution to Eq.(\ref{eq:dt_2m_2m-2})
is evidently
\begin{equation}
\bm{\left|T_{2m,t}\right\rangle }=e^{\mathbf{F}_{2m}t}\bm{\left|T_{2m,0}\right\rangle }+e^{\mathbf{F}_{2m}t}\int_{0}^{t}\left(e^{-\mathbf{F}_{2m}\tau}\mathbf{G}_{2m}\bm{\left|T_{2m-2,\tau}\right\rangle }\right)d\tau.\label{eq:T2m_int}
\end{equation}
From this, we obtain $\bm{\left|T_{2,t}\right\rangle }=e^{\mathbf{F}_{2}t}\left(\bm{\left|T_{2,0}\right\rangle }+\mathbf{F}_{2}^{-1}\mathbf{G}_{2}\right)-\mathbf{F}_{2}^{-1}\mathbf{G}_{2}$
and
\begin{equation}
\begin{array}{rl}
\bm{\left|T_{4,t}\right\rangle }= & e^{\mathbf{F}_{4}t}\bm{\left|T_{4,0}\right\rangle }+e^{\mathbf{F}_{4}t}\int_{0}^{t}\left(e^{-\mathbf{F}_{4}\tau}\mathbf{G}_{4}\bm{\left|T_{2,\tau}\right\rangle }\right)d\tau\\
= & e^{\mathbf{F}_{4}t}\bm{\left|T_{4,0}\right\rangle }+e^{\mathbf{F}_{4}t}\left\{ \mathbf{R}_{4}\int_{0}^{t}i\left(e^{-\mathbf{F}_{2}\tau}|F_{B}\rrangle\right)\otimes\left(\bm{\left|T_{2,0}\right\rangle }+\mathbf{F}_{2}^{-1}\mathbf{G}_{2}\right)d\tau+\mathbb{R}_{(4,2)}\int_{0}^{t}e^{-\mathbf{F}_{4}\tau}\left[|F_{B}\rrangle\otimes\left(\mathbf{F}_{2}^{-1}|F_{B}\rrangle\right)\right]d\tau\right\} \\
= & e^{\mathbf{F}_{4}t}\bm{\left|T_{4,0}\right\rangle }-i\mathbf{R}_{4}\left\{ \left[\left(\mathbf{I}_{2}-e^{\mathbf{F}_{2}t}\right)\mathbf{F}_{2}^{-1}|F_{B}\rrangle\right]\otimes\left[e^{\mathbf{F}_{2}t}\left(\bm{\left|T_{2,0}\right\rangle }+\mathbf{F}_{2}^{-1}\mathbf{G}_{2}\right)\right]\right\} \\
 & +\mathbb{R}_{(4,2)}\left(e^{\mathbf{F}_{4}t}-\mathbf{I}_{4}\right)\mathbf{F}_{4}^{-1}\left[|F_{B}\rrangle\otimes\left(\mathbf{F}_{2}^{-1}|F_{B}\rrangle\right)\right].
\end{array}\label{eq:T4}
\end{equation}

In Appendix A.2, we found that $\Re(\alpha_{s})\geqslant0$; therefore,
our assumed condition $\det(\mathbf{F}_{2})\neq0$ is equivalent to
the uniqueness of the non-equilibrium steady state. Due to $\mathbf{F}_{4}=\mathbf{I}_{2}\otimes\mathbf{F}_{2}+\mathbf{F}_{2}\otimes\mathbf{I}_{2}$,
if we give the eigendecomposition $\mathbf{F}_{2}=\sum_{s=1}^{4L^{2}}\lambda_{2,s}\left|\psi_{2,s}\right\rangle \left\langle \phi_{2,s}\right|$,
we know
\[
\begin{array}{rl}
 & \mathbb{R}_{(4,2)}\mathbf{F}_{4}^{-1}\left[|F_{B}\rrangle\otimes\left(\mathbf{F}_{2}^{-1}|F_{B}\rrangle\right)\right]\\
= & \mathbb{R}_{(4,2)}\stackrel[j_{1},j_{2}=1]{4L^{2}}{\sum}\frac{\left\langle \phi_{2,j_{1}}\right|F_{B}\rrangle\left\langle \phi_{2,j_{2}}\right|F_{B}\rrangle}{\lambda_{2,j_{2}}\left(\lambda_{2,j_{1}}+\lambda_{2,j_{2}}\right)}\left|\psi_{2,j_{1}}\right\rangle \otimes\left|\psi_{2,j_{2}}\right\rangle \\
= & \mathbb{R}_{(4,2)}\stackrel[j_{1},j_{2}=1]{4L^{2}}{\sum}\frac{1}{\lambda_{2,j_{1}}}\left(\frac{1}{\lambda_{2,j_{2}}}-\frac{1}{\lambda_{2,j_{1}}+\lambda_{2,j_{2}}}\right)\left\langle \phi_{2,j_{1}}\right|F_{B}\rrangle\left\langle \phi_{2,j_{2}}\right|F_{B}\rrangle\left|\psi_{2,j_{1}}\right\rangle \otimes\left|\psi_{2,j_{2}}\right\rangle \\
= & \mathbb{R}_{(4,2)}\left[\left(\mathbf{F}_{2}^{-1}|F_{B}\rrangle\right)\otimes\left(\mathbf{F}_{2}^{-1}|F_{B}\rrangle\right)\right]-\mathbb{R}_{(4,2)}\mathbf{F}_{4}^{-1}\left[\left(\mathbf{F}_{2}^{-1}|F_{B}\rrangle\right)\otimes|F_{B}\rrangle\right],
\end{array}
\]
Using Lemma 1, we find that $\mathbb{R}_{(4,2)}\mathbf{F}_{4}^{-1}\left[|F_{B}\rrangle\otimes\left(\mathbf{F}_{2}^{-1}|F_{B}\rrangle\right)\right]=\frac{1}{2}\mathbb{R}_{(4,2)}\left[\stackrel[r=1]{2}{\bigotimes}\left(\mathbf{F}_{2}^{-1}|F_{B}\rrangle\right)\right]$.
Meanwhile, since $i\left(\mathbf{I}_{2}-e^{\mathbf{F}_{2}t}\right)\mathbf{F}_{2}^{-1}|F_{B}\rrangle=\bm{\left|T_{2,t}^{T}\right\rangle }-e^{\mathbf{F}_{2}t}\bm{\left|T_{2,0}^{T}\right\rangle }$,
we can further rewrite Eq.(\ref{eq:T4}) as
\begin{equation}
\begin{array}{rl}
\bm{\left|T_{4,t}\right\rangle }= & e^{\mathbf{F}_{4}t}\bm{\left|T_{4,0}\right\rangle }-\mathbf{R}_{4}\left[\left(\bm{\left|T_{2,t}^{T}\right\rangle }-e^{\mathbf{F}_{2}t}\bm{\left|T_{2,0}^{T}\right\rangle }\right)\otimes\left(e^{\mathbf{F}_{2}t}\bm{\left|T_{2,0}\right\rangle }\right)\right]+\frac{1}{2}\mathbb{R}_{(4,2)}\left[\stackrel[r=1]{2}{\bigotimes}\left(\bm{\left|T_{2,t}^{T}\right\rangle }-e^{\mathbf{F}_{2}t}\bm{\left|T_{2,0}^{T}\right\rangle }\right)\right]\\
= & \stackrel[l=0]{2}{\sum}\frac{(-1)^{2-l}}{(2-l)!}\mathbb{R}_{(4,2l+2)}\left\{ \left[\stackrel[r=1]{2-l}{\bigotimes}\left(\bm{\left|T_{2,t}^{T}\right\rangle }-e^{\mathbf{F}_{2}t}\bm{\left|T_{2,0}^{T}\right\rangle }\right)\right]\otimes\left[e^{\mathbf{F}_{2l}t}\bm{\left|T_{2l,0}\right\rangle }\right]\right\} .
\end{array}\label{eq:T4_end}
\end{equation}
Additionally, using $\frac{d}{dt}\left(e^{-\mathbf{F}_{2}t}\bm{\left|T_{2,t}^{T}\right\rangle }-\bm{\left|T_{2,0}^{T}\right\rangle }\right)=-ie^{-\mathbf{F}_{2}t}|F_{B}\rrangle$
and Lemma 1, it is not difficult to see that
\begin{equation}
\begin{array}{rl}
 & e^{\mathbf{F}_{2m}t}\int_{0}^{t}e^{-\mathbf{F}_{2m}\tau}\mathbf{G}_{2m}\mathbb{R}_{(2m-2,2l+2)}\left\{ \left[\stackrel[r=1]{m-1-l}{\bigotimes}\left(\bm{\left|T_{2,\tau}^{T}\right\rangle }-e^{\mathbf{F}_{2}\tau}\bm{\left|T_{2,0}^{T}\right\rangle }\right)\right]\otimes\left[e^{\mathbf{F}_{2l}\tau}\bm{\left|T_{2l,0}\right\rangle }\right]\right\} d\tau\\
= & e^{\mathbf{F}_{2m}t}\int_{0}^{t}\mathbb{R}_{(2m,2l+2)}\left\{ \left(ie^{-\mathbf{F}_{2}\tau}|F_{B}\rrangle\right)\otimes\left[\stackrel[r=1]{m-1-l}{\bigotimes}\left(e^{-\mathbf{F}_{2}\tau}\bm{\left|T_{2,\tau}^{T}\right\rangle }-\bm{\left|T_{2,0}^{T}\right\rangle }\right)\right]\otimes\bm{\left|T_{2l,0}\right\rangle }\right\} d\tau\\
= & \frac{-1}{m-l}\mathbb{R}_{(2m,2l+2)}\left\{ \left[\stackrel[r=1]{m-l}{\bigotimes}\left(\bm{\left|T_{2,t}^{T}\right\rangle }-e^{\mathbf{F}_{2}t}\bm{\left|T_{2,0}^{T}\right\rangle }\right)\right]\otimes\left[e^{\mathbf{F}_{2l}t}\bm{\left|T_{2l,0}\right\rangle }\right]\right\} .
\end{array}\label{eq:int_F2m}
\end{equation}
Combining Eq.(\ref{eq:T2m_int},\ref{eq:T4_end},\ref{eq:int_F2m}),
it is easy to see that
\[
\boldsymbol{\left|T_{2m,t}\right\rangle }=\stackrel[l=0]{m}{\sum}\frac{(-1)^{m-l}}{(m-l)!}\mathbb{R}_{(2m,2l+2)}\left\{ \left[\stackrel[r=1]{m-l}{\bigotimes}\left(\boldsymbol{\left|T_{2,t}^{T}\right\rangle }-e^{\mathbf{F}_{2}t}\boldsymbol{\left|T_{2,0}^{T}\right\rangle }\right)\right]\otimes\left[e^{\mathbf{F}_{2l}t}\boldsymbol{\left|T_{2l,0}\right\rangle }\right]\right\} .
\]

\section{\label{sec:Wick}The matrix version of the Wick's theorem}

After adding subscripts to Eq.(\ref{eq:wick}), it can be specifically
written as
\[
\bm{\left|T_{2l}\right\rangle }_{j_{1},\cdots,j_{2l}}=\frac{1}{l}\stackrel[k=1]{2l}{\sum}\stackrel[s=1]{k-1}{\sum}(-1)^{k+s+1}\bm{\left|T_{2}\right\rangle }_{j_{s},j_{k}}\bm{\left|T_{2l-2}\right\rangle }_{j_{1},\cdots,j_{s-1},j_{s+1},\cdots,j_{k-1},j_{k+1},\cdots,j_{2l}},
\]
we can consider $(-1)^{k+s+1}$ as the permutation parity brought
by replacing sequence $(j_{1},j_{2},\cdots,j_{2l})$ with sequence
$(j_{s},j_{k},j_{1},\cdots,j_{s-1},j_{s+1},\cdots,j_{k-1},j_{k+1},\cdots,j_{2l})$.
By iterating this repeatedly, we get
\[
\begin{array}{rl}
\bm{\left|T_{2l}\right\rangle }_{j_{1},\cdots,j_{2l}}= & \frac{(-1)^{l}}{l!}\left[\mathbb{R}_{(2l,2)}\left(\stackrel[r=1]{l}{\bigotimes}\bm{\left|T_{2}^{T}\right\rangle }\right)\right]_{j_{1},\cdots,j_{2l}}\\
= & \underset{\overrightarrow{\sigma}}{\sum}\frac{sgn(\overrightarrow{\sigma})}{l!}\bm{\left|T_{2}\right\rangle }_{j_{\sigma_{1}},j_{\sigma_{2}}}\bm{\left|T_{2}\right\rangle }_{j_{\sigma_{3}},j_{\sigma_{4}}}\cdots\bm{\left|T_{2}\right\rangle }_{j_{\sigma_{2l-1}},j_{\sigma_{2l}}},
\end{array}
\]
 where $\sigma_{2r-1}<\sigma_{2r}$ and $sgn(\overrightarrow{\sigma})$
represents the permutation parity brought by replacing $(1,2,\cdots,2l)$
with $\overrightarrow{\sigma}=(\sigma_{1},\sigma_{2},\cdots,\sigma_{2l})$,
the sum is the summation of all permutation operations that satisfy
$\sigma_{2r-1}<\sigma_{2r}$. This is evidently the Wick's theorem.

\section{\label{sec:The-derivation-of Lquar}The derivation of the general
quartic Liouvillian superoperator $\hat{\mathcal{L}}_{quar}$}

Here, we present the general quartic Liouvillian superoperator $\hat{\mathcal{L}}_{quar}$,
where the Hamiltonian is extended from Eq.(\ref{eq:H}) by adding
$H_{add}=\stackrel[j_{1},j_{2},j_{3},j_{4}=1]{2L}{\sum}\left(U_{H}\right)_{j_{1},j_{2},j_{3},j_{4}}\omega_{j_{1}}\omega_{j_{2}}\omega_{j_{3}}\omega_{j_{4}}$
and introducing quadratic dissipative operators $L_{t,\mu}=\stackrel[j_{1},j_{2}=1]{2L}{\sum}\left(U_{L,\mu}\right)_{j_{1},j_{2}}\omega_{j_{1}}\omega_{j_{2}}$.
Since $\{\omega_{j_{1}},\omega_{j_{2}}\}=\delta_{j_{1},j_{2}}$, we
can naturally define $U_{H}$ as a fully antisymmetric tensor, i.e.,
$\left(U_{H}\right)_{\cdots,j_{r},j_{r+1},\cdots}=-\left(U_{H}\right)_{\cdots,j_{r+1},j_{r},\cdots}$,
where any exchange of adjacent indices results in a negative sign.
It is easy to see that Eq.(\ref{eq:dt_Lquad}) will become $\frac{d}{dt}|\rho\rrangle=\hat{\mathcal{L}}_{quar}|\rho\rrangle=(\mathcal{\hat{L}}_{quad}+\hat{\mathcal{L}}_{add})|\rho\rrangle$,
where $\mathcal{\hat{L}}_{quad}$ remain unchanged (see Eq.(\ref{eq:Lquad})).
The additional part $\hat{\mathcal{L}}_{add}$ is
\[
\begin{array}{rl}
\hat{\mathcal{L}}_{add}= & -i\underset{j_{1},j_{2},j_{3},j_{4}}{\sum}\left(U_{H}\right)_{j_{1},j_{2},j_{3},j_{4}}\left[\left(\omega_{j_{1}}\omega_{j_{2}}\omega_{j_{3}}\omega_{j_{4}}\right)\otimes\mathbb{I}-\mathbb{I}\otimes\left(\omega_{j_{4}}^{T}\omega_{j_{3}}^{T}\omega_{j_{2}}^{T}\omega_{j_{1}}^{T}\right)\right]\\
 & +2\underset{\mu}{\sum}\left(\underset{j_{1},j_{2}}{\sum}\left(U_{L,\mu}\right)_{j_{1},j_{2}}\omega_{j_{1}}\omega_{j_{2}}\right)\otimes\left(\underset{j_{3},j_{4}}{\sum}\left(U_{L,\mu}^{*}\right)_{j_{3},j_{4}}\omega_{j_{3}}^{T}\omega_{j_{4}}^{T}\right)\\
 & -\underset{\mu}{\sum}\left[\left(\underset{j_{1},j_{2}}{\sum}\left(U_{L,\mu}^{*}\right)_{j_{1},j_{2}}\omega_{j_{2}}\omega_{j_{1}}\right)\left(\underset{j_{3},j_{4}}{\sum}\left(U_{L,\mu}\right)_{j_{3},j_{4}}\omega_{j_{3}}\omega_{j_{4}}\right)\right]\otimes\mathbb{I}\\
 & -\underset{\mu}{\sum}\mathbb{I}\otimes\left[\left(\underset{j_{1},j_{2}}{\sum}\left(U_{L,\mu}\right)_{j_{1},j_{2}}\omega_{j_{2}}^{T}\omega_{j_{1}}^{T}\right)\left(\underset{j_{3},j_{4}}{\sum}\left(U_{L,\mu}^{*}\right)_{j_{3},j_{4}}\omega_{j_{3}}^{T}\omega_{j_{4}}^{T}\right)\right].
\end{array}
\]

Writing the fourth-order tensor $U_{H}$ as a matrix, let $\mathrm{A}_{(j_{1},j_{2}),(j_{3},j_{4})}=\left(U_{H}\right)_{j_{1},j_{2},j_{3},j_{4}}$,
here, $(j_{1},j_{2})$ are treated as composite indices representing
the row indices, and $(j_{3},j_{4})$ are treated as composite indices
representing the column indices. Using the property of $R$ described
in Eq.(\ref{eq:R}), it is easy to see that $\left(R\mathrm{A}^{T}R\right)_{(j_{1},j_{2}),(j_{3},j_{4})}=\left(U_{H}\right)_{j_{4},j_{3},j_{2},j_{1}}$.
We then denote the direct product operation that only acts at the
superoperator level as ``$\boxtimes$''. For example:
\[
\begin{array}{rl}
	& \begin{array}{ccc}
		\left[\omega_{1}\otimes\mathbb{I},\right. & \cdots, & \left.\omega_{2L}\otimes\mathbb{I}\right]\end{array}\boxtimes\begin{array}{ccc}
		\left[\omega_{1}\otimes\mathbb{I},\right. & \cdots, & \left.\omega_{2L}\otimes\mathbb{I}\right]\end{array}\\
	= & \begin{array}{ccccccc}
		\left[\omega_{1}\omega_{1}\otimes\mathbb{I},\right. & \omega_{1}\omega_{2}\otimes\mathbb{I}, & \cdots, & \omega_{1}\omega_{2L}\otimes\mathbb{I}, & \omega_{2}\omega_{1}\otimes\mathbb{I}, & \cdots, & \left.\omega_{2L}\omega_{2L}\otimes\mathbb{I}\right]\end{array}
\end{array}
\]
 instead of $\left[\left(\omega_{1}\otimes\mathbb{I}\right)\otimes\left(\omega_{1}\otimes\mathbb{I}\right)\right.,\cdots,\left(\omega_{1}\otimes\mathbb{I}\right)\otimes\left(\omega_{2L}\otimes\mathbb{I}\right),\left(\omega_{2}\otimes\mathbb{I}\right)\otimes\left(\omega_{1}\otimes\mathbb{I}\right),\cdots,\left.\left(\omega_{2L}\otimes\mathbb{I}\right)\otimes\left(\omega_{2L}\otimes\mathbb{I}\right)\right]$.
Thus, we can rewrite $\hat{\mathcal{L}}_{add}$ as
\begin{equation}
\begin{array}{rl}
\hat{\mathcal{L}}_{add}= & \left(\mathbf{\Omega}_{r}\boxtimes\mathbf{\Omega}_{r}\right)\left(-i\mathrm{A}-\underset{\mu}{\sum}|U_{L,\mu}^{\dagger}\rrangle\llangle U_{L,\mu}|\right)\left(\mathbf{\Omega}_{c}\boxtimes\mathbf{\Omega}_{c}\right)\\
 & +\left(\mathbf{\bar{\Omega}}_{r}\boxtimes\mathbf{\bar{\Omega}}_{r}\right)\left(iR\mathrm{A}^{T}R-\underset{\mu}{\sum}|U_{L,\mu}^{T}\rrangle\llangle U_{L,\mu}^{*}|\right)\left(\mathbf{\bar{\Omega}}_{c}\boxtimes\mathbf{\bar{\Omega}}_{c}\right)\\
 & +\left(\mathbf{\Omega}_{r}\boxtimes\mathbf{\Omega}_{r}\right)\left(2\underset{\mu}{\sum}|U_{L,\mu}\rrangle\llangle U_{L,\mu}^{*}|\right)\left(\mathbf{\bar{\Omega}}_{c}\boxtimes\mathbf{\bar{\Omega}}_{c}\right),
\end{array}\label{eq:Lint_w}
\end{equation}
where $\llangle U_{L,\mu}|=\left(|U_{L,\mu}\rrangle\right)^{T}$ and
\[
\mathbf{\Omega}_{r}=\left[\begin{array}{cccc}
\omega_{1}\otimes\mathbb{I}, & \omega_{2}\otimes\mathbb{I}, & \cdots, & \omega_{2L}\otimes\mathbb{I}\end{array}\right],
\]
\[
\mathbf{\bar{\Omega}}_{r}=\left[\begin{array}{cccc}
\mathbb{I}\otimes\omega_{1}^{T}, & \mathbb{I}\otimes\omega_{2}^{T}, & \cdots, & \mathbb{I}\otimes\omega_{2L}^{T}\end{array}\right],
\]
\begin{equation}
\begin{array}{cc}
\mathbf{\Omega}_{c}=\left[\begin{array}{c}
\omega_{1}\otimes\mathbb{I}\\
\omega_{2}\otimes\mathbb{I}\\
\vdots\\
\omega_{2L}\otimes\mathbb{I}
\end{array}\right], & \mathbf{\bar{\Omega}}_{c}=\left[\begin{array}{c}
\mathbb{I}\otimes\omega_{1}^{T}\\
\mathbb{I}\otimes\omega_{2}^{T}\\
\vdots\\
\mathbb{I}\otimes\omega_{2L}^{T}
\end{array}\right]\end{array}.
\end{equation}
When $1\leqslant j\leqslant L$, we have $\omega_{j}=\frac{1}{\sqrt{2}}\left(a_{j}+a_{j}^{\dagger}\right)$,
$\omega_{L+j}=\frac{i}{\sqrt{2}}\left(a_{j}-a_{j}^{\dagger}\right)$,
we stipulate $a_{j}$ are real, then $\omega_{j}^{T}=\omega_{j}$,
$\omega_{L+j}^{T}=-\omega_{L+j}$,
\begin{equation}
\begin{array}{ll}
\omega_{j}\otimes\mathbb{I}=\hat{f}_{j}, & \omega_{L+j}\otimes\mathbb{I}=\hat{f}_{L+j},\\
\mathbb{I}\otimes\omega_{j}^{T}=i\hat{f}_{3L+j}\hat{\mathcal{P}}, & \mathbb{I}\otimes\omega_{L+j}^{T}=i\hat{f}_{2L+j}\hat{\mathcal{P}},
\end{array}
\end{equation}
so we can get
\[
\mathbf{\bar{\Omega}}_{r}\boxtimes\mathbf{\bar{\Omega}}_{r}=\left(\left[\begin{array}{ccc}
\hat{f}_{2L+1}, & \cdots, & \hat{f}_{4L}\end{array}\right]\boxtimes\left[\begin{array}{ccc}
\hat{f}_{2L+1}, & \cdots, & \hat{f}_{4L}\end{array}\right]\right)\left(\left[\begin{array}{cc}
 & I_{L}\\
I_{L}
\end{array}\right]\otimes\left[\begin{array}{cc}
 & I_{L}\\
I_{L}
\end{array}\right]\right),
\]
\[
\mathbf{\bar{\Omega}}_{c}\boxtimes\mathbf{\bar{\Omega}}_{c}=\left(\left[\begin{array}{cc}
 & I_{L}\\
I_{L}
\end{array}\right]\otimes\left[\begin{array}{cc}
 & I_{L}\\
I_{L}
\end{array}\right]\right)\left(\left[\begin{array}{c}
\hat{f}_{2L+1}\\
\vdots\\
\hat{f}_{4L}
\end{array}\right]\boxtimes\left[\begin{array}{c}
\hat{f}_{2L+1}\\
\vdots\\
\hat{f}_{4L}
\end{array}\right]\right),
\]
Eq.(\ref{eq:Lint_w}) can be rewritten as
\begin{equation}
\begin{array}{rl}
\hat{\mathcal{L}}_{add}= & \stackrel[j_{1},j_{2},j_{3},j_{4}=1]{2L}{\sum}\left[\bar{A}_{(j_{1},j_{2}),(j_{3},j_{4})}\hat{f}_{j_{1}}\hat{f}_{j_{2}}\hat{f}_{j_{3}}\hat{f}_{j_{4}}\right.\\
 & +\widetilde{B}_{(j_{1},j_{2}),(j_{3},j_{4})}\hat{f}_{j_{1}}\hat{f}_{j_{2}}\hat{f}_{2L+j_{3}}\hat{f}_{2L+j_{4}}\\
 & \left.+\widetilde{A}_{(j_{1},j_{2}),(j_{3},j_{4})}\hat{f}_{2L+j_{1}}\hat{f}_{2L+j_{2}}\hat{f}_{2L+j_{3}}\hat{f}_{2L+j_{4}}\right]
\end{array}\label{eq:L_int}
\end{equation}
with
\[
\bar{A}=-i\mathrm{A}-\underset{\mu}{\sum}|U_{L,\mu}^{\dagger}\rrangle\llangle U_{L,\mu}|,
\]
\[
\widetilde{B}=2\underset{\mu}{\sum}|U_{L,\mu}\rrangle\llangle U_{L,\mu}^{*}|\left(\left[\begin{array}{cc}
 & I_{L}\\
I_{L}
\end{array}\right]\otimes\left[\begin{array}{cc}
 & I_{L}\\
I_{L}
\end{array}\right]\right),
\]
\[
\widetilde{A}=\left(\left[\begin{array}{cc}
 & I_{L}\\
I_{L}
\end{array}\right]\otimes\left[\begin{array}{cc}
 & I_{L}\\
I_{L}
\end{array}\right]\right)\left(iR\mathrm{A}^{T}R-\underset{\mu}{\sum}|U_{L,\mu}^{T}\rrangle\llangle U_{L,\mu}^{*}|\right)\left(\left[\begin{array}{cc}
 & I_{L}\\
I_{L}
\end{array}\right]\otimes\left[\begin{array}{cc}
 & I_{L}\\
I_{L}
\end{array}\right]\right).
\]

\section{Discussion on the effect brought by $\hat{\mathcal{L}}_{add}$}

Here, we will discuss the effects brought by $\hat{\mathcal{L}}_{add}$.
Evidently, $[\hat{\mathcal{L}}_{add},\hat{\mathcal{P}}]=0$. Using
Eq.(\ref{eq:DO_Dt}), we get
\[
\frac{d}{dt}T_{j_{1},j_{2},\cdots,j_{2m}}=\llangle\mathbb{I}|\left[\stackrel[l=1]{2m}{\overrightarrow{\prod}}\hat{f}_{j_{l}},\left(\mathbf{\widehat{f}}F_{+}\widehat{\mathbf{f}}^{T}\right)+\hat{\mathcal{L}}_{add}\right]|\rho(t)\rrangle.
\]
Next, let's carefully analyze the additional part $\llangle\mathbb{I}|\left[\stackrel[l=1]{2m}{\overrightarrow{\prod}}\hat{f}_{j_{l}},\hat{\mathcal{L}}_{add}\right]|\rho(t)\rrangle$
compared to before.

\subsection{Second-order correlation function}

Let's first consider the case when $m=1$, $1\leqslant j_{1},j_{2}\leqslant2L$,
we have
\begin{equation}
\left[\hat{f}_{j_{1}}\hat{f}_{j_{2}},\stackrel[s_{1},s_{2},s_{3},s_{4}=1]{2L}{\sum}\widetilde{A}_{(s_{1},s_{2}),(s_{3},s_{4})}\left(\stackrel[l=1]{4}{\overrightarrow{\prod}}\hat{f}_{2L+s_{l}}\right)\right]=0,\label{eq:Lint_3}
\end{equation}
this indicates that the last term in Eq.(\ref{eq:L_int}) has no effect
when calculating the correlation function. Considering the first term
in Eq.(\ref{eq:L_int}), using the formula $\left[\hat{f}_{j_{1}}\hat{f}_{j_{2}},\hat{f}_{j_{3}}\hat{f}_{j_{4}}\right]=\delta_{j_{2},j_{3}}\hat{f}_{j_{1}}\hat{f}_{j_{4}}-\delta_{j_{1},j_{3}}\hat{f}_{j_{2}}\hat{f}_{j_{4}}+\delta_{j_{2},j_{4}}\hat{f}_{j_{3}}\hat{f}_{j_{1}}-\delta_{j_{1},j_{4}}\hat{f}_{j_{3}}\hat{f}_{j_{2}}$,
we obtain
\begin{equation}
\begin{array}{rl}
 & [\hat{f}_{j_{1}}\hat{f}_{j_{2}},\stackrel[s_{1},s_{2},s_{3},s_{4}=1]{2L}{\sum}\bar{A}_{(s_{1},s_{2}),(s_{3},s_{4})}\hat{f}_{s_{1}}\hat{f}_{s_{2}}\hat{f}_{s_{3}}\hat{f}_{s_{4}}]\\
= & \stackrel[s_{2},s_{3},s_{4}=1]{2L}{\sum}\left(\bar{A}_{(j_{2},s_{2}),(s_{3},s_{4})}\hat{f}_{j_{1}}\hat{f}_{s_{2}}\hat{f}_{s_{3}}\hat{f}_{s_{4}}+\bar{A}_{(s_{2},j_{2}),(s_{3},s_{4})}\hat{f}_{s_{2}}\hat{f}_{j_{1}}\hat{f}_{s_{3}}\hat{f}_{s_{4}}\right.\\
 & \left.+\bar{A}_{(s_{3},s_{4}),(j_{2},s_{2})}\hat{f}_{s_{3}}\hat{f}_{s_{4}}\hat{f}_{j_{1}}\hat{f}_{s_{2}}+\bar{A}_{(s_{3},s_{4}),(s_{2},j_{2})}\hat{f}_{s_{3}}\hat{f}_{s_{4}}\hat{f}_{s_{2}}\hat{f}_{j_{1}}\right)-\left.\cdots\right|_{j_{1}\leftrightarrow j_{2}}\\
= & \stackrel[s_{2},s_{3},s_{4}=1]{2L}{\sum}\left[\left(\bar{A}-R\bar{A}\right)_{(j_{2},s_{2}),(s_{3},s_{4})}\hat{f}_{j_{1}}\hat{f}_{s_{2}}\hat{f}_{s_{3}}\hat{f}_{s_{4}}+\left(\bar{A}-\bar{A}R\right)_{(s_{3},s_{4}),(j_{2},s_{2})}\hat{f}_{s_{3}}\hat{f}_{s_{4}}\hat{f}_{j_{1}}\hat{f}_{s_{2}}\right]\\
 & +\stackrel[s_{1},s_{2}=1]{2L}{\sum}\left(\bar{A}+\bar{A}^{T}\right)_{(j_{1},j_{2}),(s_{1},s_{2})}\hat{f}_{s_{1}}\hat{f}_{s_{2}}-\left.\cdots\right|_{j_{1}\leftrightarrow j_{2}}\\
= & \stackrel[s_{2},s_{3},s_{4}=1]{2L}{\sum}\left[\left(\mathbf{I}_{2}-R\right)\left(\bar{A}+\bar{A}^{T}\right)\right]_{(j_{2},s_{2}),(s_{3},s_{4})}\hat{f}_{j_{1}}\hat{f}_{s_{2}}\hat{f}_{s_{3}}\hat{f}_{s_{4}}\\
 & +\stackrel[s_{1},s_{2}=1]{2L}{\sum}\left(\left[\left(\mathbf{I}_{2}-R\right)\bar{A}^{T}\left(\mathbf{I}_{2}-R\right)\right]_{(j_{2},s_{1}),(j_{1},s_{2})}+\left(\bar{A}+\bar{A}^{T}\right)_{(j_{1},j_{2}),(s_{1},s_{2})}\right)\hat{f}_{s_{1}}\hat{f}_{s_{2}}\\
 & -\stackrel[s_{1},s_{2}=1]{2L}{\sum}\left[\left(\mathbf{I}_{2}-R\right)\bar{A}^{T}\left(\mathbf{I}_{2}-R\right)\right]_{(j_{2},s_{1}),(s_{1},s_{2})}\hat{f}_{j_{1}}\hat{f}_{s_{2}}\\
 & -\stackrel[s_{1}=1]{2L}{\sum}\left[\left(\mathbf{I}_{2}-R\right)\bar{A}^{T}\left(\mathbf{I}_{2}-R\right)\right]_{(j_{2},s_{1}),(j_{1},s_{1})}-\left.\cdots\right|_{j_{1}\leftrightarrow j_{2}},
\end{array}\label{eq:Lint_1}
\end{equation}
where $\left.\cdots\right|_{j_{1}\leftrightarrow j_{2}}$ represents
the new term obtained by swapping the subscripts $j_{1}$ and $j_{2}$
in all preceding terms. Next, considering the second term in Eq.(\ref{eq:L_int}),
we have
\begin{equation}
\begin{array}{rl}
 & [\hat{f}_{j_{1}}\hat{f}_{j_{2}},\stackrel[s_{1},s_{2},s_{3},s_{4}=1]{2L}{\sum}\widetilde{B}_{(s_{1},s_{2}),(s_{3},s_{4})}\hat{f}_{s_{1}}\hat{f}_{s_{2}}\hat{f}_{2L+s_{3}}\hat{f}_{2L+s_{4}}]\\
= & \stackrel[s_{2},s_{3},s_{4}=1]{2L}{\sum}\left(\widetilde{B}_{(j_{2},s_{2}),(s_{3},s_{4})}\hat{f}_{j_{1}}\hat{f}_{s_{2}}\hat{f}_{2L+s_{3}}\hat{f}_{2L+s_{4}}+\widetilde{B}_{(s_{2},j_{2}),(s_{3},s_{4})}\hat{f}_{s_{2}}\hat{f}_{j_{1}}\hat{f}_{2L+s_{3}}\hat{f}_{2L+s_{4}}\right)-\left.\cdots\right|_{j_{1}\leftrightarrow j_{2}}\\
= & \stackrel[s_{2},s_{3},s_{4}=1]{2L}{\sum}\left(\widetilde{B}-R\widetilde{B}\right)_{(j_{2},s_{2}),(s_{3},s_{4})}\hat{f}_{j_{1}}\hat{f}_{s_{2}}\hat{f}_{2L+s_{3}}\hat{f}_{2L+s_{4}}+\stackrel[s_{1},s_{2}=1]{2L}{\sum}\widetilde{B}_{(j_{1},j_{2}),(s_{1},s_{2})}\hat{f}_{2L+s_{1}}\hat{f}_{2L+s_{2}}-\left.\cdots\right|_{j_{1}\leftrightarrow j_{2}}.
\end{array}
\end{equation}
Using Eq.(\ref{eq:f2f}), we have
\[
\llangle\mathbb{I}|\hat{f}_{j_{1}}\hat{f}_{s_{2}}\hat{f}_{2L+s_{3}}\hat{f}_{2L+s_{4}}=\llangle\mathbb{I}|\hat{f}_{s_{4}^{\prime}}\hat{f}_{s_{3}^{\prime}}\hat{f}_{j_{1}}\hat{f}_{s_{2}}
\]
\[
\llangle\mathbb{I}|\hat{f}_{2L+s_{1}}\hat{f}_{2L+s_{2}}=\llangle\mathbb{I}|\hat{f}_{s_{2}^{\prime}}\hat{f}_{s_{1}^{\prime}}
\]
with
\begin{equation}
s_{l}^{\prime}=\left\{ \begin{array}{cc}
L+s_{l}, & 1\leqslant s_{l}\leqslant L,\\
s_{l}-L, & L<s_{l}\leqslant2L.
\end{array}\right.
\end{equation}
So we have
\begin{equation}
\begin{array}{rl}
 & \stackrel[s_{2},s_{3},s_{4}=1]{2L}{\sum}\left(\widetilde{B}-R\widetilde{B}\right)_{(j_{2},s_{2}),(s_{3},s_{4})}\llangle\mathbb{I}|\hat{f}_{j_{1}}\hat{f}_{s_{2}}\hat{f}_{2L+s_{3}}\hat{f}_{2L+s_{4}}\\
= & \stackrel[s_{2},s_{3},s_{4}=1]{2L}{\sum}\left(\widetilde{B}R-R\widetilde{B}R\right)_{(j_{2},s_{2}),(s_{4},s_{3})}\llangle\mathbb{I}|\hat{f}_{s_{4}^{\prime}}\hat{f}_{s_{3}^{\prime}}\hat{f}_{j_{1}}\hat{f}_{s_{2}}\\
= & \stackrel[s_{2}=1]{2L}{\sum}\llangle\mathbb{I}|\left[\left(\widetilde{B}R-R\widetilde{B}R\right)_{(j_{2},s_{2}),(:)}\left(\left[\begin{array}{c}
\hat{f}_{L+1}\\
\vdots\\
\hat{f}_{2L}\\
\hat{f}_{1}\\
\vdots\\
\hat{f}_{L}
\end{array}\right]\boxtimes\left[\begin{array}{c}
\hat{f}_{L+1}\\
\vdots\\
\hat{f}_{2L}\\
\hat{f}_{1}\\
\vdots\\
\hat{f}_{L}
\end{array}\right]\right)\right]\hat{f}_{j_{1}}\hat{f}_{s_{2}}\\
= & \stackrel[s_{2},s_{3},s_{4}=1]{2L}{\sum}\left[\left(\widetilde{B}R-R\widetilde{B}R\right)\left(\left[\begin{array}{cc}
 & I_{L}\\
I_{L}
\end{array}\right]\otimes\left[\begin{array}{cc}
 & I_{L}\\
I_{L}
\end{array}\right]\right)\right]_{(j_{2},s_{2}),(s_{3},s_{4})}\llangle\mathbb{I}|\hat{f}_{s_{3}}\hat{f}_{s_{4}}\hat{f}_{j_{1}}\hat{f}_{s_{2}}\\
= & 2\stackrel[s_{2},s_{3},s_{4}=1]{2L}{\sum}\left[\left(\mathbf{I}_{2}-R\right)\bar{B}\right]_{(j_{2},s_{2}),(s_{3},s_{4})}\llangle\mathbb{I}|\hat{f}_{s_{3}}\hat{f}_{s_{4}}\hat{f}_{j_{1}}\hat{f}_{s_{2}}\\
= & 2\stackrel[s_{2},s_{3},s_{4}=1]{2L}{\sum}\left[\left(\mathbf{I}_{2}-R\right)\bar{B}\right]_{(j_{2},s_{2}),(s_{3},s_{4})}\llangle\mathbb{I}|\hat{f}_{j_{1}}\hat{f}_{s_{2}}\hat{f}_{s_{3}}\hat{f}_{s_{4}}\\
 & +2\stackrel[s_{1},s_{2}=1]{2L}{\sum}\left[\left(\mathbf{I}_{2}-R\right)\bar{B}\left(\mathbf{I}_{2}-R\right)\right]_{(j_{2},s_{1}),(j_{1},s_{2})}\llangle\mathbb{I}|\hat{f}_{s_{1}}\hat{f}_{s_{2}}\\
 & -2\stackrel[s_{1},s_{2}=1]{2L}{\sum}\left[\left(\mathbf{I}_{2}-R\right)\bar{B}\left(\mathbf{I}_{2}-R\right)\right]_{(j_{2},s_{1}),(s_{1},s_{2})}\llangle\mathbb{I}|\hat{f}_{j_{1}}\hat{f}_{s_{2}}\\
 & -2\stackrel[s_{1}=1]{2L}{\sum}\left[\left(\mathbf{I}_{2}-R\right)\bar{B}\left(\mathbf{I}_{2}-R\right)\right]_{(j_{2},s_{1}),(j_{1},s_{1})}\llangle\mathbb{I}|
\end{array}
\end{equation}
with $\bar{B}=\underset{\mu}{\sum}\left(|U_{L,\mu}\rrangle\llangle U_{L,\mu}^{\dagger}|\right)$
and

\begin{equation}
\begin{array}{rl}
 & \stackrel[s_{1},s_{2}=1]{2L}{\sum}\widetilde{B}_{(j_{1},j_{2}),(s_{1},s_{2})}\llangle\mathbb{I}|\hat{f}_{2L+s_{1}}\hat{f}_{2l+s_{2}}\\
= & \stackrel[s_{1},s_{2}=1]{2L}{\sum}\left(\widetilde{B}R\right)_{(j_{1},j_{2}),(s_{2},s_{1})}\llangle\mathbb{I}|\hat{f}_{s_{2}^{\prime}}\hat{f}_{s_{1}^{\prime}}\\
= & 2\stackrel[s_{1},s_{2}=1]{2L}{\sum}\bar{B}_{(j_{1},j_{2}),(s_{1},s_{2})}\llangle\mathbb{I}|\hat{f}_{s_{1}}\hat{f}_{s_{2}}.
\end{array}
\end{equation}
From this, we can obtain the contribution from the second term in
Eq.(\ref{eq:L_int}),
\begin{equation}
\begin{array}{rl}
 & \llangle\mathbb{I}|\left[\hat{f}_{j_{1}}\hat{f}_{j_{2}},\stackrel[s_{1},s_{2},s_{3},s_{4}=1]{2L}{\sum}\widetilde{B}_{(s_{1},s_{2}),(s_{3},s_{4})}\hat{f}_{s_{1}}\hat{f}_{s_{2}}\hat{f}_{2L+s_{3}}\hat{f}_{2L+s_{4}}\right]\\
= & 2\stackrel[s_{2},s_{3},s_{4}=1]{2L}{\sum}\left[\left(\mathbf{I}_{2}-R\right)\bar{B}\right]_{(j_{2},s_{2}),(s_{3},s_{4})}\llangle\mathbb{I}|\hat{f}_{j_{1}}\hat{f}_{s_{2}}\hat{f}_{s_{3}}\hat{f}_{s_{4}}\\
 & +2\stackrel[s_{1},s_{2}=1]{2L}{\sum}\left(\left[\left(\mathbf{I}_{2}-R\right)\bar{B}\left(\mathbf{I}_{2}-R\right)\right]_{(j_{2},s_{1}),(j_{1},s_{2})}+\bar{B}_{(j_{1},j_{2}),(s_{1},s_{2})}\right)\llangle\mathbb{I}|\hat{f}_{s_{1}}\hat{f}_{s_{2}}\\
 & -2\stackrel[s_{1},s_{2}=1]{2L}{\sum}\left[\left(\mathbf{I}_{2}-R\right)\bar{B}\left(\mathbf{I}_{2}-R\right)\right]_{(j_{2},s_{1}),(s_{1},s_{2})}\llangle\mathbb{I}|\hat{f}_{j_{1}}\hat{f}_{s_{2}}\\
 & -2\stackrel[s_{1}=1]{2L}{\sum}\left[\left(\mathbf{I}_{2}-R\right)\bar{B}\left(\mathbf{I}_{2}-R\right)\right]_{(j_{2},s_{1}),(j_{1},s_{1})}\llangle\mathbb{I}|-\left.\cdots\right|_{j_{1}\leftrightarrow j_{2}}.
\end{array}\label{eq:Lint_2}
\end{equation}

Combining Eq.(\ref{eq:L_int},\ref{eq:Lint_3},\ref{eq:Lint_1},\ref{eq:Lint_2}),
we obtain
\begin{equation}
\begin{array}{rl}
 & \llangle\mathbb{I}|\left[\hat{f}_{j_{1}}\hat{f}_{j_{2}},\hat{\mathcal{L}}_{add}\right]\\
= & \stackrel[s_{2},s_{3},s_{4}=1]{2L}{\sum}\left[\left(\mathbf{I}_{2}-R\right)\left(\bar{A}+\bar{A}^{T}+2\bar{B}\right)\right]_{(j_{2},s_{2}),(s_{3},s_{4})}\llangle\mathbb{I}|\hat{f}_{j_{1}}\hat{f}_{s_{2}}\hat{f}_{s_{3}}\hat{f}_{s_{4}}\\
 & +\stackrel[s_{1},s_{2}=1]{2L}{\sum}\left(\left[\left(\mathbf{I}_{2}-R\right)\left(\bar{A}^{T}+2\bar{B}\right)\left(\mathbf{I}_{2}-R\right)\right]_{(j_{2},s_{1}),(j_{1},s_{2})}+\left[\bar{A}+\bar{A}^{T}+2\bar{B}\right]_{(j_{1},j_{2}),(s_{1},s_{2})}\right)\llangle\mathbb{I}|\hat{f}_{s_{1}}\hat{f}_{s_{2}}\\
 & -\stackrel[s_{1},s_{2}=1]{2L}{\sum}\left[\left(\mathbf{I}_{2}-R\right)\left(\bar{A}^{T}+2\bar{B}\right)\left(\mathbf{I}_{2}-R\right)\right]_{(j_{2},s_{1}),(s_{1},s_{2})}\llangle\mathbb{I}|\hat{f}_{j_{1}}\hat{f}_{s_{2}}\\
 & -\stackrel[s_{1}=1]{2L}{\sum}\left[\left(\mathbf{I}_{2}-R\right)\left(\bar{A}^{T}+2\bar{B}\right)\left(\mathbf{I}_{2}-R\right)\right]_{(j_{2},s_{1}),(j_{1},s_{1})}\llangle\mathbb{I}|-\left.\cdots\right|_{j_{1}\leftrightarrow j_{2}}.
\end{array}
\end{equation}
We label $\left(\mathbf{I}_{2}-R\right)\left(\bar{A}+\bar{A}^{T}+2\bar{B}\right)=\overline{C}$,
since $\stackrel[s_{2},s_{3},s_{4}=1]{2L}{\sum}\overline{C}_{(j_{2},s_{2}),(s_{3},s_{4})}\llangle\mathbb{I}|\hat{f}_{j_{1}}\hat{f}_{s_{2}}\hat{f}_{s_{3}}\hat{f}_{s_{4}}=\underset{s_{1},s_{2},s_{3},s_{4}}{\sum}\delta_{j_{1},s_{1}}\overline{C}_{(s_{1},s_{2}),(s_{3},s_{4})}\llangle\mathbb{I}|\hat{f}_{s_{1}}\hat{f}_{s_{2}}\hat{f}_{s_{3}}\hat{f}_{s_{4}}$,
we can divide all combinations of $s_{1},s_{2},s_{3},s_{4}$ into
two parts: combinations where $s_{1},s_{2},s_{3},s_{4}$ are all different,
and combinations where there are repeated numbers among $s_{1},s_{2},s_{3},s_{4}$.
We denote these as $\neq_{(s_{1},s_{2},s_{3},s_{4})}$ and $=_{(s_{1},s_{2},s_{3},s_{4})}$,
respectively. Then we have
\begin{equation}
\begin{array}{rl}
 & \stackrel[s_{2},s_{3},s_{4}=1]{2L}{\sum}\overline{C}_{(j_{2},s_{2}),(s_{3},s_{4})}\llangle\mathbb{I}|\hat{f}_{j_{1}}\hat{f}_{s_{2}}\hat{f}_{s_{3}}\hat{f}_{s_{4}}\\
= & \underset{\neq_{(s_{1},s_{2},s_{3},s_{4})}}{\sum}\delta_{j_{1},s_{1}}\overline{C}_{(j_{2},s_{2}),(s_{3},s_{4})}\llangle\mathbb{I}|\hat{f}_{s_{1}}\hat{f}_{s_{2}}\hat{f}_{s_{3}}\hat{f}_{s_{4}}\\
 & +\underset{=_{(s_{1},s_{2},s_{3},s_{4})}}{\sum}\delta_{j_{1},s_{1}}\overline{C}_{(j_{2},s_{2}),(s_{3},s_{4})}\llangle\mathbb{I}|\hat{f}_{s_{1}}\hat{f}_{s_{2}}\hat{f}_{s_{3}}\hat{f}_{s_{4}}.
\end{array}
\end{equation}
 It is easy to see that the presence of terms $\underset{\neq_{(s_{1},s_{2},s_{3},s_{4})}}{\sum}\delta_{j_{1},s_{1}}\overline{C}_{(j_{2},s_{2}),(s_{3},s_{4})}\llangle\mathbb{I}|\hat{f}_{s_{1}}\hat{f}_{s_{2}}\hat{f}_{s_{3}}\hat{f}_{s_{4}}$
causes the second-order correlation function to be non-closed. Therefore,
the closure of the second-order correlation function is equivalent
to
\begin{equation}
\underset{\neq_{(s_{1},s_{2},s_{3},s_{4})}}{\sum}\delta_{j_{1},s_{1}}\overline{C}_{(j_{2},s_{2}),(s_{3},s_{4})}\llangle\mathbb{I}|\hat{f}_{s_{1}}\hat{f}_{s_{2}}\hat{f}_{s_{3}}\hat{f}_{s_{4}}-\left.\cdots\right|_{j_{1}\leftrightarrow j_{2}}=0.
\end{equation}
We will later prove that this condition will manifest as
\begin{equation}
\left\{ \begin{array}{c}
U_{H}=0,\\
\Im\left(\underset{\mu}{\sum}U_{a,\mu}\otimes U_{a,\mu}^{*}\right)=0,
\end{array}\right.\label{eq:closure two}
\end{equation}
 with $U_{a,\mu}=U_{L,\mu}-U_{L,\mu}^{T}$.

Since
\begin{equation}
\begin{array}{rl}
\underset{=_{(s_{1},s_{2},s_{3},s_{4})}}{\sum}= & \stackrel[s_{1},s_{2},s_{3},s_{4}=1]{2L}{\sum}\left[\left(\delta_{s_{1},s_{2}}+\delta_{s_{1},s_{3}}+\delta_{s_{1},s_{4}}+\delta_{s_{2},s_{3}}+\delta_{s_{2},s_{4}}+\delta_{s_{3},s_{4}}\right)\right.\\
 & \left.-2\left(\delta_{s_{1},s_{2}}\delta_{s_{2},s_{3}}+\delta_{s_{1},s_{2}}\delta_{s_{2},s_{4}}+\delta_{s_{1},s_{3}}\delta_{s_{3},s_{4}}+\delta_{s_{2},s_{3}}\delta_{s_{3},s_{4}}\right)+3\delta_{s_{1},s_{2}}\delta_{s_{2},s_{3}}\delta_{s_{3},s_{4}}\right],
\end{array}
\end{equation}
we have
\begin{equation}
\begin{array}{rl}
 & \underset{=_{(s_{1},s_{2},s_{3},s_{4})}}{\sum}\delta_{j_{1},s_{1}}\overline{C}_{(j_{2},s_{2}),(s_{3},s_{4})}\hat{f}_{s_{1}}\hat{f}_{s_{2}}\hat{f}_{s_{3}}\hat{f}_{s_{4}}\\
= & \frac{1}{2}\stackrel[s_{1},s_{2}=1]{2L}{\sum}\left\{ \left[\left(R\overline{C}\right)_{(j_{1},j_{2}),(s_{1},s_{2})}+\left(\overline{C}R-\overline{C}\right)_{(j_{2},s_{1}),(j_{1},s_{2})}\right]\hat{f}_{s_{1}}\hat{f}_{s_{2}}\right.\\
 & \left.+\left[\left(\overline{C}-\overline{C}R\right)_{(j_{2},s_{1}),(s_{1},s_{2})}+\overline{C}_{(j_{2},s_{2}),(s_{1},s_{1})}\right]\hat{f}_{j_{1}}\hat{f}_{s_{2}}\right\} -\frac{1}{4}\overline{C}_{(j_{2},j_{1}),(j_{1},j_{1})}.
\end{array}
\end{equation}
When the second-order correlation function is closed, we have
\begin{equation}
\begin{array}{rl}
 & \llangle\mathbb{I}|\left[\hat{f}_{j_{1}}\hat{f}_{j_{2}},\hat{\mathcal{L}}_{add}\right]\\
= & \stackrel[s_{1},s_{2}=1]{2L}{\sum}\left(C_{\alpha}\right)_{(j_{1},s_{1}),(s_{2},j_{2})}\llangle\mathbb{I}|\hat{f}_{s_{1}}\hat{f}_{s_{2}}\\
 & +\stackrel[s_{1},s_{2}=1]{2L}{\sum}\left(C_{\beta}\right)_{(j_{2},s_{2}),(s_{1},s_{1})}\llangle\mathbb{I}|\hat{f}_{j_{1}}\hat{f}_{s_{2}}\\
 & +\stackrel[s_{1},s_{2}=1]{2L}{\sum}\left(C_{\beta}\right)_{(j_{1},s_{2}),(s_{1},s_{1})}\llangle\mathbb{I}|\hat{f}_{s_{2}}\hat{f}_{j_{2}}+\left(C_{\gamma}\right)_{j_{1},j_{2}}\llangle\mathbb{I}|
\end{array}\label{eq:second order}
\end{equation}
with
\[
\begin{array}{l}
C_{\alpha}=\left(\mathbf{I}_{2}-R\right)\left(\bar{B}+\bar{B}^{T}\right)\left(\mathbf{I}_{2}-R\right),\\
\left(C_{\beta}\right)_{(j_{2},s_{2}),(s_{1},s_{1})}=\frac{1}{2}\overline{C}_{(j_{2},s_{2}),(s_{1},s_{1})}+\frac{1}{2}\left[\left(\mathbf{I}_{2}-R\right)\left(\bar{A}-\bar{A}^{T}-2\bar{B}\right)\left(\mathbf{I}_{2}-R\right)\right]_{(j_{2},s_{1}),(s_{1},s_{2})},\\
\left(C_{\gamma}\right)_{j_{1},j_{2}}=\frac{1}{4}\left(\overline{C}_{(j_{1},j_{2}),(j_{2},j_{2})}-\overline{C}_{(j_{2},j_{1}),(j_{1},j_{1})}\right)+\stackrel[s_{1}=1]{2L}{\sum}\left\{ \left[\left(\mathbf{I}_{2}-R\right)\left(\bar{B}^{T}-\bar{B}\right)\left(\mathbf{I}_{2}-R\right)\right]_{(j_{1},s_{1}),(s_{1},j_{2})}-\frac{1}{2}\overline{C}_{(j_{1},j_{2}),(s_{1},s_{1})}\right\} .
\end{array}
\]
Using Eq.(\ref{eq:closure two}), we can express it in terms of $U_{a,\mu}=U_{L,\mu}-U_{L,\mu}^{T}$,
i.e.,
\begin{equation}
\begin{array}{l}
\begin{array}{l}
C_{\alpha}=\underset{\mu}{\sum}\left(|U_{a,\mu}\rrangle\llangle U_{a,\mu}^{\dagger}|+|U_{a,\mu}^{\dagger}\rrangle\llangle U_{a,\mu}|\right),\\
C_{\beta}=\frac{1}{2}\underset{\mu}{\sum}\left(|U_{a,\mu}\rrangle\llangle U_{L,\mu}^{\dagger}|-|U_{a,\mu}^{\dagger}\rrangle\llangle U_{L,\mu}|\right)-\Re\left[\underset{\mu}{\sum}U_{a,\mu}\otimes U_{a,\mu}^{*}\right],\\
\left(C_{\gamma}\right)_{j_{1},j_{2}}=\Re\left\{ \underset{\mu}{\sum}\left[\left(U_{a,\mu}\right)_{j_{1},j_{2}}\left(\frac{1}{2}(U_{L,\mu}^{*})_{j_{1},j_{1}}+\frac{1}{2}(U_{L,\mu}^{*})_{j_{2},j_{2}}-Tr(U_{L,\mu}^{*})\right)\right]\right\} .
\end{array}\end{array}\label{eq:C_alp_bet_gam}
\end{equation}

\subsection{Closure condition for the second-order correlation function}

Here, let's analyze carefully the closure condition for the second-order
correlation function, which is
\[
\underset{\neq_{(s_{1},s_{2},s_{3},s_{4})}}{\sum}\delta_{j_{1},s_{1}}\overline{C}_{(j_{2},s_{2}),(s_{3},s_{4})}\llangle\mathbb{I}|\hat{f}_{s_{1}}\hat{f}_{s_{2}}\hat{f}_{s_{3}}\hat{f}_{s_{4}}-\left.\cdots\right|_{j_{1}\leftrightarrow j_{2}}=0.
\]
Expanding $\underset{\neq_{(s_{1},s_{2},s_{3},s_{4})}}{\sum}$ specifically,
we have
\begin{equation}
\underset{\begin{array}{c}
s_{2}<s_{3}<s_{4}\\
s_{2},s_{3},s_{4}\neq j_{1}
\end{array}}{\sum}\left[\left(\overline{C}-\overline{C}R\right)_{(j_{2},s_{2}),(s_{3},s_{4})}+\left(\overline{C}-\overline{C}R\right)_{(j_{2},s_{3}),(s_{4},s_{2})}+\left(\overline{C}-\overline{C}R\right)_{(j_{2},s_{4}),(s_{2},s_{3})}\right]\llangle\mathbb{I}|\hat{f}_{j_{1}}\hat{f}_{s_{2}}\hat{f}_{s_{3}}\hat{f}_{s_{4}}-\left.\cdots\right|_{j_{1}\leftrightarrow j_{2}}=0,\label{eq:closure before}
\end{equation}
when $j_{1}=j_{2}$, it is obviously true. Therefore, we only need
to consider the case when $j_{1}\neq j_{2}$. In this case, we first
choose $s_{2}<s_{3}<s_{4}$, $s_{2}\neq j_{1},j_{2}$, $s_{3}\neq j_{1},j_{2}$,
$s_{4}\neq j_{1},j_{2}$, when we observe the coefficient in front
of $\llangle\mathbb{I}|\hat{f}_{j_{1}}\hat{f}_{s_{2}}\hat{f}_{s_{3}}\hat{f}_{s_{4}}$,
we get
\[
\left(\overline{C}-\overline{C}R\right)_{(j_{2},s_{2}),(s_{3},s_{4})}+\left(\overline{C}-\overline{C}R\right)_{(j_{2},s_{3}),(s_{4},s_{2})}+\left(\overline{C}-\overline{C}R\right)_{(j_{2},s_{4}),(s_{2},s_{3})}=0.
\]
Additionally, if we select $s_{2}=j_{2}<s_{3}<s_{4}$ and $s_{3}\neq j_{1},j_{2}$,
$s_{4}\neq j_{1},j_{2}$ in the summation part in front of Eq.(\ref{eq:closure before}),
while selecting $s_{2}=j_{1}<s_{3}<s_{4}$ and $s_{3}\neq j_{1},j_{2}$,
$s_{4}\neq j_{1},j_{2}$ in $\left.\cdots\right|_{j_{1}\leftrightarrow j_{2}}$,
and then observe the coefficient in front of $\llangle\mathbb{I}|\hat{f}_{j_{1}}\hat{f}_{j_{2}}\hat{f}_{s_{3}}\hat{f}_{s_{4}}$,
we obtain
\[
\left(\overline{C}-\overline{C}R\right)_{(j_{2},j_{2}),(s_{3},s_{4})}+\left(\overline{C}-\overline{C}R\right)_{(j_{2},s_{3}),(s_{4},j_{2})}+\left(\overline{C}-\overline{C}R\right)_{(j_{2},s_{4}),(j_{2},s_{3})}+\left.\cdots\right|_{j_{1}\leftrightarrow j_{2}}=0,
\]
since we previously set $j_{1}\neq j_{2}$, and this closure condition
should hold for all possible $j_{1}$ and $j_{2}$, we have
\[
\left(\overline{C}-\overline{C}R\right)_{(j_{2},j_{2}),(s_{3},s_{4})}+\left(\overline{C}-\overline{C}R\right)_{(j_{2},s_{3}),(s_{4},j_{2})}+\left(\overline{C}-\overline{C}R\right)_{(j_{2},s_{4}),(j_{2},s_{3})}=0.
\]
Similarly, we obtain
\[
\left(\overline{C}-\overline{C}R\right)_{(j_{2},s_{2}),(j_{2},s_{4})}+\left(\overline{C}-\overline{C}R\right)_{(j_{2},j_{2}),(s_{4},s_{2})}+\left(\overline{C}-\overline{C}R\right)_{(j_{2},s_{4}),(s_{2},j_{2})}=0
\]
and
\[
\left(\overline{C}-\overline{C}R\right)_{(j_{2},s_{2}),(s_{3},j_{2})}+\left(\overline{C}-\overline{C}R\right)_{(j_{2},s_{3}),(j_{2},s_{2})}+\left(\overline{C}-\overline{C}R\right)_{(j_{2},j_{2}),(s_{2},s_{3})}=0.
\]

Therefore, we can say that Eq.(\ref{eq:closure before}) is equivalent
to
\begin{equation}
	\forall\ s_{2}<s_{3}<s_{4},\ \left(\overline{C}-\overline{C}R\right)_{(j_{2},s_{2}),(s_{3},s_{4})}+\left(\overline{C}-\overline{C}R\right)_{(j_{2},s_{3}),(s_{4},s_{2})}+\left(\overline{C}-\overline{C}R\right)_{(j_{2},s_{4}),(s_{2},s_{3})}=0,\label{eq:closure_middle}
\end{equation}
if we exchange the indices $s_{2}$ and $s_{3}$, it becomes $\forall\ s_{3}<s_{2}<s_{4}$,
\[
\begin{array}{rl}
 & \left(\overline{C}-\overline{C}R\right)_{(j_{2},s_{3}),(s_{2},s_{4})}+\left(\overline{C}-\overline{C}R\right)_{(j_{2},s_{2}),(s_{4},s_{3})}+\left(\overline{C}-\overline{C}R\right)_{(j_{2},s_{4}),(s_{3},s_{2})}\\
= & -\left(\overline{C}-\overline{C}R\right)_{(j_{2},s_{2}),(s_{3},s_{4})}-\left(\overline{C}-\overline{C}R\right)_{(j_{2},s_{3}),(s_{4},s_{2})}-\left(\overline{C}-\overline{C}R\right)_{(j_{2},s_{4}),(s_{2},s_{3})}\\
= & 0,
\end{array}
\]
that is $\forall\ s_{3}<s_{2}<s_{4},\ \left(\overline{C}-\overline{C}R\right)_{(j_{2},s_{2}),(s_{3},s_{4})}+\left(\overline{C}-\overline{C}R\right)_{(j_{2},s_{3}),(s_{4},s_{2})}+\left(\overline{C}-\overline{C}R\right)_{(j_{2},s_{4}),(s_{2},s_{3})}=0$,
this indicates that in Eq.(\ref{eq:closure_middle}), the indices
$s_{2}$ and $s_{3}$ have the same status. Similarly, by selecting
other exchange methods for the indices $s_{2}$, $s_{3}$ and $s_{4}$,
we can see that they all have the same status in Eq.(\ref{eq:closure_middle}),
therefore, their order is not important, and Eq.(\ref{eq:closure_middle})
can be rewritten as
\begin{equation}
	\forall\ s_{2}\neq s_{3},\ s_{2}\neq s_{4},\ s_{3}\neq s_{4},\ \left(\overline{C}-\overline{C}R\right)_{(j_{2},s_{2}),(s_{3},s_{4})}+\left(\overline{C}-\overline{C}R\right)_{(j_{2},s_{3}),(s_{4},s_{2})}+\left(\overline{C}-\overline{C}R\right)_{(j_{2},s_{4}),(s_{2},s_{3})}=0.\label{eq:closure_middle2}
\end{equation}

When $s_{2}=s_{3}$,
\[
\begin{array}{rl}
 & \left(\overline{C}-\overline{C}R\right)_{(j_{2},s_{2}),(s_{3},s_{4})}+\left(\overline{C}-\overline{C}R\right)_{(j_{2},s_{3}),(s_{4},s_{2})}+\left(\overline{C}-\overline{C}R\right)_{(j_{2},s_{4}),(s_{2},s_{3})}\\
= & \left(\overline{C}-\overline{C}R\right)_{(j_{2},s_{2}),(s_{2},s_{4})}+\left(\overline{C}-\overline{C}R\right)_{(j_{2},s_{2}),(s_{4},s_{2})}+\left(\overline{C}-\overline{C}R\right)_{(j_{2},s_{4}),(s_{2},s_{2})}\\
= & 0,
\end{array}
\]
this means that when $s_{2}=s_{3}$, Eq.(\ref{eq:closure_middle2})
naturally holds. Similarly, when $s_{2}=s_{4}$ and $s_{3}=s_{4}$,
Eq.(\ref{eq:closure_middle2}) also naturally holds. Therefore, Eq.(\ref{eq:closure_middle2})
is equivalent to
\begin{equation}
\forall\ s_{1},s_{2},s_{3},s_{4},\ C_{(s_{1},s_{2}),(s_{3},s_{4})}+C_{(s_{1},s_{3}),(s_{4},s_{2})}+C_{(s_{1},s_{4}),(s_{2},s_{3})}=0\label{eq:closure_end}
\end{equation}
with $C=\overline{C}-\overline{C}R$. Since $\left(U_{H}\right)_{j_{1},j_{2},j_{3},j_{4}}=\mathrm{A}_{(j_{1},j_{2}),(j_{3},j_{4})}$
is a fully antisymmetric tensor, we know that $C=-8i\mathrm{A}+\underset{\mu}{\sum}\left(|U_{a,\mu}\rrangle\llangle U_{a,\mu}^{\dagger}|-|U_{a,\mu}^{\dagger}\rrangle\llangle U_{a,\mu}|\right)$
with $U_{a,\mu}=U_{L,\mu}-U_{L,\mu}^{T}$, Eq.(\ref{eq:closure_end})
can be rewritten as
\[
-24i\mathrm{A}+\underset{\mu}{\sum}\left(|U_{a,\mu}\rrangle\llangle U_{a,\mu}^{\dagger}|-|U_{a,\mu}^{\dagger}\rrangle\llangle U_{a,\mu}|\right)+\underset{\mu}{\sum}\left(U_{a,\mu}\otimes U_{a,\mu}^{*}-U_{a,\mu}^{\dagger}\otimes U_{a,\mu}^{T}\right)+\underset{\mu}{\sum}\left(U_{a,\mu}\otimes U_{a,\mu}^{\dagger}-U_{a,\mu}^{\dagger}\otimes U_{a,\mu}\right)R=0.
\]
Using $\left(U_{a,\mu}\otimes U_{a,\mu}^{\dagger}\right)R=R\left(U_{a,\mu}^{\dagger}\otimes U_{a,\mu}\right)$
and $U_{a,\mu}^{T}=-U_{a,\mu}$, we can decompose it into symmetric
and antisymmetric parts, yielding
\[
\left\{ \begin{array}{c}
\mathrm{A}=\frac{1}{12}\Im\left(\underset{\mu}{\sum}U_{a,\mu}\otimes U_{a,\mu}^{*}\right),\\
\Im\left(\underset{\mu}{\sum}U_{a,\mu}\otimes U_{a,\mu}^{*}\right)=\Im\left(\underset{\mu}{\sum}|U_{a,\mu}^{*}\rrangle\llangle U_{a,\mu}|\right).
\end{array}\right.
\]
Here, the second condition, using the first condition and $\left(U_{H}\right)_{j_{1},j_{2},j_{3},j_{4}}=\mathrm{A}_{(j_{1},j_{2}),(j_{3},j_{4})}$,
can be written as $\left(U_{H}\right)_{j_{1},j_{2},j_{3},j_{4}}=\left(U_{H}\right)_{j_{1},j_{3},j_{2},j_{4}}$,
since we have defined $U_{H}$ as a fully antisymmetric tensor, we
have $U_{H}=0$. From this, we deduce that the closure condition for
the second-order correlation function is
\begin{equation}
\left\{ \begin{array}{c}
U_{H}=0,\\
\Im\left(\underset{\mu}{\sum}U_{a,\mu}\otimes U_{a,\mu}^{*}\right)=0.
\end{array}\right.\label{eq:closure_two}
\end{equation}

\subsection{Higher-order correlation function}

Understanding the closure condition for the second-order correlation
function, we can now consider higher-order correlation functions.
We have
\[
\llangle\mathbb{I}|\left[\stackrel[l=1]{2m}{\overrightarrow{\prod}}\hat{f}_{j_{l}},\hat{\mathcal{L}}_{add}\right]|\rho(t)\rrangle=\stackrel[k=1]{m}{\sum}\llangle\mathbb{I}|\left(\stackrel[l=1]{2k-2}{\overrightarrow{\prod}}\hat{f}_{j_{l}}\right)\left[\hat{f}_{j_{2k-1}}\hat{f}_{j_{2k}},\hat{\mathcal{L}}_{add}\right]\left(\stackrel[l=2k+1]{2m}{\overrightarrow{\prod}}\hat{f}_{j_{l}}\right)|\rho(t)\rrangle.
\]
First, let us consider one of the terms, namely the $\llangle\mathbb{I}|\left(\stackrel[l=1]{2k-2}{\overrightarrow{\prod}}\hat{f}_{j_{l}}\right)\left[\hat{f}_{j_{2k-1}}\hat{f}_{j_{2k}},\hat{\mathcal{L}}_{add}\right]$
part. Referring to the derivation process of Eq.(\ref{eq:Lint_1},\ref{eq:Lint_2}),
as well as the resulting Eq.(\ref{eq:second order}) after the closure
of the second-order correlation function, it is easy to see that when
the second-order correlation function is closed (Eq.(\ref{eq:closure_two})
holds), we have
\begin{equation}
\begin{array}{rl}
 & \llangle\mathbb{I}|\left(\stackrel[l=1]{2k-2}{\overrightarrow{\prod}}\hat{f}_{j_{l}}\right)\left[\hat{f}_{j_{2k-1}}\hat{f}_{j_{2k}},\hat{\mathcal{L}}_{add}\right]\\
= & 2\stackrel[s_{2},s_{3},s_{4}=1]{2L}{\sum}\left[\left(\mathbf{I}_{2}-R\right)\bar{B}\right]_{(j_{2k},s_{2}),(s_{3},s_{4})}\llangle\mathbb{I}|\left[\hat{f}_{s_{3}}\hat{f}_{s_{4}},\left(\stackrel[l=1]{2k-2}{\overrightarrow{\prod}}\hat{f}_{j_{l}}\right)\right]\hat{f}_{j_{2k-1}}\hat{f}_{s_{2}}\\
 & +2\stackrel[s_{1},s_{2}=1]{2L}{\sum}\bar{B}_{(j_{2k-1},j_{2k}),(s_{1},s_{2})}\llangle\mathbb{I}|\left[\hat{f}_{s_{1}}\hat{f}_{s_{2}},\left(\stackrel[l=1]{2k-2}{\overrightarrow{\prod}}\hat{f}_{j_{l}}\right)\right]\\
 & -\left.\cdots\right|_{j_{2k-1}\leftrightarrow j_{2k}}\\
 & +\stackrel[s_{1},s_{2}=1]{2L}{\sum}\left(C_{\alpha}\right)_{(j_{2k-1},s_{1}),(s_{2},j_{2k})}\llangle\mathbb{I}|\left(\stackrel[l=1]{2k-2}{\overrightarrow{\prod}}\hat{f}_{j_{l}}\right)\hat{f}_{s_{1}}\hat{f}_{s_{2}}\\
 & +\stackrel[s_{1},s_{2}=1]{2L}{\sum}\left(C_{\beta}\right)_{(j_{2k},s_{2}),(s_{1},s_{1})}\llangle\mathbb{I}|\left(\stackrel[l=1]{2k-2}{\overrightarrow{\prod}}\hat{f}_{j_{l}}\right)\hat{f}_{j_{2k-1}}\hat{f}_{s_{2}}\\
 & +\stackrel[s_{1},s_{2}=1]{2L}{\sum}\left(C_{\beta}\right)_{(j_{2k-1},s_{2}),(s_{1},s_{1})}\llangle\mathbb{I}|\left(\stackrel[l=1]{2k-2}{\overrightarrow{\prod}}\hat{f}_{j_{l}}\right)\hat{f}_{s_{2}}\hat{f}_{j_{2k}}\\
 & +\left(C_{\gamma}\right)_{j_{2k-1},j_{2k}}\llangle\mathbb{I}|\left(\stackrel[l=1]{2k-2}{\overrightarrow{\prod}}\hat{f}_{j_{l}}\right).
\end{array}\label{eq:high_Lint}
\end{equation}
It should be noted that $\left.\cdots\right|_{j_{2k-1}\leftrightarrow j_{2k}}$
here only represents the new term obtained by exchanging $j_{2k-1}$
and $j_{2k}$ in the first two terms. Given
\[
\begin{array}{rl}
 & \left[\hat{f}_{s_{1}}\hat{f}_{s_{2}},\left(\stackrel[l=1]{2k-2}{\overrightarrow{\prod}}\hat{f}_{j_{l}}\right)\right]\\
= & \stackrel[r=1]{k-1}{\sum}\left(\stackrel[l=1]{2r-2}{\overrightarrow{\prod}}\hat{f}_{j_{l}}\right)\left[\hat{f}_{s_{1}}\hat{f}_{s_{2}},\hat{f}_{j_{2r-1}}\hat{f}_{j_{2r}}\right]\left(\stackrel[l=2r+1]{2k-2}{\overrightarrow{\prod}}\hat{f}_{j_{l}}\right)\\
= & \stackrel[r=1]{k-1}{\sum}\left(\stackrel[l=1]{2r-2}{\overrightarrow{\prod}}\hat{f}_{j_{l}}\right)\left(\delta_{s_{2},j_{2r-1}}\hat{f}_{s_{1}}\hat{f}_{j_{2r}}+\delta_{s_{2},j_{2r}}\hat{f}_{j_{2r-1}}\hat{f}_{s_{1}}-\delta_{s_{1},j_{2r-1}}\hat{f}_{s_{2}}\hat{f}_{j_{2r}}-\delta_{s_{1},j_{2r}}\hat{f}_{j_{2r-1}}\hat{f}_{s_{2}}\right)\left(\stackrel[l=2r+1]{2k-2}{\overrightarrow{\prod}}\hat{f}_{j_{l}}\right),
\end{array}
\]
we obtain
\begin{equation}
\begin{array}{rl}
 & \stackrel[s_{1},s_{2}=1]{2L}{\sum}\bar{B}_{(j_{2k-1},j_{2k}),(s_{1},s_{2})}\left[\hat{f}_{s_{1}}\hat{f}_{s_{2}},\left(\stackrel[l=1]{2k-2}{\overrightarrow{\prod}}\hat{f}_{j_{l}}\right)\right]\\
= & \stackrel[r=1]{k-1}{\sum}\left(\stackrel[l=1]{2r-2}{\overrightarrow{\prod}}\hat{f}_{j_{l}}\right)\stackrel[s_{1}=1]{2L}{\sum}\left[\left(\bar{B}R-\bar{B}\right)_{(j_{2k-1},j_{2k}),(j_{2r-1},s_{1})}\hat{f}_{s_{1}}\hat{f}_{j_{2r}}+\left(\bar{B}R-\bar{B}\right)_{(j_{2k-1},j_{2k}),(j_{2r},s_{1})}\hat{f}_{j_{2r-1}}\hat{f}_{s_{1}}\right]\left(\stackrel[l=2r+1]{2k-2}{\overrightarrow{\prod}}\hat{f}_{j_{l}}\right),
\end{array}\label{eq:B_bar}
\end{equation}
since $\bar{B}=\underset{\mu}{\sum}\left(|U_{L,\mu}\rrangle\llangle U_{L,\mu}^{\dagger}|\right)$,
if we denote $U_{a,\mu}=U_{L,\mu}-U_{L,\mu}^{T}$, Eq.(\ref{eq:B_bar})
can be rewritten as
\[
\begin{array}{rl}
 & \stackrel[s_{1},s_{2}=1]{2L}{\sum}\bar{B}_{(j_{2k-1},j_{2k}),(s_{1},s_{2})}\left[\hat{f}_{s_{1}}\hat{f}_{s_{2}},\left(\stackrel[l=1]{2k-2}{\overrightarrow{\prod}}\hat{f}_{j_{l}}\right)\right]\\
= & \stackrel[r=1]{k-1}{\sum}\stackrel[s_{1}=1]{2L}{\sum}\underset{\mu}{\sum}\left(\stackrel[l=1]{2r-2}{\overrightarrow{\prod}}\hat{f}_{j_{l}}\right)\left(U_{a,\mu}^{*}\right)_{j_{2r-1},s_{1}}\hat{f}_{s_{1}}\hat{f}_{j_{2r}}\left(\stackrel[l=2r+1]{2k-2}{\overrightarrow{\prod}}\hat{f}_{j_{l}}\right)\left(U_{L,\mu}\right)_{j_{2k-1},j_{2k}}\\
 & +\stackrel[r=1]{k-1}{\sum}\stackrel[s_{1}=1]{2L}{\sum}\underset{\mu}{\sum}\left(\stackrel[l=1]{2r-2}{\overrightarrow{\prod}}\hat{f}_{j_{l}}\right)\left(U_{a,\mu}^{*}\right)_{j_{2r},s_{1}}\hat{f}_{j_{2r-1}}\hat{f}_{s_{1}}\left(\stackrel[l=2r+1]{2k-2}{\overrightarrow{\prod}}\hat{f}_{j_{l}}\right)\left(U_{L,\mu}\right)_{j_{2k-1},j_{2k}}\\
= & \stackrel[r=1]{2k-2}{\sum}\stackrel[s_{1}=1]{2L}{\sum}\underset{\mu}{\sum}\left(\stackrel[l=1]{r-1}{\overrightarrow{\prod}}\hat{f}_{j_{l}}\right)\left(U_{a,\mu}^{*}\right)_{j_{r},s_{1}}\hat{f}_{s_{1}}\left(\stackrel[l=r+1]{2k-2}{\overrightarrow{\prod}}\hat{f}_{j_{l}}\right)\left(U_{L,\mu}\right)_{j_{2k-1},j_{2k}}.
\end{array}
\]
Thus, we rewrite the first two terms of Eq.(\ref{eq:high_Lint}) as
\[
\begin{array}{rl}
 & 2\stackrel[s_{2},s_{3},s_{4}=1]{2L}{\sum}\left[\left(\mathbf{I}_{2}-R\right)\bar{B}\right]_{(j_{2k},s_{2}),(s_{3},s_{4})}\llangle\mathbb{I}|\left[\hat{f}_{s_{3}}\hat{f}_{s_{4}},\left(\stackrel[l=1]{2k-2}{\overrightarrow{\prod}}\hat{f}_{j_{l}}\right)\right]\hat{f}_{j_{2k-1}}\hat{f}_{s_{2}}\\
 & +2\stackrel[s_{1},s_{2}=1]{2L}{\sum}\bar{B}_{(j_{2k-1},j_{2k}),(s_{1},s_{2})}\llangle\mathbb{I}|\left[\hat{f}_{s_{1}}\hat{f}_{s_{2}},\left(\stackrel[l=1]{2k-2}{\overrightarrow{\prod}}\hat{f}_{j_{l}}\right)\right]\\
= & 2\stackrel[s_{1},s_{2}=1]{2L}{\sum}\underset{\mu}{\sum}\stackrel[r=1]{2k-2}{\sum}\llangle\mathbb{I}|\left(\stackrel[l=1]{r-1}{\overrightarrow{\prod}}\hat{f}_{j_{l}}\right)\left(U_{a,\mu}^{*}\right)_{j_{r},s_{1}}\hat{f}_{s_{1}}\left(\stackrel[l=r+1]{2k-1}{\overrightarrow{\prod}}\hat{f}_{j_{l}}\right)\left(U_{a,\mu}\right)_{j_{2k},s_{2}}\hat{f}_{s_{2}}\\
 & +2\stackrel[s_{1}=1]{2L}{\sum}\underset{\mu}{\sum}\stackrel[r=1]{2k-2}{\sum}\llangle\mathbb{I}|\left(\stackrel[l=1]{r-1}{\overrightarrow{\prod}}\hat{f}_{j_{l}}\right)\left(U_{a,\mu}^{*}\right)_{j_{r},s_{1}}\hat{f}_{s_{1}}\left(\stackrel[l=r+1]{2k-2}{\overrightarrow{\prod}}\hat{f}_{j_{l}}\right)\left(U_{L,\mu}\right)_{j_{2k-1},j_{2k}},
\end{array}
\]
so we have
\[
\begin{array}{rl}
 & \llangle\mathbb{I}|\left[\stackrel[l=1]{2m}{\overrightarrow{\prod}}\hat{f}_{j_{l}},\hat{\mathcal{L}}_{add}\right]|\rho(t)\rrangle\\
= & 2\stackrel[k=1]{m}{\sum}\stackrel[s_{1},s_{2}=1]{2L}{\sum}\underset{\mu}{\sum}\stackrel[r=1]{2k-2}{\sum}\llangle\mathbb{I}|\left(\stackrel[l=1]{r-1}{\overrightarrow{\prod}}\hat{f}_{j_{l}}\right)\left(U_{a,\mu}^{*}\right)_{j_{r},s_{1}}\hat{f}_{s_{1}}\left(\stackrel[l=r+1]{2k-1}{\overrightarrow{\prod}}\hat{f}_{j_{l}}\right)\left(U_{a,\mu}\right)_{j_{2k},s_{2}}\hat{f}_{s_{2}}\left(\stackrel[l=2k+1]{2m}{\overrightarrow{\prod}}\hat{f}_{j_{l}}\right)|\rho(t)\rrangle\\
 & +2\stackrel[k=1]{m}{\sum}\stackrel[s_{1}=1]{2L}{\sum}\underset{\mu}{\sum}\stackrel[r=1]{2k-2}{\sum}\llangle\mathbb{I}|\left(\stackrel[l=1]{r-1}{\overrightarrow{\prod}}\hat{f}_{j_{l}}\right)\left(U_{a,\mu}^{*}\right)_{j_{r},s_{1}}\hat{f}_{s_{1}}\left(\stackrel[l=r+1]{2k-2}{\overrightarrow{\prod}}\hat{f}_{j_{l}}\right)\left(U_{L,\mu}\right)_{j_{2k-1},j_{2k}}\left(\stackrel[l=2k+1]{2m}{\overrightarrow{\prod}}\hat{f}_{j_{l}}\right)|\rho(t)\rrangle\\
 & -\left.\cdots\right|_{j_{2k-1}\leftrightarrow j_{2k}}\\
 & +\stackrel[k=1]{m}{\sum}\stackrel[s_{1},s_{2}=1]{2L}{\sum}\left(C_{\alpha}\right)_{(j_{2k-1},s_{1}),(s_{2},j_{2k})}\llangle\mathbb{I}|\left(\stackrel[l=1]{2k-2}{\overrightarrow{\prod}}\hat{f}_{j_{l}}\right)\hat{f}_{s_{1}}\hat{f}_{s_{2}}\left(\stackrel[l=2k+1]{2m}{\overrightarrow{\prod}}\hat{f}_{j_{l}}\right)|\rho(t)\rrangle\\
 & +\stackrel[k=1]{m}{\sum}\stackrel[s_{1},s_{2}=1]{2L}{\sum}\left(C_{\beta}\right)_{(j_{2k},s_{2}),(s_{1},s_{1})}\llangle\mathbb{I}|\left(\stackrel[l=1]{2k-2}{\overrightarrow{\prod}}\hat{f}_{j_{l}}\right)\hat{f}_{j_{2k-1}}\hat{f}_{s_{2}}\left(\stackrel[l=2k+1]{2m}{\overrightarrow{\prod}}\hat{f}_{j_{l}}\right)|\rho(t)\rrangle\\
 & +\stackrel[k=1]{m}{\sum}\stackrel[s_{1},s_{2}=1]{2L}{\sum}\left(C_{\beta}\right)_{(j_{2k-1},s_{2}),(s_{1},s_{1})}\llangle\mathbb{I}|\left(\stackrel[l=1]{2k-2}{\overrightarrow{\prod}}\hat{f}_{j_{l}}\right)\hat{f}_{s_{2}}\hat{f}_{j_{2k}}\left(\stackrel[l=2k+1]{2m}{\overrightarrow{\prod}}\hat{f}_{j_{l}}\right)|\rho(t)\rrangle\\
 & +\stackrel[k=1]{m}{\sum}\left(C_{\gamma}\right)_{j_{2k-1},j_{2k}}\llangle\mathbb{I}|\left(\stackrel[l=1]{2k-2}{\overrightarrow{\prod}}\hat{f}_{j_{l}}\right)\left(\stackrel[l=2k+1]{2m}{\overrightarrow{\prod}}\hat{f}_{j_{l}}\right)|\rho(t)\rrangle\\
= & 2\stackrel[k=1]{m}{\sum}\stackrel[s_{1},s_{2}=1]{2L}{\sum}\underset{\mu}{\sum}\stackrel[r=1]{2k-2}{\sum}\llangle\mathbb{I}|\left(\stackrel[l=1]{r-1}{\overrightarrow{\prod}}\hat{f}_{j_{l}}\right)\left(U_{a,\mu}^{*}\right)_{j_{r},s_{1}}\hat{f}_{s_{1}}\left(\stackrel[l=r+1]{2k-2}{\overrightarrow{\prod}}\hat{f}_{j_{l}}\right)*\\
 & \left[\hat{f}_{j_{2k-1}}\left(U_{a,\mu}\right)_{j_{2k},s_{2}}\hat{f}_{s_{2}}+\left(U_{a,\mu}\right)_{j_{2k-1},s_{2}}\hat{f}_{s_{2}}\hat{f}_{j_{2k}}\right]\left(\stackrel[l=2k+1]{2m}{\overrightarrow{\prod}}\hat{f}_{j_{l}}\right)|\rho(t)\rrangle\\
 & +\stackrel[k=1]{m}{\sum}\stackrel[s_{1},s_{2}=1]{2L}{\sum}\left(C_{\alpha}\right)_{(j_{2k-1},s_{1}),(s_{2},j_{2k})}\llangle\mathbb{I}|\left(\stackrel[l=1]{2k-2}{\overrightarrow{\prod}}\hat{f}_{j_{l}}\right)\hat{f}_{s_{1}}\hat{f}_{s_{2}}\left(\stackrel[l=2k+1]{2m}{\overrightarrow{\prod}}\hat{f}_{j_{l}}\right)|\rho(t)\rrangle\\
 & +\stackrel[k=1]{m}{\sum}\stackrel[s_{1},s_{2}=1]{2L}{\sum}\left(C_{\beta}\right)_{(j_{2k},s_{2}),(s_{1},s_{1})}\llangle\mathbb{I}|\left(\stackrel[l=1]{2k-2}{\overrightarrow{\prod}}\hat{f}_{j_{l}}\right)\hat{f}_{j_{2k-1}}\hat{f}_{s_{2}}\left(\stackrel[l=2k+1]{2m}{\overrightarrow{\prod}}\hat{f}_{j_{l}}\right)|\rho(t)\rrangle\\
 & +\stackrel[k=1]{m}{\sum}\stackrel[s_{1},s_{2}=1]{2L}{\sum}\left(C_{\beta}\right)_{(j_{2k-1},s_{2}),(s_{1},s_{1})}\llangle\mathbb{I}|\left(\stackrel[l=1]{2k-2}{\overrightarrow{\prod}}\hat{f}_{j_{l}}\right)\hat{f}_{s_{2}}\hat{f}_{j_{2k}}\left(\stackrel[l=2k+1]{2m}{\overrightarrow{\prod}}\hat{f}_{j_{l}}\right)|\rho(t)\rrangle\\
 & +\stackrel[k=1]{m}{\sum}\left(C_{\gamma}\right)_{j_{2k-1},j_{2k}}\llangle\mathbb{I}|\left(\stackrel[l=1]{2k-2}{\overrightarrow{\prod}}\hat{f}_{j_{l}}\right)\left(\stackrel[l=2k+1]{2m}{\overrightarrow{\prod}}\hat{f}_{j_{l}}\right)|\rho(t)\rrangle\\
= & \left(\mathbf{F}_{add,2m}\bm{\left|T_{2m}\right\rangle }\right)_{j_{1},j_{2},\cdots,j_{2m}}+\stackrel[k=1]{m}{\sum}\left(C_{\gamma}\right)_{j_{2k-1},j_{2k}}\left(\bm{\left|T_{2m-2}\right\rangle }\right)_{j_{1},\cdots,j_{2k-2},j_{2k+1},\cdots,j_{2m}}\\
= & \left(\mathbf{F}_{add,2m}\bm{\left|T_{2m}\right\rangle }\right)_{j_{1},j_{2},\cdots,j_{2m}}+\stackrel[k=1]{m}{\sum}\left\{ \left[R_{2k}\otimes\mathbf{I}_{2m-2k}\right]^{2}\left[C_{\gamma}\otimes\bm{\left|T_{2m-2}\right\rangle }\right]\right\} _{j_{1},j_{2},\cdots,j_{2m}},
\end{array}
\]
where
\[
\begin{array}{rl}
\mathbf{F}_{add,2m}= & \stackrel[k=1]{m}{\sum}\mathbf{I}_{2k-2}\otimes\left(F_{\alpha}+F_{\beta}\otimes\mathbf{I}_{1}+\mathbf{I}_{1}\otimes F_{\beta}\right)\otimes\mathbf{I}_{2m-2k}\\
 & +2\stackrel[k=1]{m}{\sum}\underset{\mu}{\sum}\left(\stackrel[r=1]{2k-2}{\sum}\mathbf{I}_{r-1}\otimes U_{a,\mu}^{*}\otimes\mathbf{I}_{2k-r-2}\right)\otimes\left(U_{a,\mu}\otimes\mathbf{I}_{1}+\mathbf{I}_{1}\otimes U_{a,\mu}\right)\otimes\mathbf{I}_{2m-2k},
\end{array}
\]
\[
\left\{ \begin{array}{l}
\left(F_{\alpha}\right)_{(j_{2k-1},j_{2k}),(s_{1},s_{2})}=\left(C_{\alpha}\right)_{(j_{2k-1},s_{1}),(s_{2},j_{2k})},\\
\left(F_{\beta}\right)_{j_{2k-1},s_{2}}=\stackrel[s_{1}=1]{2L}{\sum}\left(C_{\beta}\right)_{(j_{2k-1},s_{2}),(s_{1},s_{1})}.
\end{array}\right.
\]
To maintain consistent notation, we denote $F_{\gamma}=C_{\gamma}$,
further using Eq.(\ref{eq:C_alp_bet_gam}), we can obtain
\[
\left\{ \begin{array}{l}
F_{\alpha}=2\Re\left[\underset{\mu}{\sum}\left(U_{a,\mu}\otimes U_{a,\mu}^{*}\right)\right],\\
F_{\beta}=\Re\left[\underset{\mu}{\sum}\left(U_{a,\mu}U_{a,\mu}^{*}+Tr(U_{t,\mu}^{*})*U_{a,\mu}\right)\right],\\
\left(F_{\gamma}\right)_{j_{1},j_{2}}=\Re\left\{ \underset{\mu}{\sum}\left[\left(U_{a,\mu}\right)_{j_{1},j_{2}}\left(\frac{1}{2}(U_{L,\mu}^{*})_{j_{1},j_{1}}+\frac{1}{2}(U_{L,\mu}^{*})_{j_{2},j_{2}}-Tr(U_{L,\mu}^{*})\right)\right]\right\} .
\end{array}\right.
\]
With the help of Eq.(\ref{eq:closure_two}), we can rewrite $\mathbf{F}_{add,2m}$
as
\[
\mathbf{F}_{add,2m}=\underset{\mu}{\sum}\mathbf{U_{a,\mu}^{(2m)}}\left(\mathbf{U_{a,\mu}^{(2m)}}\right)^{*}+\Re\left[\underset{\mu}{\sum}\left(Tr(U_{L,\mu}^{*})*\mathbf{U_{a,\mu}^{(2m)}}\right)\right]
\]
with $\mathbf{U_{a,\mu}^{(2m)}}=\stackrel[k=1]{2m}{\sum}\mathbf{I}_{k-1}\otimes U_{a,\mu}\otimes\mathbf{I}_{2m-k}$.
Thus, we obtain the dynamical equation satisfied by the higher-order
correlation function when the second-order correlation function is
closed as
\[
\frac{d}{dt}\bm{\left|T_{2m,t}\right\rangle }=\mathbf{F}_{M,2m}\bm{\left|T_{2m,t}\right\rangle }+\mathbf{G}_{M,2m}\bm{\left|T_{2m-2,t}\right\rangle }
\]
with $\mathbf{F}_{M,2m}=\mathbf{F}_{2m}+\mathbf{F}_{add,2m}$, $\mathbf{G}_{M,2m}=\mathbf{G}_{2m}+\stackrel[k=1]{m}{\sum}\left[R_{2k}\otimes\mathbf{I}_{2m-2k}\right]^{2}\left(|F_{\gamma}\rrangle\otimes\mathbf{I}_{2m-2}\right)$.

\end{widetext}

\end{document}